\documentclass[10pt,preprint2]{emulateapj}

\usepackage{amsmath}
\usepackage{amsfonts}
\usepackage{amssymb}

\usepackage{array,multirow,makecell}

\slugcomment{Draft version of \today }

\shorttitle{High-$z$ in MACS0717}
\shortauthors{N. Laporte et al.}

\begin{document}


\title{Young Galaxy Candidates in the Hubble Frontier Fields:\\
III. MACSJ0717.5+3745}


\author{N. Laporte\altaffilmark{1}, L. Infante\altaffilmark{1}, P. Troncoso Iribarren\altaffilmark{1,16}, W. Zheng\altaffilmark{2}, A. Molino\altaffilmark{5,6},  F. E. Bauer\altaffilmark{1,3,4}, D. Bina\altaffilmark{9}, Tom Broadhurst\altaffilmark{18,19}, I. Chilingarian\altaffilmark{10,17}, S. Garcia\altaffilmark{1,16}, S. Kim\altaffilmark{1,16}, R. Marques-Chaves\altaffilmark{7,8}, J. Moustakas\altaffilmark{15},R. Pell\'o\altaffilmark{9},  I. P\'erez-Fournon\altaffilmark{7,8}, X. Shu\altaffilmark{11,12} , A. Streblyanska\altaffilmark{7,8} and  A. Zitrin\altaffilmark{13,14}   }


\altaffiltext{1}{Instituto de Astrof\'isica and Centro de Astroingenier\'ia, Facultad de F\'isica, Pontificia Universidad
Cat\'olica de Chile, Vicu\~na Mackenna 4860, 7820436 Macul, Santiago, Chile}
\altaffiltext{2}{Department of Physics and Astronomy, Johns Hopkins University, Baltimore, MD 21218}
\altaffiltext{3}{Millennium Institute of Astrophysics, Vicu\~na Mackenna 4860, 7820436 Macul, Santiago, Chile}
\altaffiltext{4}{Space Science Institute, Boulder, CO 80301}
\altaffiltext{5}{Instituto de Astronom\'ia, Geof\'isica e Ci\^encias Atmosf\'ericas, Universidade de S\~ao Paulo, Cidade Universit\'aria, 05508-090, S\~ao Paulo, Brazil}
\altaffiltext{6}{Instituto de Astrof\'isica de Andaluc\'a - CSIC, Glorieta de la Astronom\'ia, s/n. E-18008, Granada, Spain}
\altaffiltext{7}{ Instituto de Astrof\'{\i}sica de Canarias (IAC), E-38200 La Laguna, Tenerife, Spain.}
\altaffiltext{8}{Departamento de Astrof\'{\i}sica, Universidad de La Laguna (ULL), E-38205 La Laguna, Tenerife, Spain}
\altaffiltext{9}{IRAP, CNRS - 14 Avenue Edouard Belin - F-31400 Toulouse, France}
\altaffiltext{10}{Smithsonian Astrophysical Observatory, 60 Garden St. MS09, Cambridge MA 02138, USA}
\altaffiltext{11}{CAS Key Laboratory for Research in Galaxies and Cosmology, Department of Astronomy, University of Science and Technology of China,
Hefei, Anhui 230026, China}
\altaffiltext{12}{CEA Saclay, DSM / Irfu / Service d'Astrophysique, Orme des Merisiers, F-91191 Gif-sur-Yvette Cedex, France}
\altaffiltext{13}{Cahill Center for Astronomy and Astrophysics, California Institute of Technology, MC 249-17, Pasadena, CA 91125, USA}
\altaffiltext{14}{Hubble Fellow}
\altaffiltext{15}{Department of Physics and Astronomy, Siena College, Loudonville, NY 12211}
\altaffiltext{16}{Centro de Astro-Ingeniería, Pontificia Universidad Católica de Chile, Santiago, Chile}
\altaffiltext{17}{Sternberg Astronomical Institute, Moscow State University, 13 Universitetsky prospect, Moscow, 119992, Russia}
\altaffiltext{18}{Department of Theoretical Physics, University of Basque Country UPV/EHU, Bilbao, Spain}
\altaffiltext{19}{IKERBASQUE, Basque Foundation for Science, Bilbao, Spain}

\begin{abstract}
In this paper we present the results of our search for and study of $z \gtrsim 6$ galaxy candidates behind the third \textit{Frontier Fields} (FF) cluster, MACSJ0717.5+3745, and its parallel field, combining data from \textit{Hubble} and \textit{Spitzer}. We select 39 candidates using the Lyman Break technique, for which the clear non-detection in optical make the extreme mid-$z$ interlopers hypothesis unlikely.  We also take benefit from $z \gtrsim 6$ samples selected using previous Frontier Fields datasets of Abell 2744 and MACS0416 to improve the constraints on the properties of very high-redshift objects. We compute the redshift and the physical properties, such emission lines properties, star formation rate, reddening, and stellar mass for all Frontier Fields objects from their spectral energy distribution using templates including nebular emission lines. We study the relationship between several physical properties and confirm the trend already observed in previous surveys for evolution of star formation rate with galaxy mass, and between the size and the UV luminosity of our candidates. 
The analysis of the evolution of the UV Luminosity Function with redshift seems more compatible with an evolution of density. Moreover, no robust $z\ge$8.5 object is selected behind the cluster field, and few $z$$\sim$9 candidates have been selected in the two previous datasets from this legacy survey, suggesting a strong evolution in the number density of galaxies between $z$$\sim$8 and 9. Thanks to the use of the lensing cluster, we study the evolution of the star formation rate density produced by galaxies with L$>$0.03L$^{\star}$, and confirm the strong decrease observed between $z$$\sim$8 and 9. 
\end{abstract}



\section{Introduction}
One of the most intriguing challenges of the coming decade is undoubtedly the search for the first stars and galaxies that appeared a few hundreds million years after the Big-Bang. During the last ten years major advances have been made in the quest of the first galaxies in our Universe, thanks to the commissioning of new facilities such as the WFC3/HST \citep{2011ApJS..193...27W}, WIRCam/CFHT \citep{2004SPIE.5492..978P},  MOSFIRE/Keck \citep{2012SPIE.8446E..0JM} or X-Shooter/VLT \citep{2011A&A...536A.105V}, and the arrival of extremely deep surveys as for example the \textit{Hubble Ultra Deep Field} \citep{2006AJ....132.1729B}, \textit{Cluster  Lensing And Supernova survey with Hubble} (CLASH - \citealt{2012ApJS..199...25P}) or the \textit{Brightest of Reionizing Galaxies Survey} (BoRG - \citealt{2011ApJ...727L..39T}). Among all the results achieved, one can mention the great leap forward in the number of $z \gtrsim 6.5$ sources known that count in several hundreds at $z \sim 7$ (\citealt{2010ApJ...709L..16O}, \citealt{2010ApJ...725.1587B}, \citealt{2013ApJ...768..196S}), hundred at $z \sim 8$ (\citealt{2012ApJ...760..108B}, \citealt{2013ApJ...777L..19L}, \citealt{2012ApJ...761..177Y}) and dozens at $z \gtrsim 8.5$ (\citealt{2013MNRAS.432.2696M}, \citealt{2014ApJ...786..108O}), with the most distant spectroscopically confirmed galaxy at $z$$=$8.68 \citep{2015arXiv150702679Z}, and the highest photometrically selected galaxy at $z\sim$11 \citep{2013ApJ...762...32C}  . 

The main interest of studying the first galaxies is to constrain the role they played during the reionization of the Universe. This period corresponds to the reionization of neutral hydrogen in the early Universe by UV photons (e.g. \citealt{2013ASSL..396...45Z}). The end of this phenomenon is relatively well defined by observations of quasars at $5.9 \leq z \leq 6.4$ (\citealt{2015MNRAS.447..499M}, \citealt{2013MNRAS.428.3058S}). The most likely sources of reionization are primeval galaxies, however the contribution from galaxies detected in current surveys is not sufficient to match the ionizing background required to reionize the Universe at $z \sim 6$ (\citealt{1999ApJ...514..648M}, \citealt{2015arXiv150501846D}). But recent studies have demonstrated that abundant fainter galaxies, below the detection limits of current instruments, may have played a crucial role in this process \citep{2015arXiv150308228B}. One way to start studying these fainter objects before the arrival of future extremely large telescopes is to harness gravitational lensing which amplifies their light \citep{2011A&ARv..19...47K}. Several studies have already demonstrated the interest of using galaxy clusters to detect the faintest objects during the first billion years of the Universe (\citealt{2010A&A...509A.105M}, \citealt{2012Natur.489..406Z}, \citealt{2015ApJ...801...44Z}), but the number of faint sources is not sufficient to give robust constraints on their properties during the epoch of reionization. 

The number of relatively bright objects, however, starts to be sufficient to at least study the bright-end of the UV Luminosity Function (LF) and its evolution over the first billion years of the Universe. The study of the luminosity distribution of galaxies at lower redshift confirms that the UV LF is well fitted by a \citet{1976ApJ...203..297S} function \citep{2012A&A...539A..31C}. However analysis of several deep blank fields suggested that the bright-part of the UV LF at $z > 6$ deviates from the standard shape (\citealt{2014MNRAS.440.2810B}, \citealt{2014arXiv1410.5439F}) with an over-density of bright objects. This could be explained by a decrease of the Active Galactic Nucleus (AGN) feedback that usually suppresses star formation in these galaxies, limiting their growth, and thus the number of very massive (and bright) galaxies. If this over-density of bright objects in the early Universe is confirmed, it could demonstrate that the role of AGN at such epochs is likely to be less important than at low-redshift \citep{2013A&A...556A..55I} and could be a crucial key to improve our understanding in the reionization process. But other deep blank fields are needed to validate this conclusion. 

In September 2013, the new flagship program of the \textit{Hubble Space Telescope}, namely the \textit{Frontier Fields} (FF), started observations \citep{2014AAS...22325401L}. Thanks to the HST design, two fields for each of the six clusters planned for this program, are observed simultaneously: one centered on a gravitationally lensed cluster  and the second, ``Parallel field", located a few arcmins from the main field. The combination of these two types of fields allow to study the most distant star-forming objects in the early Universe over a large range of luminosities. To date, four clusters have been completed (namely Abell 2744, MACSJ0416.1-2403, MACSJ0717.5+3745 and MACS1149.5+2223) and the analysis of the two first datasets already proved the great potential of this project. For example, one of the most distant objects currently known ($z \sim 10$)  was selected from the FF images and shows multiple images that strongly confirm its photometric redshift \citep{2014ApJ...793L..12Z}. Dozens of objects have already been studied and led to an improvement of the constraints on the faint-end slope of the UV LF (\citealt{2014ApJ...795...93Z}, \citealt{2014ApJ...786...60A}, \citealt{2014A&A...562L...8L}, \citealt{2014arXiv1409.1228O}, \citealt{2014arXiv1412.1472M}, \citealt{2015ApJ...799...12I}, \citealt{2015ApJ...800...18A}, \citealt{2015A&A...575A..92L}, \citealt{2015ApJ...804..103K}). More recently,  \citet{Infante2015} published the discovery of a strongly amplified $z$$\sim$10 candidate ($\mu$$\sim$20) probing, for the first time, the extreme faint-end of the UV LF at $z$$\sim$10.

In this paper, we present samples selected in MACSJ0717.5+3745 cluster and parallel fields, and we combine them with similar studies made in Abell 2744 (\citealt{2014ApJ...795...93Z}, \citealt{2015ApJ...804..103K}) and MACS0416 \citep{Infante2015} to obtain a uniform sample and to add robust constraints on the UV LF over the redshift range covered by this legacy program. The organization is as follows: in section \ref{data} we describe the dataset; in section \ref{selection} the criteria we used to select candidates that are described in section \ref{sample}; and in section \ref{discussion} we estimate the contamination rate of our samples (sec. \ref{contaminants}), computed the shape of UV LF and the evolution of the SFRd as seen from half of the FF observations (sec. \ref{LF}). Throughout this paper, we use a concordance cosmology ($\Omega_M = 0.3$, $\Omega_{\Lambda} = 0.7$ and $H_0 = 70$ km/s/Mpc) and all magnitudes are quoted in the AB system \citep{1983ApJ...266..713O}.

\section{Data properties}
\label{data}
The FF project is carried out using HST Director's Discretionary Time and will use 840 orbits during Cycle 21, 22 and 23 with six strong-lensing galaxy clusters as the main targets. For each cluster the final dataset is composed of 3 images from ACS/HST (F435W, F606W and F814W) and 4 images from WFC3/HST (F105W, F125W, F140W and F160W) reaching depths of $\sim$29 mag at 5$\sigma$
in a 0\farcs4 diameter aperture. In this study, we used the final data release on MACSJ0717.5+3745 ($z = 0.551$, \citealt{2004ApJ...609L..49E}, \citealt{2013ApJ...777...43M}) made public on April 1$^{st}$ 2015. This third cluster in the FF list has been observed by HST through several observing programs, mainly related to CLASH (ID: 12103, PI: M. Postman) and the FFs (ID: 13498, PI: J. Lotz). We measured the depth of each image using non-overlapping empty 0\farcs2 radius apertures distributed over the field.

We matched the HST data with deep Spitzer/IRAC images obtained from observations  (ID: 90259)  carried out from August 2013 to January 2015 combined with archival data from November 2007 to June 2013. We merged all the raw files using MOPEX tasks, and obtained a final image of 449ksec in each band reaching a 5$\sigma$ magnitude of AB$\sim$25.6. Table \ref{data.properties} displays exposure time, depth and filters properties of the dataset we used. 


\begin{table*}
\caption{Properties of the \textit{HST} and \textit{Spitzer} data.}             
\label{data.properties}      
\centering                          
\begin{tabular}{c | c c c | c c | c c }        
\hline\hline                 
Filter & $\lambda_{central}$ & $\Delta\lambda$ & Instrument & t$^{C}_{exp}$ & m$_{C}$(5$\sigma$)   	     &  t$^{P}_{exp}$ & m$_{P}$(5$\sigma$)   \\    
         &  [$\mu$m]  & [nm]                   &   	  & [ks]          &     [AB]      	     &  [ks]          &     [AB]            \\          
 \hline                        
 F435 W	& 0.431 & 72.9   & ACS    &  54.5	     &     29.1	     &  45.7	     &	     29.3        \\
 F606W	& 0.589 & 156.5 & ACS    &  33.5	     &     29.3	     &	  25.0	     &   29.4        \\
 F814W	& 0.811 & 165.7 & ACS    &  129.9	   &   	    29.2    	     &  105.5	     &   29.5  \\
 F105W	& 1.050 &	300.0 & WFC3 &  67.3		 &   28.4	  	     &	 79.9 	     &	28.8 \\
 F125W	& 1.250 &	300.0 & WFC3 &  33.1		 &   28.4	  	     &  34.2	     &		28.4\\
 F140W	& 1.400 & 400.0 & WFC3 &  27.6		 &   28.4	  	     &  34.2	     &	28.6	 \\
 F160W	& 1.545 & 290.0 & WFC3 &  66.1		 &   28.4	  	     &  79.9	     &	29.0	 \\
\hline
 3.6	& 3.550 & 750.0   & IRAC &	449 & 25.6  	     & - 	     & -\\
 4.5 	& 4.493 & 1015.0 & IRAC &  449 & 25.6  	     &  	-     & -\\
\hline
\hline                                   
\end{tabular}

Columns: (1) filter ID, (2) filter central wavelength, (3) filter FWHM, (4) Instrument, (5, 6) exposure time and 5$\sigma$ depth in a 0\farcs2 radius aperture for {\it HST} data and 1\farcs4 radius aperture for IRAC images for the cluster centered field, (7, 8) same as column 5 and 6 but for the parallel field. $P$ stands for parallel field and $C$ for cluster field.
\end{table*}



\section {Source extraction }

We used SExtractor (version 2.19.5, \citealt{1996A&AS..117..393B}) to extract sources from our images with extraction parameters defined in \citet{2015A&A...575A..92L}. WFC3 catalogs were built on double image mode using a sum of NIR data as the detection image, and then matched to single image mode ACS catalogs with TOPCAT \citep{2005ASPC..347...29T} in order to avoid any false detections at optical wavelength. Non-detections were measured on the original images, whereas colors were measured on psf-matched data using Tiny Tim models \citep{2011SPIE.8127E..0JK}.We measured colors in SExtractor MAG\_AUTO apertures defined with Kron\_fact=1.2 and min\_radius=1.7, and we applied aperture corrections using SExractor MAG\_AUTO with default parameters (Kron\_fact =2.5 and min\_radius=3.5) in the F160W band as reference. Error bars were estimated from the noise measured in several empty 0.4\arcsec diameter apertures distributed around each candidate. 

Because we are using extraction parameters defined to select small and faint objects, our catalogs contain several false detections, such as pixels in the haloes of bright galaxies, pixels in high background level regions, etc. Thus visual inspection is needed to remove all these non-real sources. We also confirmed the non-detection of all our candidates on optical stacked images.

\section{Selection of high-$z$ candidates}
\label{selection}

\begin{figure*}
   \centering
           \includegraphics[width=8.2cm]{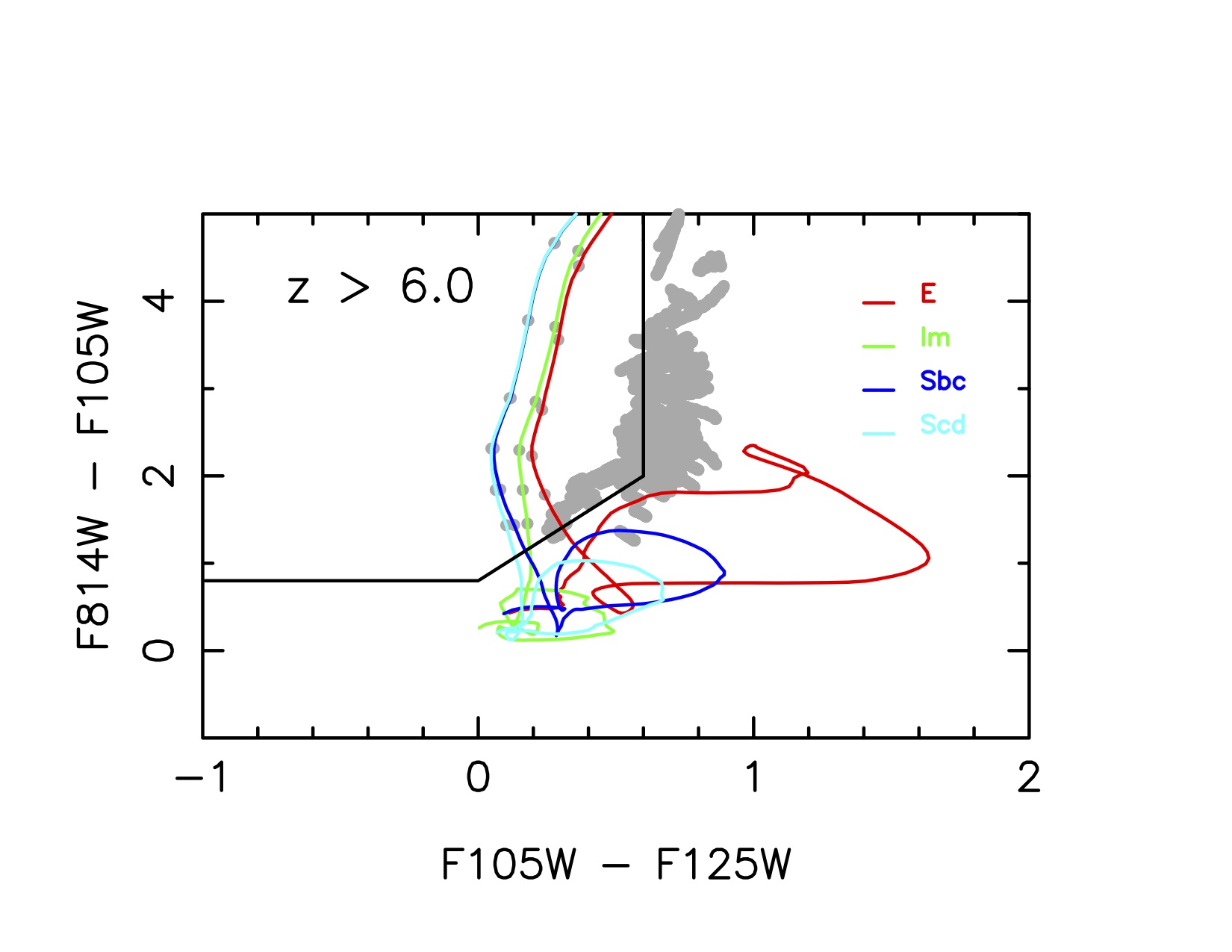}
           \includegraphics[width=8.2cm]{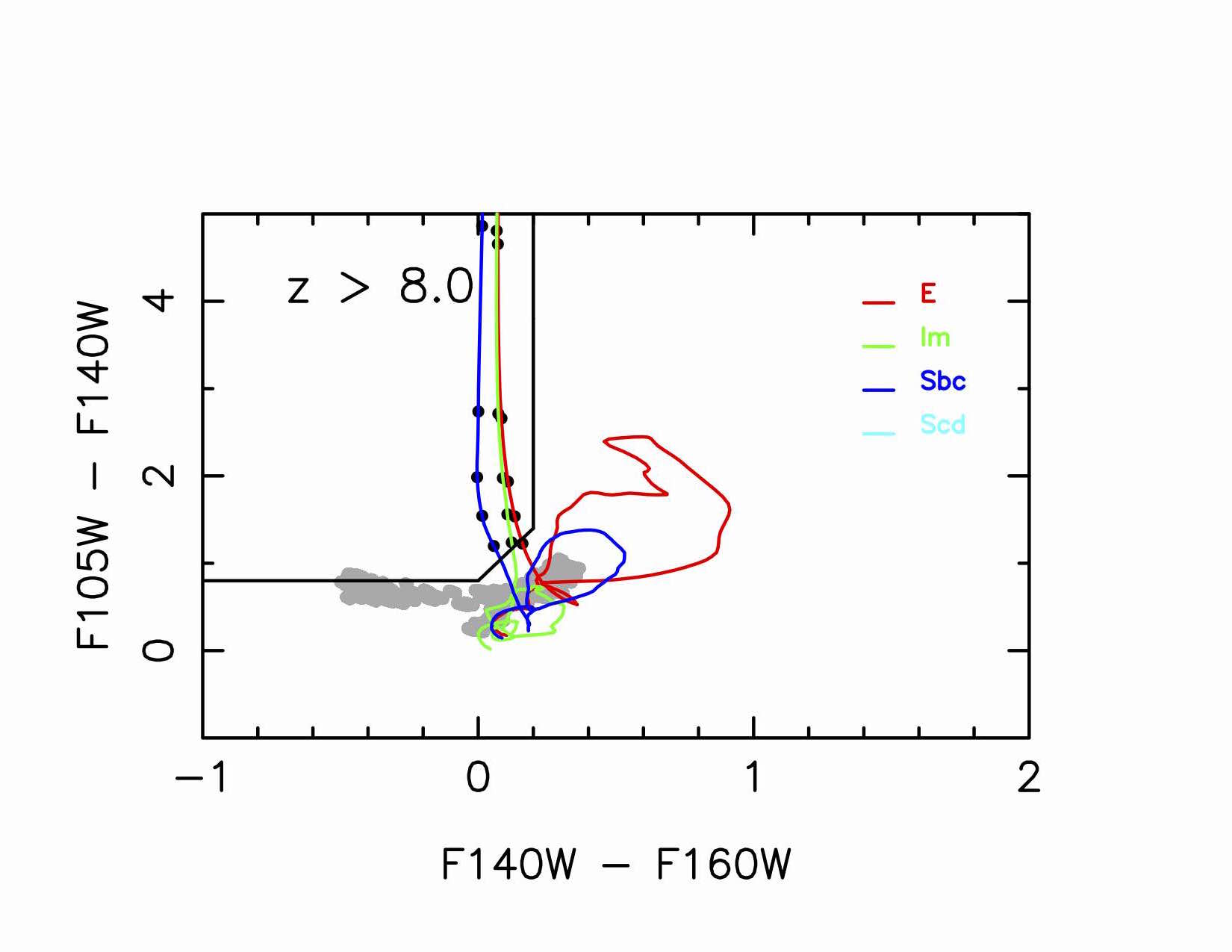}
      \caption{\label{color} Color criteria we defined to select sources at $z \gtrsim 6$ from the evolution of standard templates (see references in text). Grey dots show the expected colors of objects as a function of redshift starting from $z = 6$ on the left panel and $z=8$ on the right panel with a step of $dz$=0.2. The grey dots show the expected colors of L, M and T dwarfs from 225 spectra (see references in the text). Our color criteria are shown by the region limited by the black lines.}
   \end{figure*}

One of the most popular methods used to select objects at very high-$z$ in photometric data is the Lyman Break technique \citep{1999ApJ...519....1S}, combining non-detection in images bluewards of the Lyman break and color selection in filters redwards of the break. The selection window was computed using color evolution of standard templates (\citealt{2003MNRAS.344.1000B}, \citealt{1980ApJS...43..393C}, \citealt{1996ApJ...467...38K}, \citealt{2007ApJ...663...81P}), and defined criteria for several redshift intervals: $z \gtrsim 6$ and $ z \gtrsim 8$ (Fig. \ref{color}). To select  $z \gtrsim 6$ objects, the color criteria we used are: \\
F814W - F105W $>$ 0.8 \\
F814W - F105W $>$ 0.8 + 2.0$\times$(F125W-F140W) \\
F105W - F125W $<$ 0.6. \\
The $ z \gtrsim 8$ selection criteria are defined as below: \\
F105W - F140W $>$ 0.8 \\
F140W - F160W $<$ 0.2 \\
F105W - F140W $>$ 0.8 + 3$\times$(F140W - F160W) \\
We used the selection criteria defined by Infante et al. (2015) to select $z \gtrsim 10$ candidates: \\
F125W - F160W $>$ 0.8 \\
For each redshift interval explored, non-detection criteria are required in all the bands bluewards of the Lyman break, such as $m(F435W, F606W, F814W) > m(2\sigma)$ to select $z \gtrsim 8$ objects. Moreover, to limit spurious selection, we imposed a detection in at least two consecutive bands at more than $5\sigma$, such as $m(F125W, F140W) < m(5\sigma)$ for $z \gtrsim 8$ objects,  leading to a break of at least $\sim$2 mag that should help to remove extreme mid-$z$ interlopers \citep{2012MNRAS.425L..19H}. The reason for such care is that verification of these techniques holds to $z$$\sim$5.5, but has yet to be strongly proven at $z$$>$6-6.5, and thus the selection of faint candidates without such breaks may be dangerous and lead to an overestimation of the number of objects. We prefer to build a robust sample.

\subsection {Confirming optical non-detection: $\chi^2_{opt}$}

The visual inspection of our candidates could still allow into our samples objects that are extremely faint in optical bands, and that could not be at such high-redshift (see Sec. \ref{contaminants}). In order to limit/remove this kind of interloper, we applied the optical $\chi^2$ method defined in \citet{2011ApJ...737...90B} by: 
\begin{equation}
\chi^2_{opt}=\sum\limits_{i=1}^n SGN(f_i) \Big( \frac{f_i}{\sigma_i}\Big)^2
\end{equation}
where $f_i$ is the flux measured in band $i$, $\sigma_i$ is the uncertainty on $f_i$ and $SGN(f_i) = 1$ if $f_i > 0$ or  $SGN(f_i) = -1$ if $f_i < 0$. 

\noindent To estimate the $\chi^2_{opt}$ limit above which a candidate should be considered as detected in the optical bands, we measured the optical flux in 1000 empty 0.4'' diameter apertures distributed over the selection area, and we computed for each aperture the $\chi^2_{opt}$. We then added with the IRAF \textit{mkobjects} routine sources that are detected at $\sim$2$\sigma$ in the optical bands, and computed for each source its $\chi^2_{opt}$. We compared the $\chi^2_{opt}$ distribution for each sample (empty apertures and faint objects) and deduced the $\chi^2_{opt}$ limit from the value where the probability to get an object with a faint detection in the optical is higher than the probability to get a non-detected source. We estimated  this limit to be 0.2, therefore all sources with $\chi^2_{opt} > 0.2$ will be considered as most likely contaminants. However because of the intracluster light \citep{2015MNRAS.448.1162D}, we used a  $\chi^2_{lim}$ that was a function of the position over the region covered by HST in the cluster field.

 
Among the 28 $z \gtrsim 6$ objects that fulfill the selection criteria in the cluster field, 14 satisfied also the $\chi^2_{opt}$ criteria, and are considered to be good candidates in the following. For the parallel field sample, only 25 sources over the 42 selected have a $\chi^2_{opt}$ consistent with a real non-detection. All these objects, including those considered as detected at optical wavelengths, are presented on Table \ref{cluster_sample}, \ref{parallel_z8_sample} and \ref{parallel_z6_sample}.

\subsection {Longer Wavelength Constraints}
We used the deep 3.5 and 4.5$\mu$m/IRAC images described in section \ref{data} to add SED constraints at longer wavelengths. We performed aperture photometry within a circle of 2\arcsec4 radius and considered as "blended" all objects for which more than 2 objects are inside this aperture. For the remaining objects we follow the method described in \citet{2014ApJ...795...93Z} using GALFIT \citep{2010AJ....139.2097P}. We then used the aperture correction factors defined in \citet{2008PASP..120.1233H} in order to obtain the total photometry.

In this way, we can add SED-constraints at longer wavelengths for $\approx$72\% of our sample. For the remaining objects, we estimated their physical properties using \textit{HST} informations only (see below).

\section{Properties of our samples}
\label{sample}
In order to improve the size of our high-$z$ sample, we combined candidates described in the previous selection with objects selected following the same methods in the two first FF dataset: Abell 2744 (\citealt{2014ApJ...795...93Z}, \citealt{2015ApJ...804..103K}) and MACS0416 \citep{Infante2015}. In the following, we estimate the physical properties of all these objects (redshift, SFR, stellar mass, reddening, size) using the same methods in order to get homogeneous results (Tab. \ref{properties_macs0417} and \ref{table_pp}). 

\subsection{Photometric redshift and emission lines}
The SEDs of our candidates are constrained by at least 7 measurements (from F435W to F160W) including robust non-detection at short wavelengths. For more than 70\% of our sample, we added constraints on the SEDs within the IRAC wavelength range, making the estimation of their properties more robust. These  properties have been deduced by SED-fitting and using two different approaches: $\chi^2$ minimization with \textit{Hyperz} \citep{2000A&A...363..476B}\footnote{v12.3 available at: http://userpages.irap.omp.eu/\textasciitilde rpello/newhyperz} and using Bayesian probability with \textit{BPZ} \citep{2000ApJ...536..571B}. We run \textit{Hyperz} with a standard library templates including nebular emission lines (\citealt{1997A&A...326..950F}, \citealt{1998ApJ...509..103S}, \citealt{2003MNRAS.344.1000B}, \citealt{1980ApJS...43..393C}, \citealt{1996ApJ...467...38K}, \citealt{2007ApJ...663...81P}) and allowing as parameter space: $z \in [0.0: 12.0]$, $A_v \in [0.0: 3.0]$ mag, following the reddening law defined in \citet{2000ApJ...533..682C}. Uncertainties on photometric redshift are deduced from the 1$\sigma$ confidence interval (Table \ref{photo_z}).  BPZ was ran spanning a redshift range $z \in [0.0001,12.0]$ with a  resolution $\Delta$$z$ = 0.01, applying no priors to the Likelihoods functions and using an interpolation factor of 9 among contiguous templates. We used the new library of galaxy models in BPZ2.0 (described in \citealt{2014MNRAS.441.2891M} ) composed by five templates for elliptical galaxies, two for spiral galaxies and four for starburst galaxies along with emission lines and dust extinction. Opacity of the intergalactic medium is applied as described in \citet{1995ApJ...441...18M} or both \textit{Hyperz} and \textit{BPZ}.

   \begin{figure*}
   \centering
           \includegraphics[width=16.cm]{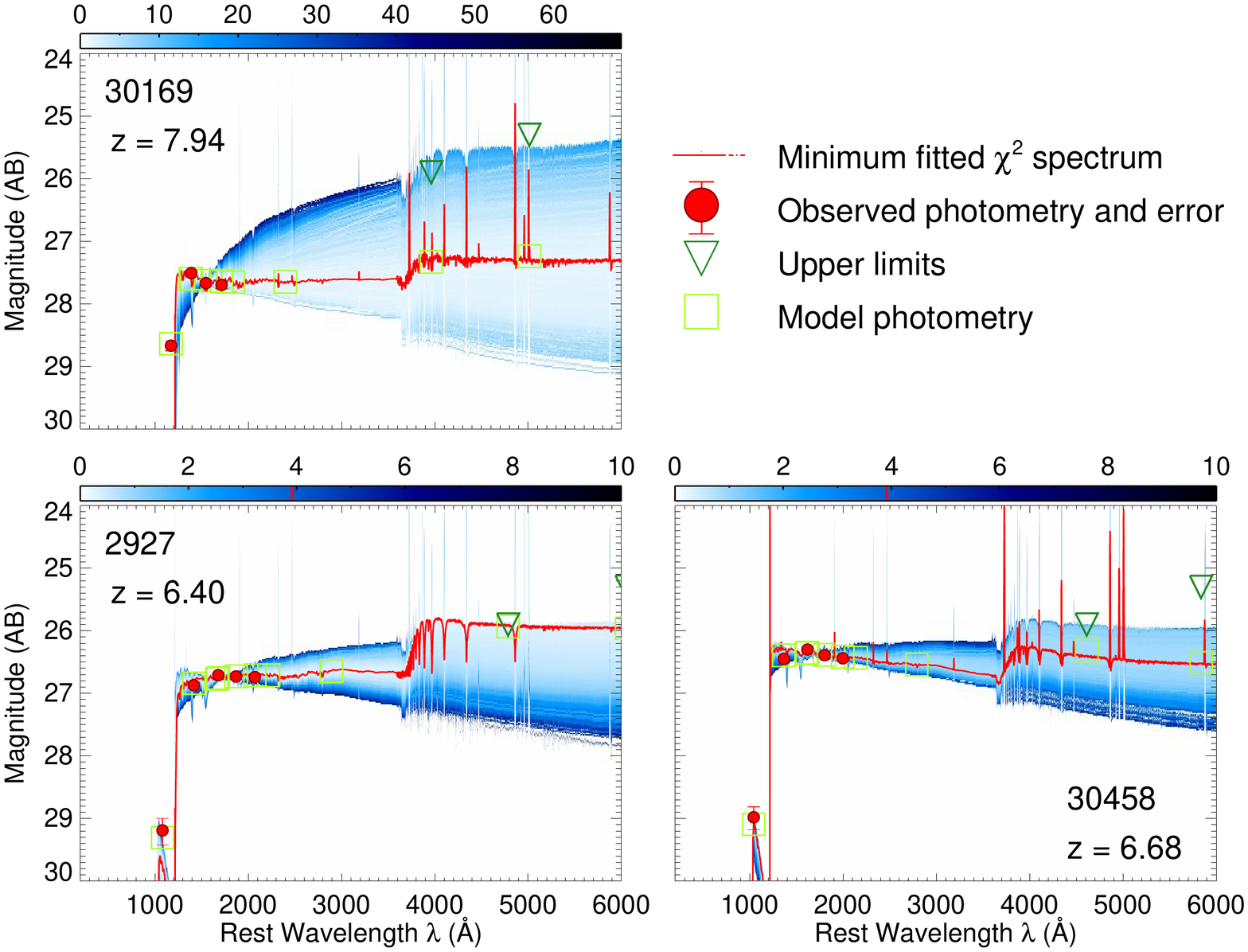}
      \caption{\label{SED_fitting_results} Example of SED-fitting results using \textit{iSEDfit} \citep{2013ApJ...767...50M} for 3 objects among our MACS0717 sample (cluster and parallel fields). Non-detections are plotted at 3$\sigma$ with green triangles, the red lines display the best fit, the green squares are photometric magnitudes of the best fit and the blue region shows several models we used to fit the SED. The color-bar indicates is a $\chi^2$ scale indicating the quality of the fit.}
   \end{figure*}

For most of our sample, photometric redshifts computed from the $\chi^2$ minimization method are consistent with those computed using the Bayesian approach, especially for all objects selected in the cluster field. About $\approx$30\% of our candidates have 1$\sigma$ error bars that disagree, but only four objects ($\approx$10\% of our sample) disagree on the nature of the candidates (from high-$z$ to low-$z$, \#44317, \#50815, \#58730, \#70084). In the following, we consider these four objects as high-$z$ candidates since they satisfied the color-color selection and they fulfilled the optical $\chi^2$ criteria. For the remaining objects, fitted with both approaches as high-$z$, the difference between the 1$\sigma$ confidence intervalle is not surprising regarding the redshift range of our objects

\begin{figure}
   \centering
           \includegraphics[width=7.cm]{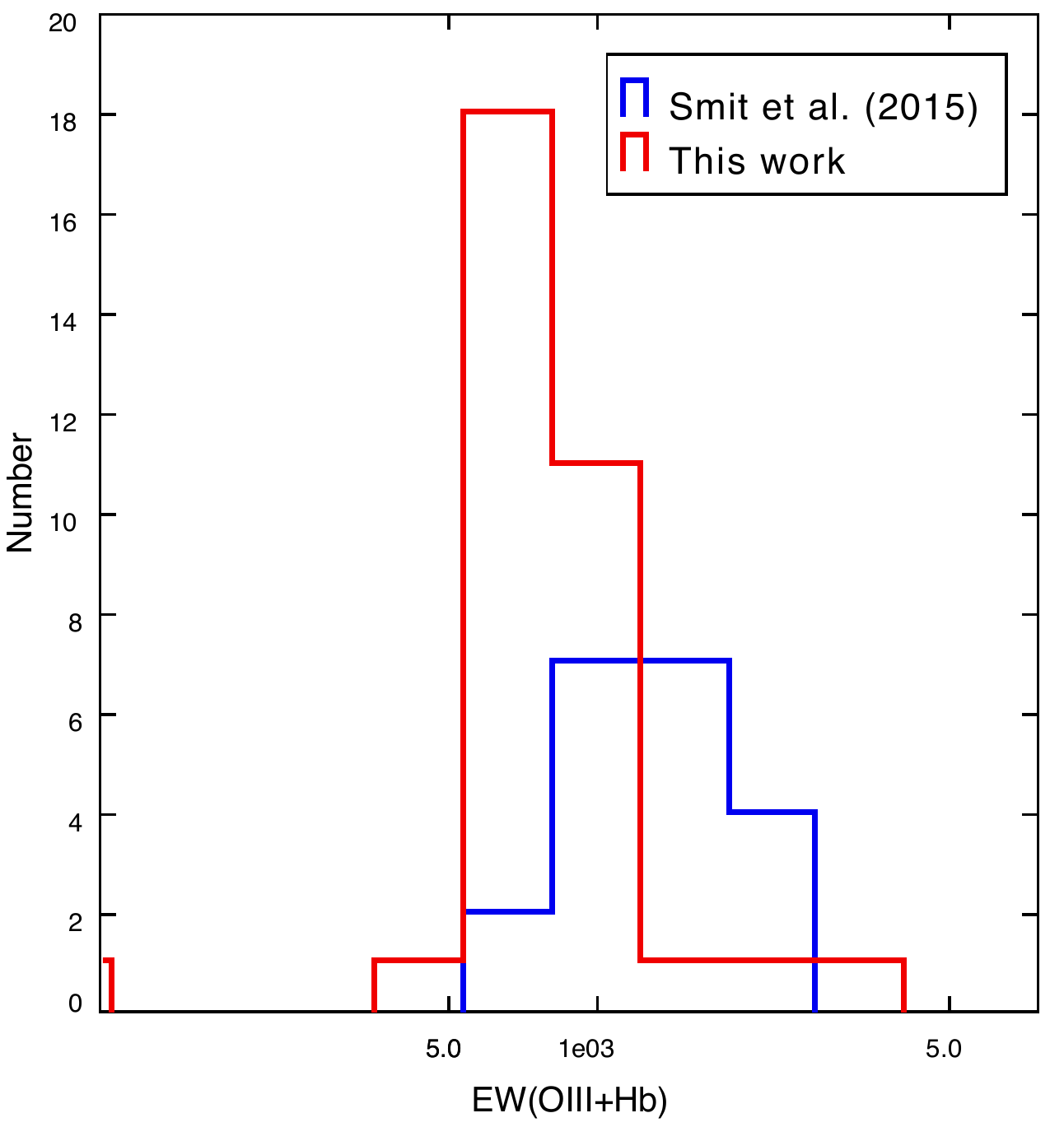}
      \caption{\label{ew} Distribution of the estimated equivalent widths of [OIII]+H$\beta$ for $z\sim$7 objects selected in the three first Frontier Fields dataset (red) compared with the distribution for 20 Lyman Break Galaxies (blue) discussed in \citet{2015ApJ...801..122S}}
   \end{figure}

By adopting templates which include nebular emission lines, we can estimate the equivalent width of the [OIII] and H$\beta$ lines and compare these values to what has been previously found at such high-redshifts in order to check the quality of our SED-fitting results. Figure \ref{ew} shows the distribution of the $z$$\sim$7 Frontier Fields candidates compared to the distribution of the 20 $z$$\sim$7 Lyman Break Galaxies discussed in \citet{2015ApJ...801..122S}. The equivalent widths of the [OIII] and H$\beta$ lines measured in Frontier Fields candidates are in excellent agreement with what has been estimated in \citet{2015ApJ...801..122S}, \citet{2015arXiv150600854R} and \citet{2013ApJ...777L..19L} 

\subsection{Magnification}
\label{magnification}
Within the framework of the FF project, several groups have provided amplification maps built using different assumptions on mass models (\citealt{2014MNRAS.444..268R}, \citealt{2014ApJ...797...48J}, \citealt{2015ApJ...801...44Z}, \citealt{2015ApJ...800...38G}, \citealt{2011MNRAS.417..333M}) . 

We estimated the amplification of our candidates by averaging all these models and uncertainties from the standard deviation. All the objects selected in the cluster field have magnification ranging from 1.8 to 7.0 (Tab. \ref{photo_z}). The parallel field can not be considered as a real blank field since the cluster mass still plays a role on the light amplification at such distances from the cluster core. Among all the models, only one covers the parallel field \citep{2011MNRAS.417..333M} but with a low resolution. According to this model, we fixed the amplification of candidates selected in the parallel field to $\mu$=1.1. 

We computed the effective surface covered by the three first FF clusters using the amplification map released by the CATS team matched to our detection images. We then masked all the bright objects in the field and computed for each amplification range the area effectively covered by the data. We estimated an effective surface covered by the 3 first FF dataset of $\approx$16 arcmin$^2$. 

We also used \textit{Lenstool} (\citealt{1996ApJ...471..643K}, \citealt{2007NJPh....9..447J}, \citealt{2009MNRAS.395.1319J}) to search for multiple images of our candidates using the CATS models \citep{2014MNRAS.444..268R}. According to this model, 10 objects among our samples could be multiple imaged (\#9313, \#13963, \#21962, \#25550, \#25990. \#29413, \#30458, \#33447, \#46005, \#46206) but none of these images are detected on FF data, most of them are outside of the Field of view covered by HST, several are located at the positions of brights objects on the region covered by FF data. 

\subsection{Stellar mass, Age, reddening, UV slope and SFR}
We used  \textit{iSEDfit} code \citep{2013ApJ...767...50M} and follow the method described in \citet{Infante2015} to generate 100 000 models including dust, nebular emission lines and assuming an initial mass function from 0.1 to 100 M$_{\odot}$. Uniform priors were adopted to estimate the following parameters: the stellar metallicity, the galaxy age, the Star Formation timescale, and rest-frame V-band attenuation. We fix the redshift to the Bayesian value given by BPZ. The results are reported on Table \ref{properties_macs0417} and example results from \textit{iSEDfit} are shown in Fig. \ref{SED_fitting_results}. Error bars on each physical parameters included the uncertainties in magnification, which we estimated from all models available for Frontier Fields clusters (see section \ref{magnification}).

\begin{figure}
   \centering
           \includegraphics[width=10.cm]{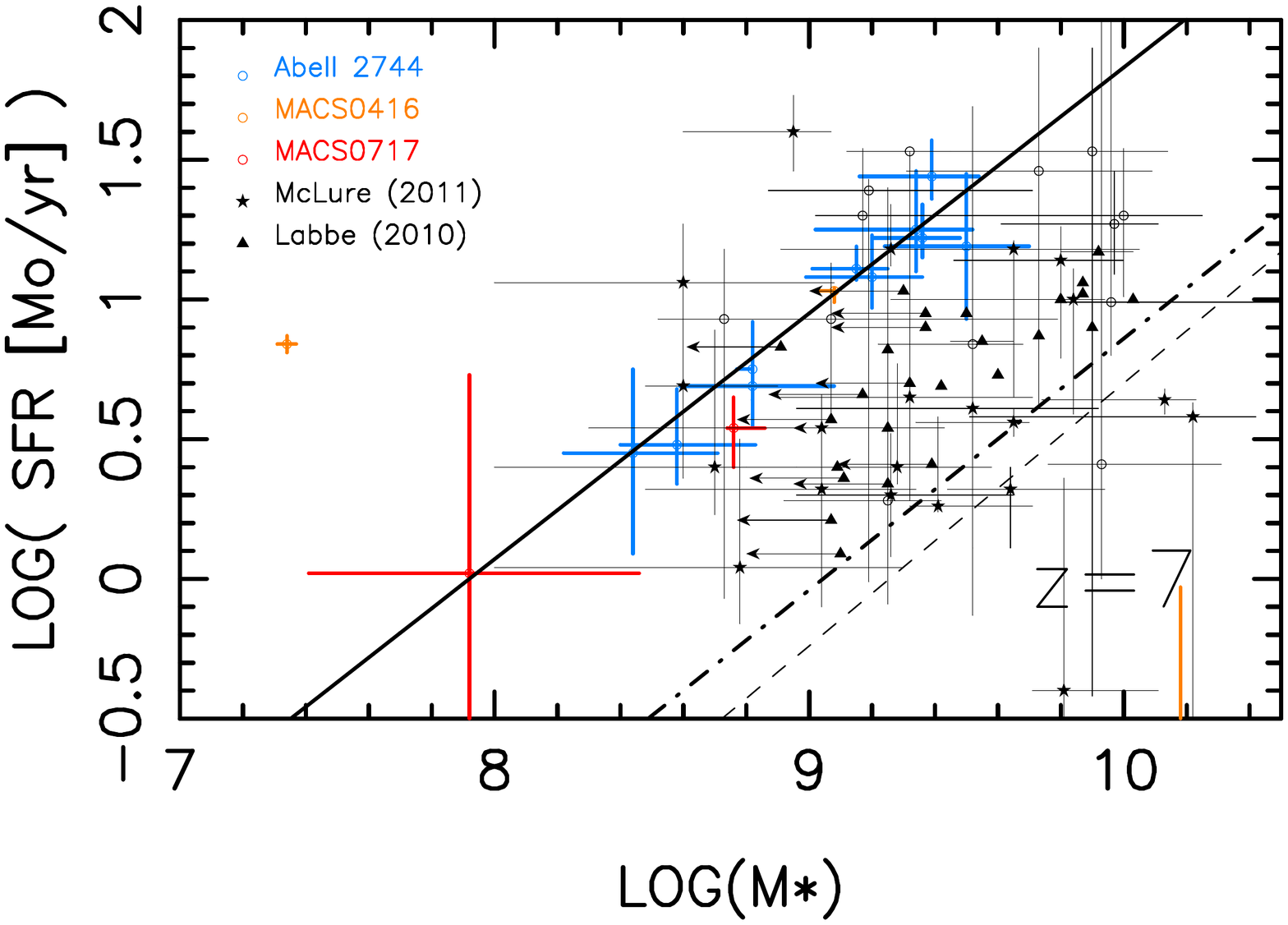}
                 \caption{\label{M_SFR} Evolution of the SFR as a function of the stellar mass for all of the $z\sim$7 candidates selected in the three first Frontier Fields dataset (Abell 2744 in blue, MACS0416 in grey and MACS0717 in red). We over-plotted objects (black points) analyzed in \citet{2010ApJ...716L.103L} and \citet{2011MNRAS.418.2074M}.  The solid line shows the best fit of the SFR-M$^{\star}$ relation deduced by $\chi^2$ minimization using Frontier Fields selected candidates. The dashed line displays the parameterization deduced from SDSS galaxy at $z$$\sim$1 \citep{2007A&A...468...33E} and the dotted-dashed line shows the relation published by \citet{2010ApJ...716L.103L} }  
   \end{figure}

We used the large sample of $z$$\sim$7 and 8 candidates identified on Frontier Fields images to study the relationship between the SFR and the galaxy mass that has been extensively investigated at lower redshift (e.g. \citealt{2008MNRAS.385..147D}, \citealt{2010MNRAS.407..830M}, \citealt{2013MNRAS.429..302C},\citealt{2014A&A...563A..81D}, \citealt{2013A&A...549A...4S}). In order to add robust constraints on the evolution of the SFR as a function of stellar mass, we only used in our analysis objects detected in at least one of the IRAC images. Among all the $z$$>$6 objects selected in Frontier Fields dataset, only 18 are detected at 3.6 and/or 4.5$\mu$m. Figure \ref{M_SFR} shows the distribution of these candidates in the (M$^{\star}$, SFR) plane, along with several $z$$\sim$7 objects previously analyzed in \citet{2010ApJ...716L.103L} and \citet{2011MNRAS.418.2074M}. As expected, the luminosity range covered by the Frontier Fields dataset is larger than previous surveys allowing to add more constraints at lower stellar mass. We used a $\chi^2$ minimization method to fit the evolution of the two properties and found an evolution at $z\sim$7 given by: 
\begin{equation}
\log [SFR] = (0.88\pm0.44)\log [M^{\star}] - (6.97\pm3.95)
\end{equation}
where error bars represent the 1$\sigma$ confidence interval. This evolution is consistent with the trend observed by \citet{2010ApJ...716L.103L}. As already demonstrated in \citet{2015A&A...581A..54T}, the relation between SFR and stellar mass currently observed over a large range of redshift seems higher than previous estimates of the main sequence by \citet{2007A&A...468...33E} (cf. Figure \ref{M_SFR}). However, in our selection criteria we requested to have a detection in at least two NIR filters (UV rest-frame) that could biais our sample by selecting sources with the highest SFR in this redshift range. Therefore the trend observed in this \textit{Frontier Fields} sample should be considered as an upper limit of the evolution.

\begin{figure*}
   \centering
           \includegraphics[width=5.9cm]{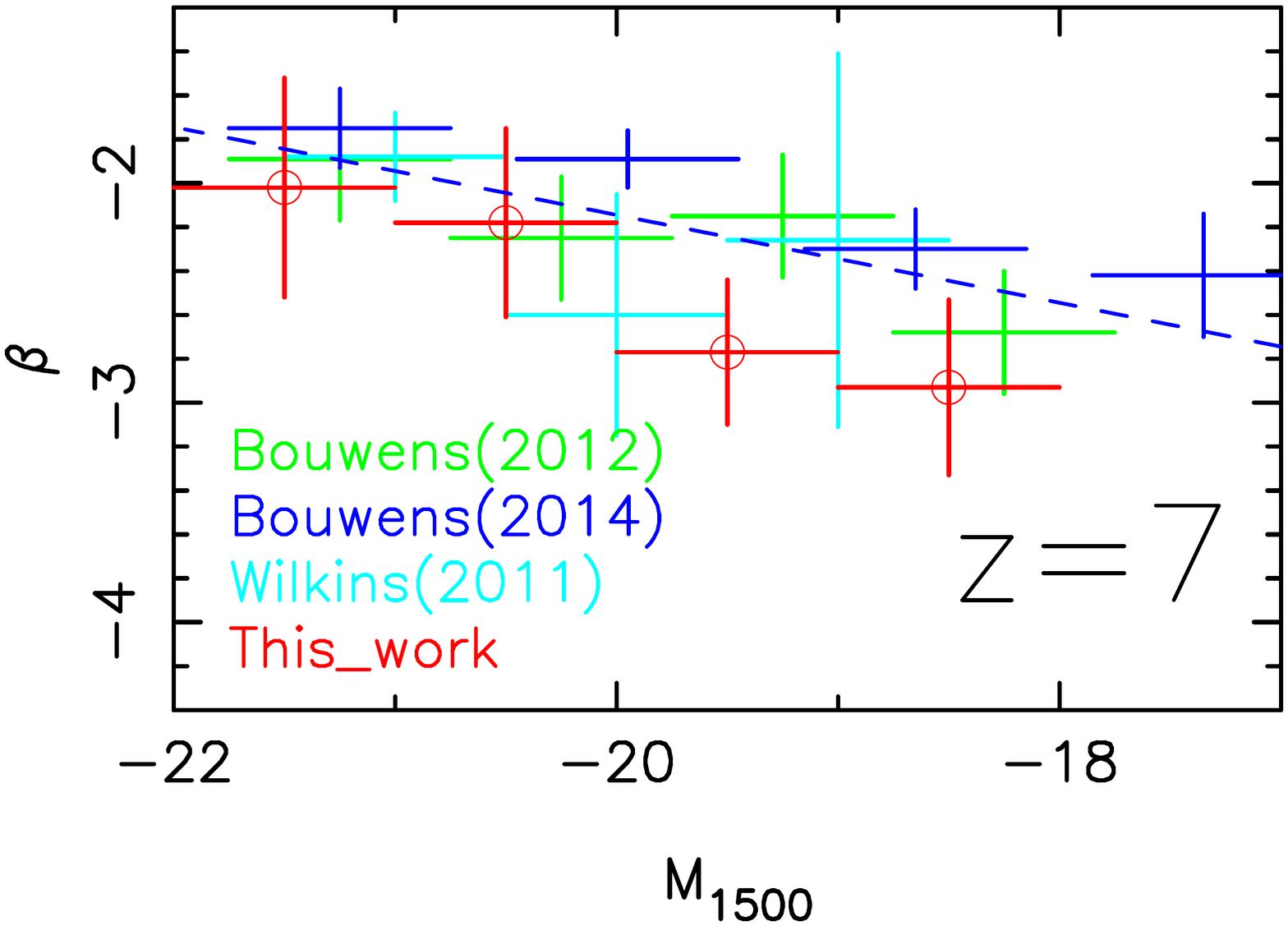}
           \includegraphics[width=5.9cm]{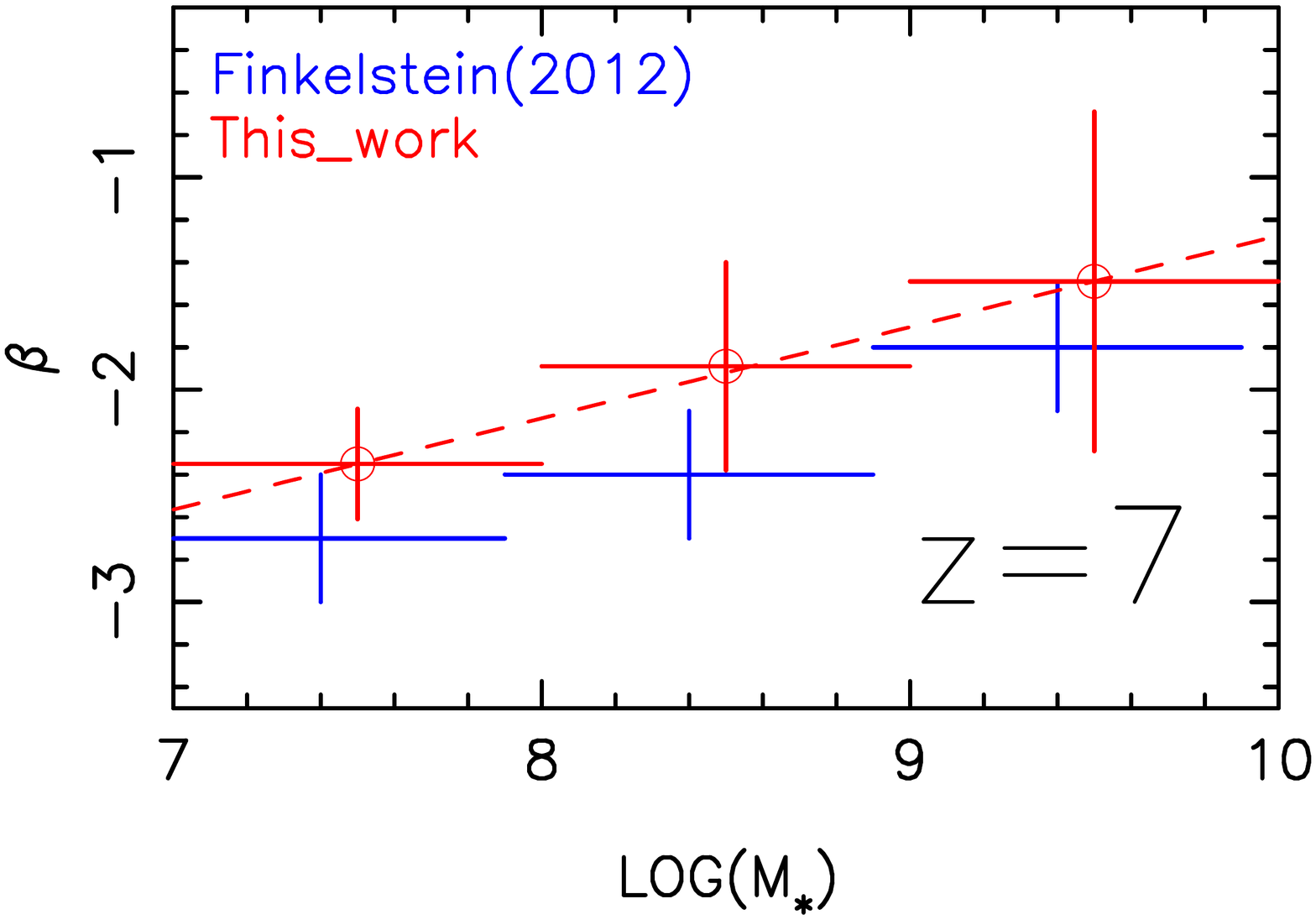}
           \includegraphics[width=5.9cm]{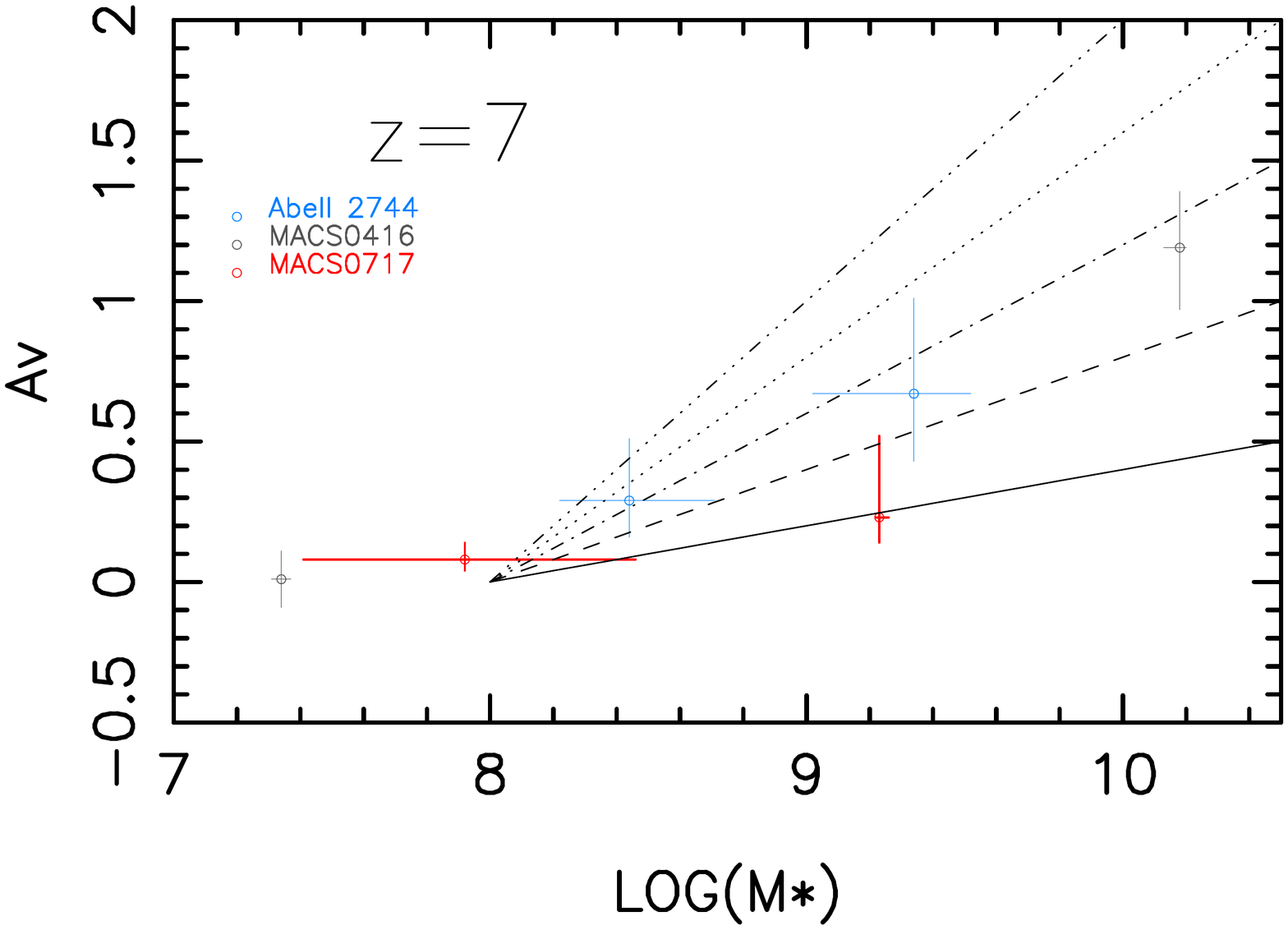}

                 \caption{\label{beta_slope} \textit{(left)} Evolution of the UV slope as a function of the UV luminosity deduced from all $z\sim$7 candidates selected in the first three Frontier Fields (red points) compared with previous findings (\citealt{2011MNRAS.417..717W},  \citealt{2012ApJ...754...83B} and \citealt{2015ApJ...803...34B}). The dashed line shows the evolution computed by  \citet{2015ApJ...803...34B} from a sample of $\approx$200 galaxies. \textit{(middle)} Evolution of the UV slope as a function of the stellar mass computed from objects detected in at least one IRAC band (red points) compared to the evolution found in \citet{2012ApJ...756..164F}. \textit{(right)}  Evolution of the reddening as a function of galaxy mass for all  the candidates selected in the three first Frontier Fields (Abell 2744 in blue, MACS0416 in grey and MACS0717 in red). We also plot the trend observed by \citet{2010A&A...515A..73S} ($A_v$$=$$\log(\frac{M_{\star}}{10^8M_{\odot}})^n$) assuming several values of $n$ (0.2, solid line, 0.4, dashed line, 0.6, dotted-dashed line, 0.8, dotted-line and 1.0 the triple dotted-dashed line). }
   \end{figure*}

The large sample of $z$$\sim$7 and 8 candidates allows us to study the evolution of the UV slope (hereafter $\beta$ slope), and therefore the reddening, as a function of stellar mass. We estimated the $\beta$ slope following equation 1 of \citet{2013MNRAS.432.3520D} and the corresponding error bars were estimated from photometric errors. Figure \ref{beta_slope} displays the evolution of the UV slope as a function of luminosity for all $z$$\sim$7 candidates selected in the Frontier Fields survey. The evolution is compatible with previous findings published in \citet{2011MNRAS.417..717W},  \citet{2012ApJ...754...83B} and \citet{2015ApJ...803...34B}. We also plot the evolution of $\beta$ as a function of the stellar mass, but only for candidates detected in at least one IRAC band. We compared this evolution with results published in \citet{2012ApJ...756..164F} and found similar evolution.  We also studied the relationship between the star formation rate and galaxy mass in $z$$\sim$7 candidates detected in at least one IRAC band. This sample seems to follow the trend observed for $z$$\sim$2 galaxies (\citealt{2007A&A...468...33E}, \citealt{2007ApJ...670..156D}). A similar trend has been deduced from the analysis of $\approx$1700 LBGs at $z$$\sim$3-6 (\citealt{2014A&A...563A..81D}, \citealt{2013A&A...549A...4S}) with a better stellar mass coverage. From our sample of IRAC detected $z\sim$7 galaxies, we confirm that massive galaxies seem more affected by dust attenuation than smaller galaxies. In order to test this result at such high-redshift, we need to strongly increase the number of $z\ge$7 candidates bright enough to be detected in IRAC data.  

The star formation history in very high-redshift galaxies can be studied through the specific SFR (sSFR), the ratio between the SFR and the stellar mass of a given galaxy. We used the sample of $z$$\sim$7 and 8 Frontier Fields candidates to estimate the sSFR at such high-redshift. As before, we only used galaxy candidates that are detected in at least one IRAC band, in order to have a more robust estimate of their stellar mass. To study the evolution of this quantity as a function of redshift, we only considered objects that have a stellar mass within the interval 2.5-7.5$\times$10$^9$ M$_{\odot}$. Errors bars were obtained by adding quadratically the errors on the SFR and the stellar mass (Figure \ref{M_SFR}). We modified the parameterization found from the VUDS survey \citep{2015A&A...581A..54T} for galaxies at $z$$>$2.4 as follows :
\begin{equation}
sSFR  = 0.2\times (1+z)^{1.2}
\end{equation}
Our values are in perfect agreement with previous findings at lower redshift (\citealt{2015A&A...581A..54T},  \citealt{2014ApJ...781...34G} and \citealt{2013ApJ...763..129S}).

   \begin{figure}
   \centering
           \includegraphics[width=10.cm]{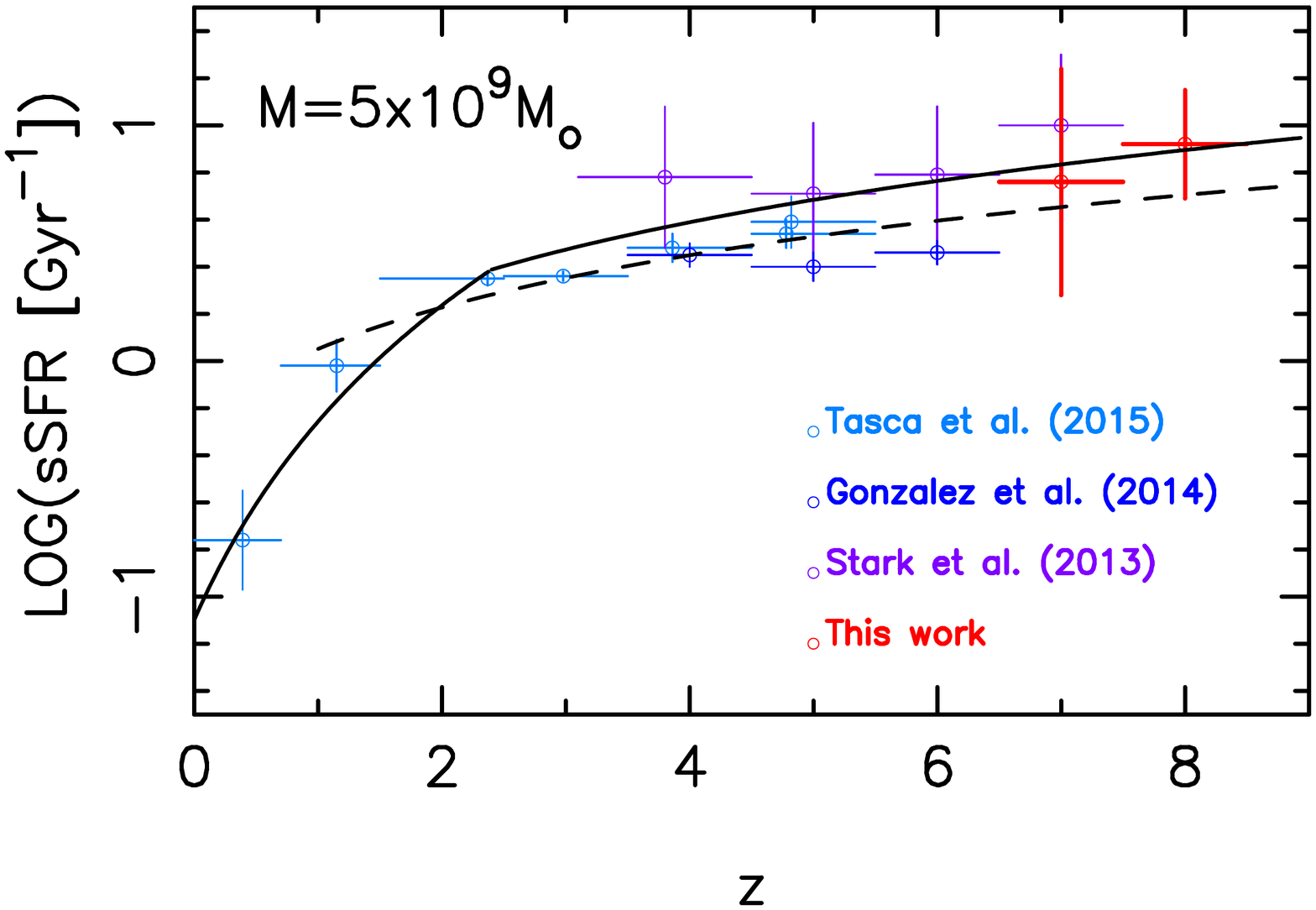}
      \caption{\label{sSFR} Specific star formation rate (sSFR) as a function of redshift for galaxies with stellar mass as of M$^{\star}$$\sim$5$\times$10$^9$M$_{\odot}$. We compare the sSFR we deduce from Frontier Fields candidates detected in at least one IRAC band with results published in \citet{2015A&A...581A..54T},  \citet{2014ApJ...781...34G} and \citet{2013ApJ...763..129S}.  The solid line shows an updated version of the parameterization discussed in \citet{2015A&A...581A..54T} and the dashed line displays the evolution found by \citet{2014ApJ...781...34G} } 
   \end{figure}

\subsection{Size}
\label{sec.size}
\begin{figure}
   \centering
           \includegraphics[width=8.5cm]{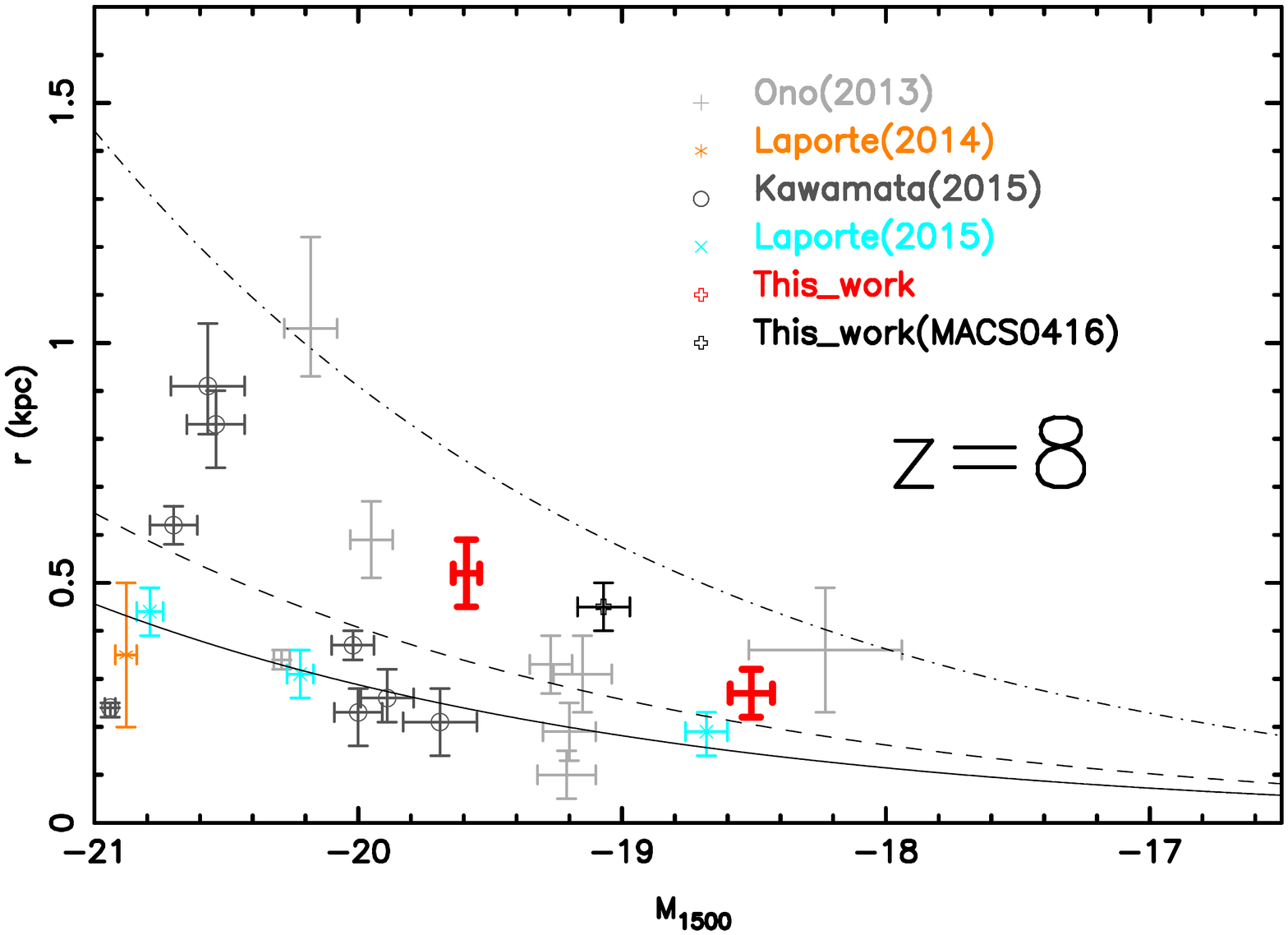}
      \caption{\label{size} Evolution of the size of $z\sim$8 candidates selected behind the three first Frontier Fields as a function of the UV Luminosity: Abell2744 (\citealt{2015ApJ...804..103K} and \citealt{2014A&A...562L...8L}), MACS0416 (\citealt{2015A&A...575A..92L}, and Infante et al. 2015) and MACS0717 (this paper). We compared this evolution with results from the HUDF 2012 campaign \citep{2013ApJ...777..155O}. We over-plotted several size-luminosity relation using different assumptions on the SFR densities (10 - solid line - 5 - dashed line - 1 - dotted dashed line - M$_{\odot}/yr$/kpc$^2$ ).}
   \end{figure}

The size of our objects was computed using the SExtractor FLUX\_RADIUS and setting the flux fraction parameter to 0.5 in order to get the half light radius. We corrected the size for PSF broadening following the method described in \citet{2010ApJ...709L..16O}: $r = \sqrt{r^2_{SEx} - r^2_{psf}}$, where $r_{SEx}$ is the half light radius and $r_{psf}$ the PSF of the F140W image. We also took into account the amplification of the light by the cluster making the observed size larger. We used the scale factor between the size on the sky and the physical size computed from \citet{2006PASP..118.1711W}. Recent studies took benefit from HST image quality to study the evolution of the size of $z\sim$8 objects selected in FF datasets as a function of the UV luminosity (e.g. \citealt{2015ApJ...804..103K}, \citealt{2015A&A...575A..92L}, \citealt{2014A&A...562L...8L}). Figure \ref{size} displays this evolution and shows that our $z\sim$8 objects are consistent with the trend observed by previous authors. 

   \begin{figure}
   \centering
           \includegraphics[width=8.5cm]{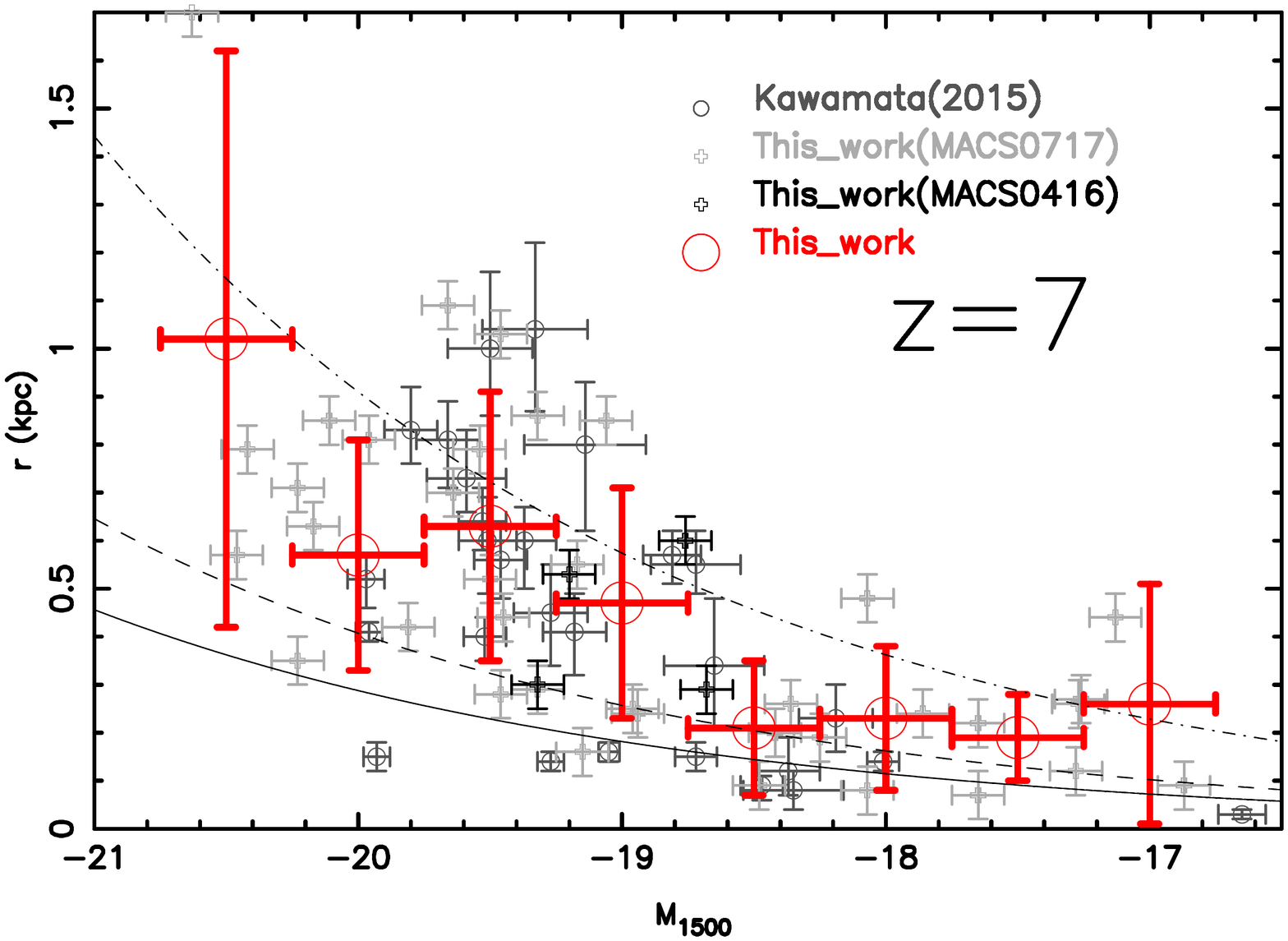}
      \caption{\label{size_z7} Evolution of the size of $z\sim$6-7 candidates selected behind the 3 first Frontier Fields as a function of the UV Luminosity: Abell2744 (\citealt{2015ApJ...804..103K} ), MACS0416 (\citealt{2015A&A...575A..92L} and this work based on \citealt{Infante2015} samples) and MACS0717 (this paper). The red points show the average radius per bin of 0.5 M$_{1500}$, error bars are the standard deviation. We over-plotted several size-luminosity relation using different assumptions on the SFR densities (10 - solid line - 5 - dashed line - 1 - dotted dashed line - M$_{\odot}/yr$/kpc$^2$  ).}
   \end{figure}

We applied the same method to compute the size of $z \sim 6-7$ objects in our sample and in those published in \citet{Infante2015}. We also used results from \citet{2015ApJ...804..103K} to study the size-luminosity relation at this redshift range as seen by the three first FF datasets. We took benefit from the large number of $z \sim 6-7$ candidates already selected in the FF data (cluster and parallel fields) to compute an average evolution (Fig. \ref{size_z7}). We used equation 4 from \citet{2013ApJ...777..155O} to constrain the SFR densities for these objects, but we failed to obtain strong constraints due to the large uncertainties on radius, that can only be reduced by further increasing the number of $z \sim 6-7$ objects. Nevertheless,  The distribution of the Frontier Fields selected candidates in the (r, M$_{1500}$) plane is consistent with previous results published at $z$$\sim$7 (\citealt{2015ApJ...808....6H}, \citealt{2010ApJ...709L..21O} and \citealt{2014arXiv1409.1832C})

Recently \citet{2014arXiv1409.1832C} claimed no-evolution in the size of Lyman-Break galaxies with redshift. We tried to investigate this conclusion using our sample of Frontier Fields selected objects at $z$$\sim$7 and 8. The averaged size of (0.3-1)L$^{\star}_{z=3}$ galaxies at $z$$\sim$7 is $r_{z=7}$(kpc)=0.80$\pm$0.18, which is similar to the value computed from HUDF objects \citep{2010ApJ...709L..21O}. In the same way, we estimated the mean size for (0.3-1)L$^{\star}_{z=3}$ objects at $z\sim$8 as $r_{z=8}(kpc)$=0.45$\pm$0.15, which is also consistent with previous results. Therefore we cannot exclude evolution in the size of Lyman-Break galaxies between $z$$\sim$7 and 8, although the number of $z$$\sim$8 candidates selected in the three first Frontier Fields clusters is still insufficient to draw any firm conclusions. 

\section{Discussion}
\label{discussion}

\subsection{Comparison of our sample with previous studies}
MACSJ0717.5+3745 is part of the CLASH survey, and several searches for high-$z$ objects have been done using shallower HST data. As shown by \citet{2014ApJ...795..126B} and \citet{2014ApJ...792...76B} no F140W$<$ 27.5 $z\sim$9 and 10 candidates has been selected behind this lensing cluster. We confirm this result and push the limits deeper by one magnitude in F140W. 

We took benefit from deeper ACS data to check the non-detection of previous high-$z$ candidates. \citet{2014ApJ...792...76B} published 15 candidates with photometric redshifts $>$5.5. One object, namely MACS0717-0247, is clearly detected in all ACS FF images and thus is no longer a good high-$z$ candidate, and MACS0717-0844 is detected in F606W, explaining why it is not in our sample. Moreover, among the 15 objects, 3 are out of the field of view covered by the FF images (MACS0717-0145, MACS0717-0166, MACS0717-0390). We recovered the following objects MACS0717-0234 (\#46206), MACS0717-0859 (\#33447, confirmed at $z = 6.39$ by spectroscopy in \citealt{2014ApJ...783L..12V}), MACS0717-1077 (\#30458), MACS0717-1730 (\#21962, confirmed at $z = 6.39$ by spectroscopy in \citealt{2014ApJ...783L..12V}) and MACS0717-1991 (\#15440). Therefore all the good $z > 6$ objects published in \citet{2014ApJ...792...76B}  located in the FF area are also in our selection, the difference between the two samples comes from the depth difference between CLASH and FF datasets, thus we only added fainter candidates and excluded the CLASH candidates that were detected in the deep ACS taken for FF.

The size of our sample of $z$$\sim$6-7 candidates is comparable to those built using previous FF data (e.g. \citealt{2015ApJ...804..103K}, \citealt{2015ApJ...800...18A}). However, the number of $z$$\ge$8 objects in the cluster field strongly differs from what has been found by previous authors (\citealt{Infante2015}, \citealt{2015A&A...575A..92L},  \citealt{2014arXiv1412.1472M}) suggesting either strong influence of cosmic variance or that our selection criteria are not well suited to select very high-$z$ objects (see discussion in Section \ref{deficit}).

We also took benefit from preliminary results of the GLASS survey \citep{2014ApJ...782L..36S} around MACS0717. In that paper, authors combined three different selection methods in order to reduce incompleteness of their sample, and retained  21 objects using CLASH data on that cluster \citep{2012ApJS..199...25P}. They added to their samples, the 15 objects selected by  \citet{2014ApJ...792...76B} and already discussed above. For the 6 remaining objects, 3 are out of the field of view covered by FF images, two are clearly detected on F606W and could not be at such high-redshift (\#1492 and \#1656) and one object does not fulfill the color-criteria we requested (\#1841). We compared our $z$$\ge$7 selected candidates with samples recently published in \citet{Schmidt15}, and noticed that the only ``good'' dropout they selected in MACS0717 (MACS0717-00908) is clearly detected in all ACS images, suggesting a low-$z$ solution for that object and explaining why it is not in our sample. Moreover, they detected emission line for 3 $z\sim$7 candidates in MACS0717 field: two have already been discussed in \citet{2014ApJ...783L..12V} and MACS0717-00370 displays a line at a signa-to-noise ratio of $\sim$3 ($f_{Ly\alpha}$=1.9$\times$10$^{-17}$ erg/s/cm$^2$). This last source is also included in our $z$$\ge$6 sample (\#13963) with a photometric redshift ranging from 6.3 to 6.8 (1$\sigma$confidence interval, see hereafter for details). Assuming that the detected emission line is Ly$\alpha$, this would place this object at $z$=6.51. For future spectroscopic follow-up, it is interesting to note that for the remaining objects no emission line has been detected for all these objects within the framework of the GLASS survey pushing the flux limit for Lyman-$\alpha$ down to 1.0$\times$10$^{-17}$erg/s/cm$^2$ at 2$\sigma$.

\subsection{Contamination of the samples}
\label{contaminants}
Among all the possible sources of contaminants in a high-$z$ samples, the most likely are the low-mass stars, the transient objects, the SNe or the low-$z$ interlopers. In the following section we discuss the contamination rate of our sample by several types of sources.

Low-mass stars have colors that could enter our selection criteria but they should be unresolved on single-epoch HST data. We computed expected colors for low mass stars from a set of 225 stellar templates of M, L and T dwarfs (\citealt{2006ApJ...637.1067B}, \citealt{2004AJ....127.2856B}, \citealt{2008ApJ...681..579B}, \citealt{2007ApJ...658..617B}, \citealt{2006ApJ...639.1095B}, \citealt{2004ApJ...604L..61C}, \citealt{2010ApJS..190..100K}, \citealt{2006ApJ...639.1114R}, \citealt{2007AJ....133.2320S}, \citealt{2006AJ....131.2722C}, \citealt{2007AJ....134.1162L}, \citealt{2006AJ....132.2074M}, \citealt{2009AJ....137..304S}, \citealt{2007ApJ...655..522L}, \citealt{2006AJ....131.1007B}). As shown on Figure \ref{color}, the selection windows we defined to select $ z \gtrsim 6$ objects exclude the large part of low-mass star colors. However, we noted that 34\% of M, L and T dwarfs we simulated have colors consistent with $z\sim$6 objects but only 2\% of these stars have colors that fulfill the criteria defined for $z\sim$8 objects.  

\noindent However to remove the stellar hypothesis for all our candidates, we first check the SExtractor stellarity parameter and then measure their size on the HST images using the SExtractor half light radius for each object. For the cluster sample, excluding two objects that display a stellarity of $\sim$0.4 (\#25990 and \#46005), all the candidates have a CLASS\_STAR parameter $<$0.1, meaning that all our candidates have a morphology inconsistent with a star.  Moreover all our objects are resolved on the F160W image making the star hypothesis unlikely. The same conclusion could be made for the parallel field with all objects have a stellarity parameter $< 0.1$. We measured the size of our objects in section \ref{sec.size}, and showed that 5 objects among the two samples are unresolved on the HST images: 1 $z \sim 8$ candidate (\#39832) and 4 $z \sim 7$ objects (\#3119, \#6576, \#66722, \#91692).

\subsection{Completeness of the selection method}
The method we used to extract and select the very high-$z$ candidates implies incompleteness. In other words we are not selecting all $z\gtrsim6$ objects that are effectively in our dataset, and thus are missing some of them. One of the goals of a very high-$z$ study is to keep this incompleteness small and to take it into account in all statistical analyses. 

In our case the incompleteness is due to the extraction method and the selection criteria described in Sec. \ref{selection}. We applied a 6 steps correction procedure summarized below: 
\begin{itemize}
\item We estimated colors of $\approx$700 000 $z\gtrsim$5 galaxies from standard templates (see references in Sec. \ref{sample}) with magnitude ranging from 22 to 32 in the filter following the position of the Lyman break, and redshifts from 5 to 12. 
\item We generated several lists of positions from an image constructed from the detection image where all objects have been masked. 
\item For each object in our mock catalogue, we associated a size assuming a log-normal distribution with a mean value of 0.15\arcsec and a sigma of 0.07\arcsec \citep{2010ApJ...709L..21O} in the source plane, and lensed it through the cluster.
\item We then add all these objects on the real images using the \textit{mkobjects} routine of IRAF
\item We apply the extraction method and selection criteria we used to select real objects
\item Incompleteness levels are deduced by comparing the list of objects we selected at the end with the input list of mock objects.
\end{itemize}

The conclusion of this analysis is that we reach a $\sim$70\% completeness at m$_{1500}^{RF}\sim$29.5 AB for our two selection functions. 

\subsection{Constraints on the UV LF}
\label{LF}
One of the main goals of the FF legacy program is to constrain the evolution of the galaxies during the first billion years of the Universe through the evolution of the UV LF, and especially by adding robust constraints at faint luminosities. We deduced number densities of our photometric samples by taking into account uncertainties on each redshift. Indeed, over the redshift interval covered by this survey uncertainties on photometric redshifts are not negligible. We applied a standard Monte-Carlo method based on the redshift probability distribution (e.g. \citealt{2015A&A...575A..92L}) what we summarize as :
\begin{itemize}
\item At each iteration, we assign a photometric redshift based on the redshift probability distribution
\item We compute the UV luminosity based on this redshift and the SED of each object
\item We repeat the previous steps $N$ times, to obtain a sample with $N$ times the size of the original sample but with the same distribution in redshift
\item We distribute objects into redshift and magnitude bins (e.g. $z$$\pm$0.5 with $z=$7, 8, 9 and 10), divide the number of objects by the number of iterations $N$ and the volume explored estimated from the detection picture. 
\item Error bars include statistical uncertainties and Cosmic Variance \citep{2008ApJ...676..767T}
\end{itemize}
We deduced upper limits based on Poisson statistics. The resulting number densities are presented on Table \ref{LF_densities}.

\begin{figure}
   \centering
           \includegraphics[width=9.5cm]{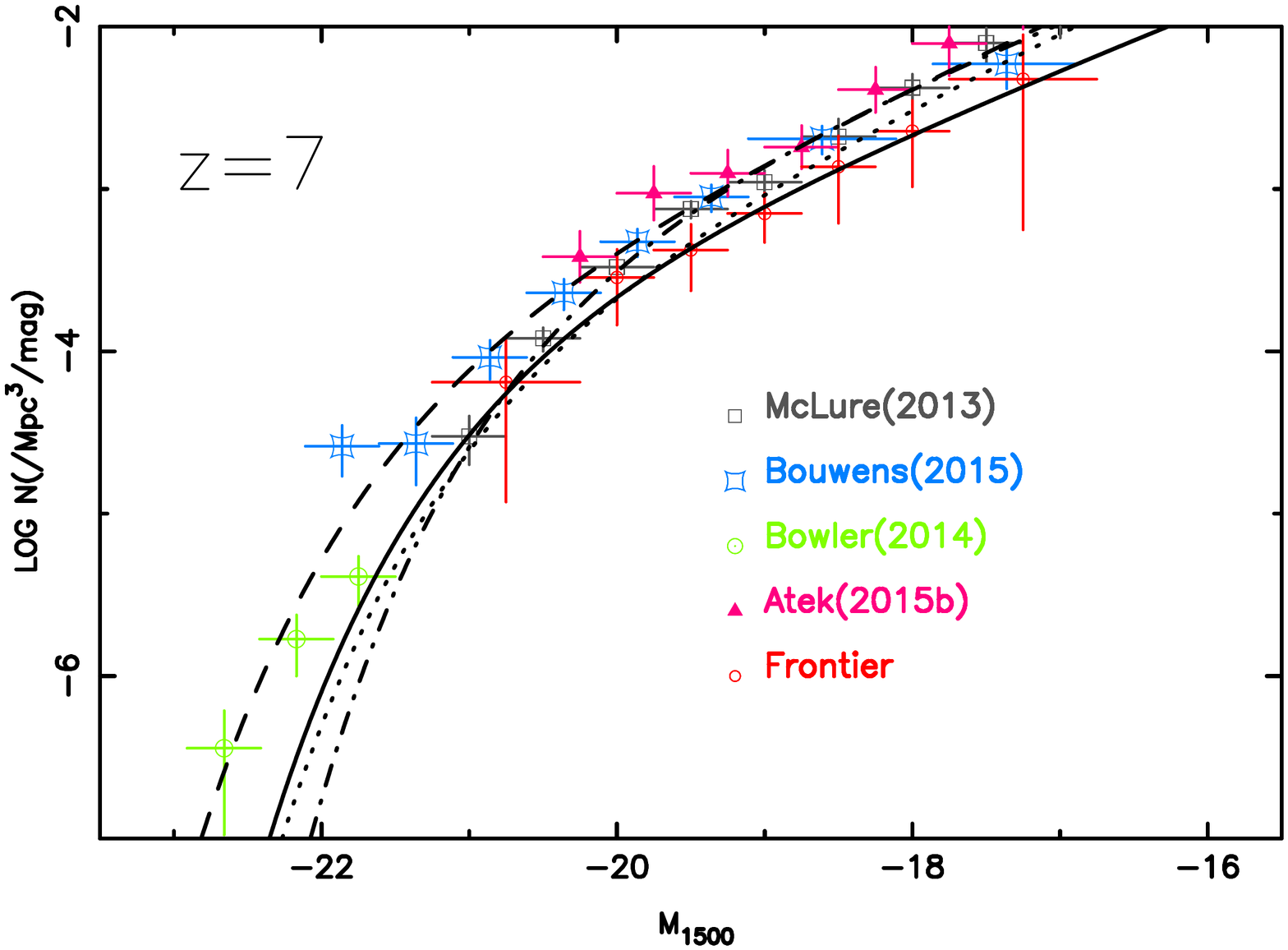}
      \caption{\label{LF_z7} UV Luminosity Function at $z$$\sim$7 computed using the first half of the Frontier Fields data. Number densities estimated from this study are in red, we over-plotted results from others groups using others datasets (\citealt{2015arXiv150906764A}, \citealt{2015ApJ...803...34B}, \citealt{2013MNRAS.432.2696M}, \citealt{2014MNRAS.440.2810B}).The solid line displays the parametrization we deduced from this study, the dot-dashed line shows the shape published by \citet{2015ApJ...803...34B}, the dashed line is from \citet{2014MNRAS.440.2810B} and the dotted line from \citet{2013MNRAS.432.2696M}  }
   \end{figure}

\begin{figure}
   \centering
           \includegraphics[width=9.5cm]{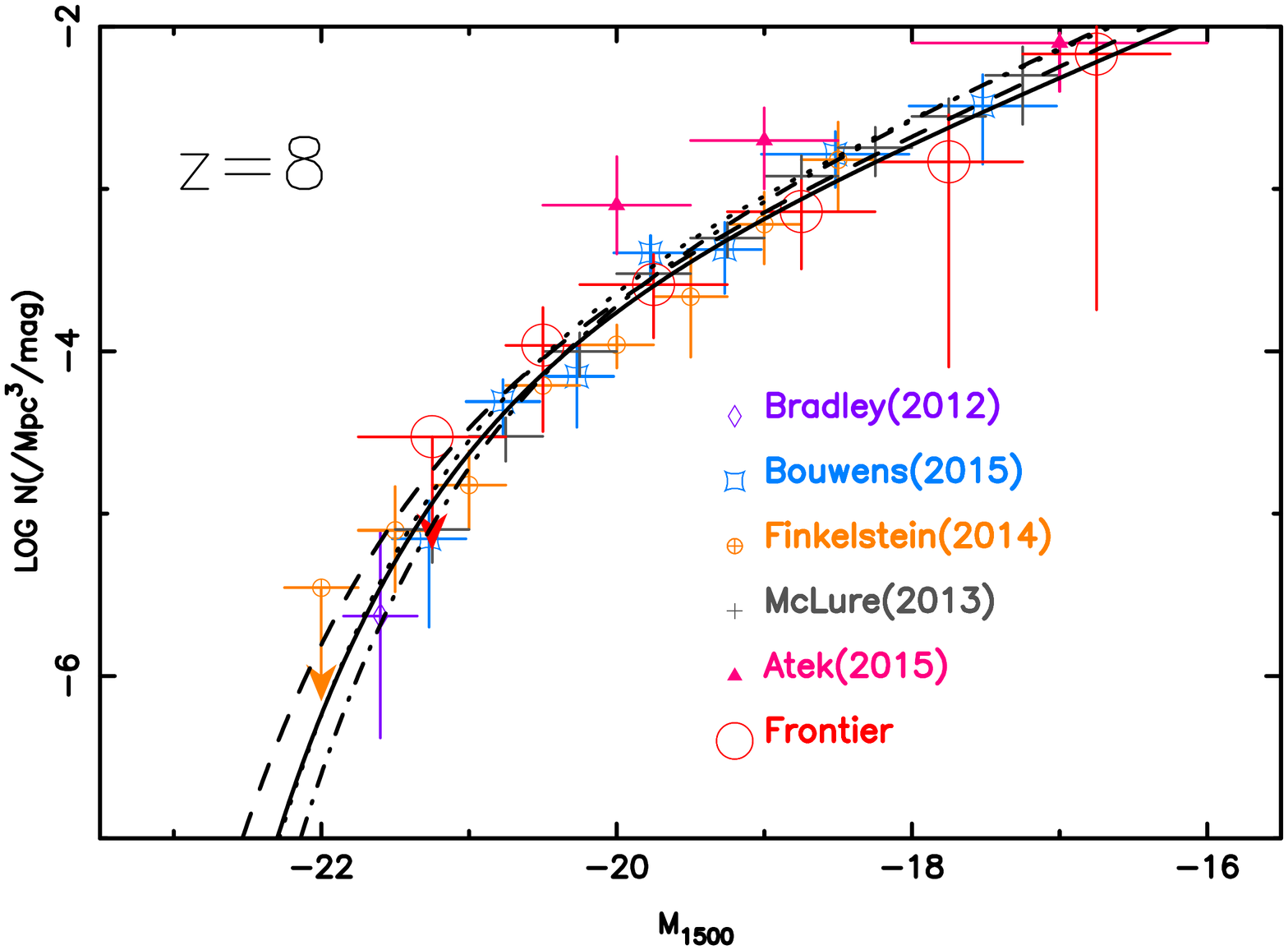}
      \caption{\label{LF_z8} UV Luminosity Function at $z$$\sim$8 computed using the first half of the Frontier Fields data. Number densities estimated from this study are in red, we over-plotted results from others groups using others datasets (\citealt{2015ApJ...803...34B}, \citealt{2013MNRAS.432.2696M}, \citealt{2012ApJ...760..108B}).The solid line displays the parametrization we deduced from this study, the dot-dashed line shows the shape published by \citet{2015ApJ...803...34B}, the dashed line is from \citet{2012ApJ...760..108B} and the dotted line from \citet{2013MNRAS.432.2696M}  }
   \end{figure}

\begin{figure}
   \centering   
           \includegraphics[width=9.5cm]{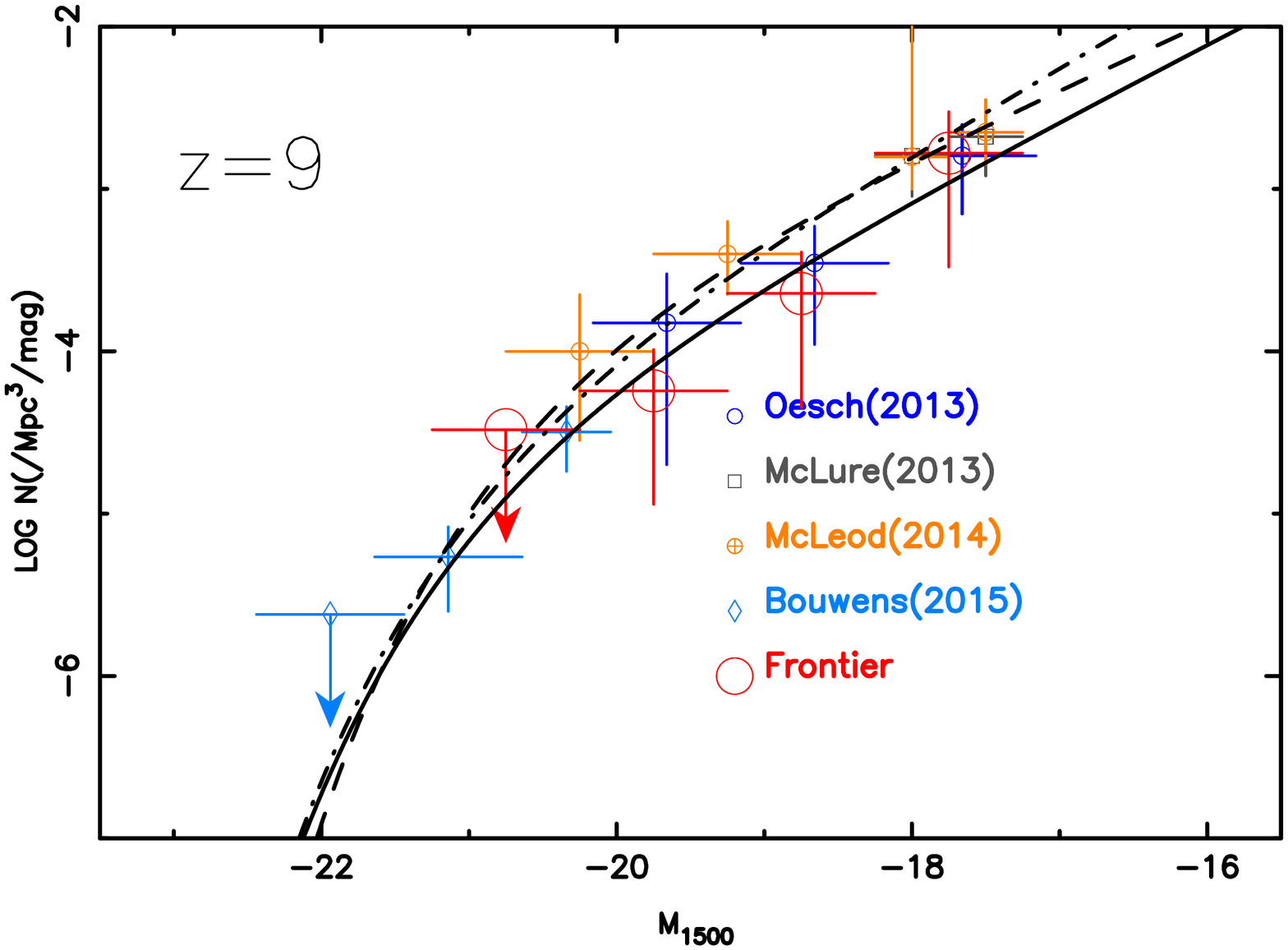}
      \caption{\label{LF_z9} UV Luminosity Function at $z$$\sim$9 computed using the first half of the Frontier Fields data. Number densities estimated from this study are in red, we over-plotted results from others groups using others datasets (\citealt{2015arXiv150601035B}, \citealt{2013MNRAS.432.2696M}, \citealt{2014arXiv1412.1472M}, \citealt{2013ApJ...773...75O} and \citealt{2011MNRAS.414.1455L}).The solid line displays the parametrization we deduced from this study, the dot-dashed line shows the shape published by \citet{2013MNRAS.432.2696M} and the dashed line is from \citet{2015arXiv150601035B}  }
   \end{figure}

\begin{figure}
   \centering   
           \includegraphics[width=9.5cm]{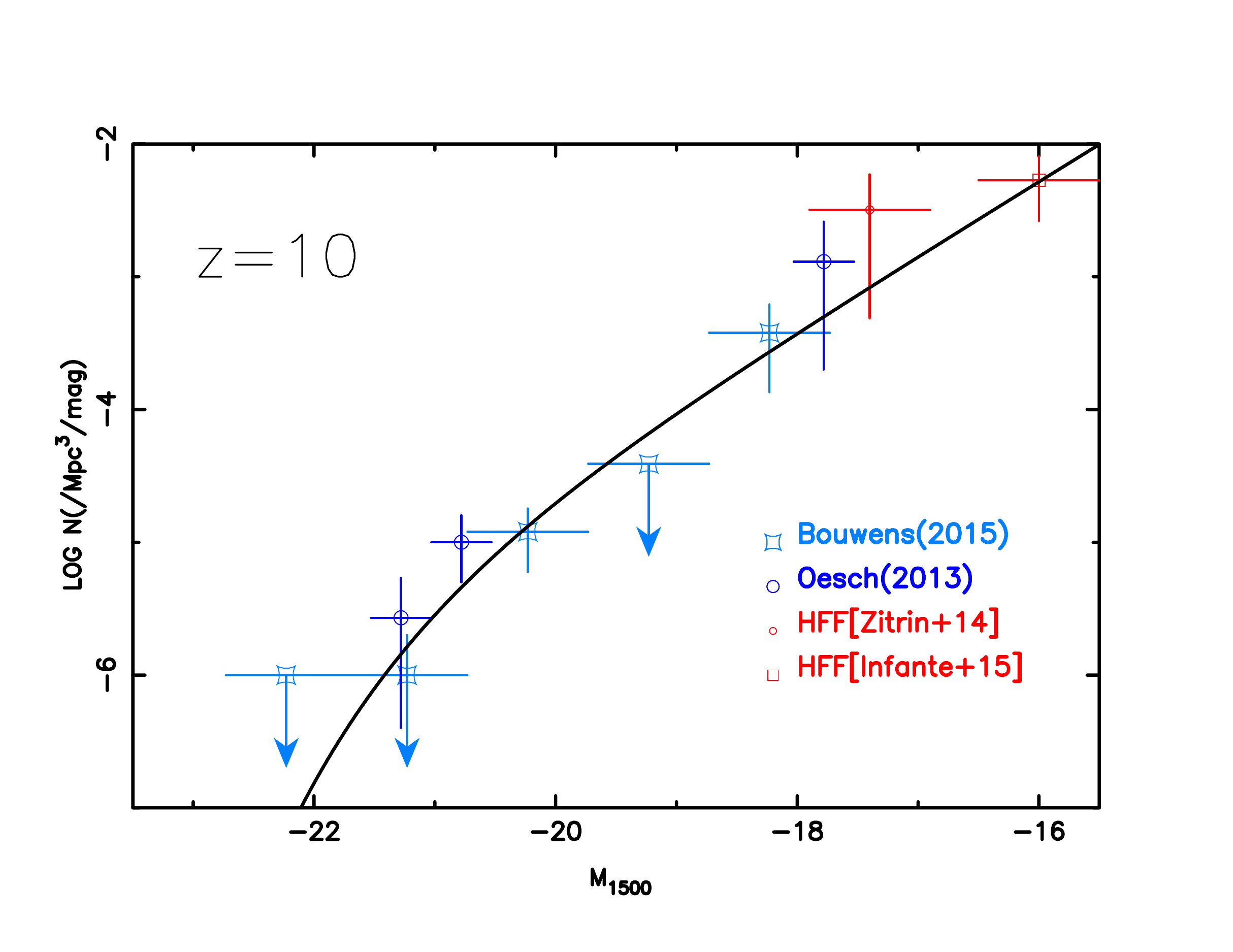}
      \caption{\label{LF_z10} UV Luminosity Function at $z$$\sim$10 computed using the first half of the Frontier Fields data. No $z\sim$10 objects has been selected in the last Frontier Fields dataset, we computed number densities based on previous $z$$\sim$10 candidates selected on the two first Frontier Fields dataset (\citealt{} and \citealt{Infante2015}). The solid line displays the parametrization published in \citet{2015ApJ...803...34B}  }
   \end{figure}

\begin{figure*}
   \centering
           \includegraphics[width=8.5cm]{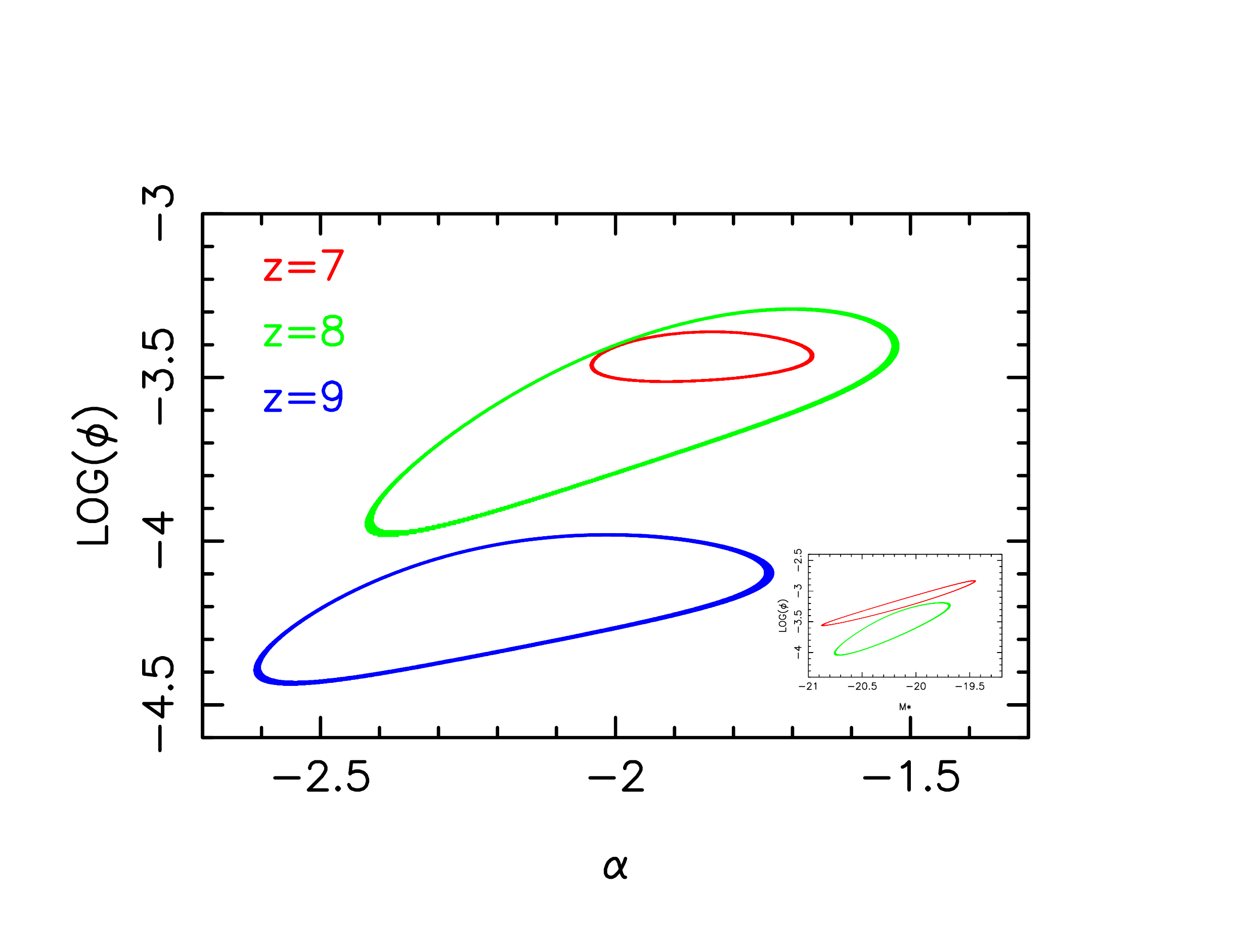} 
 	  \includegraphics[width=8.5cm]{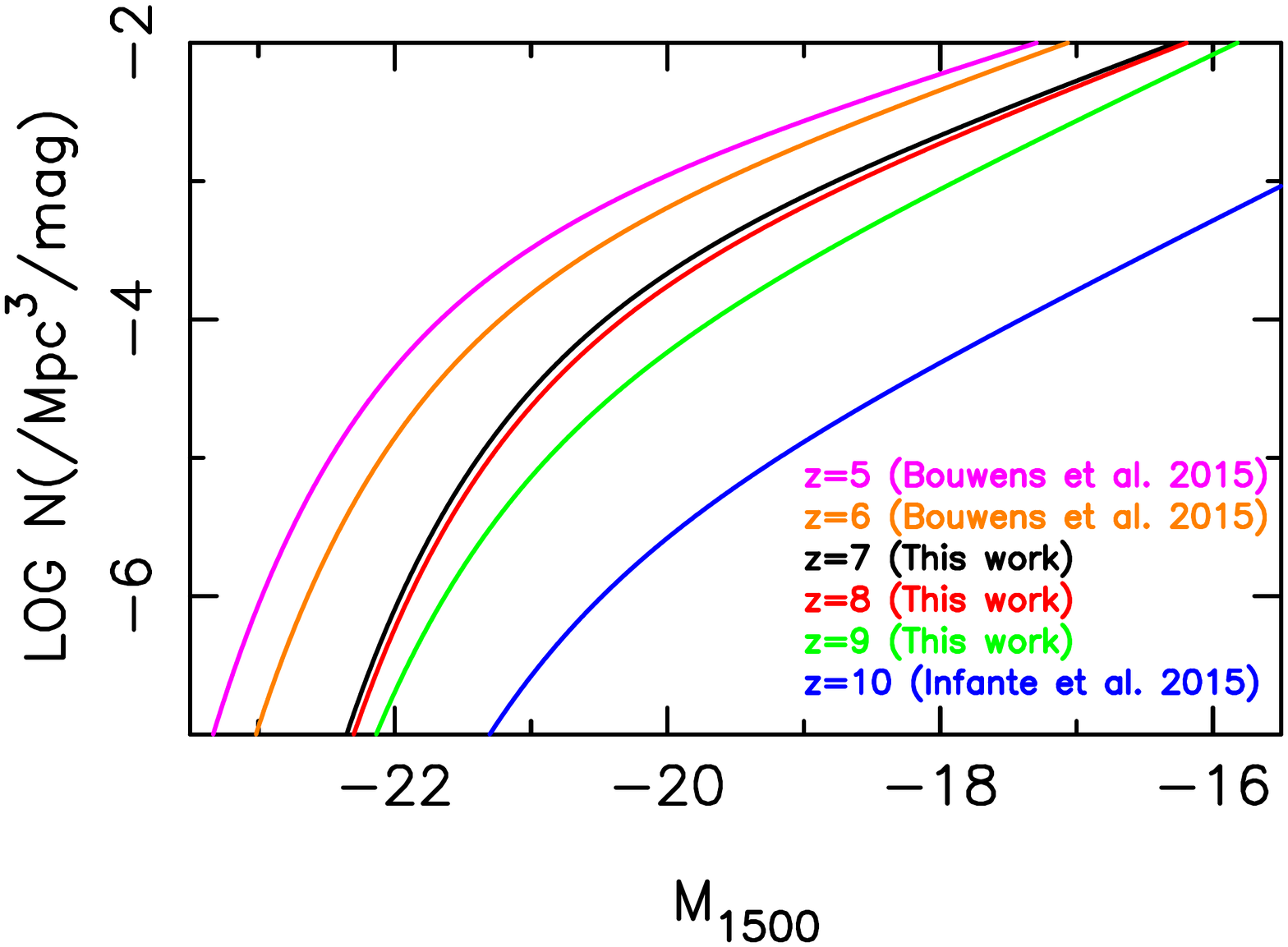} 

      \caption{\label{LF_evol} \textit{(left)} 1$\sigma$ confidence intervals on the Schechter parameterization we deduced from number densities computed using all selected objects in the three first Frontier Fields. It shows a strong evolution between $z\sim$8 and 9 of the $\Phi^{\star}$ parameter. The smaller panel shows the 1$\sigma$ confidence intervals for M$^{\star}$ and $\Phi^{\star}$ confirming an evolution of $\Phi^{\star}$ parameter. } \textit{(right)} Evolution of the UV LF found in this study at $z\sim$7, 8 and 9. For comparison purpose we over-plotted the shape of the UV LF published in \citet{2015ApJ...803...34B} at $z\sim$5 and 6 and \citet{Infante2015} at $z\sim$10. 
   \end{figure*}

In order to study the evolution of the shape of the UV LF, we adopted the Schechter parameterization \citep{1976ApJ...203..297S} and estimated the three parameters, so-called M$^{\star}$, $\Phi^{\star}$ and $\alpha$, using a $\chi^2$ minimization method and previous published densities covering other luminosities ranges. Table \ref{LF_parameters} presents the parameterization for the redshift range covered in this study and Figures \ref{LF_z7}, \ref{LF_z8}, \ref{LF_z9}, \ref{LF_z10} show the shape of the UV LF at $z\sim$7, 8, 9 and 10, respectively.  With half of the full FF data, we are probing the faint-end of the UV LF up to the highest redshift and confirm the shape found by previous studies. However, it appears that the evolution between $z$$\sim$8 and 9 is stronger than what has been previously observed (see Fig. \ref{LF_evol}), suggesting a deficit of $z$$\sim$8.5 objects (see Sec. \ref{deficit}). The evolution of the 1$\sigma$ confidence intervals from $z\sim$7 to 9 shows a clear evolution in $\Phi^{\star}$ as already noticed by \citet{2015ApJ...803...34B} with relatively small evolution in $\alpha$ (see Fig.  \ref{LF_evol}).





\subsection{A deficit of $z$$>$8.5 galaxies ?}
\label{deficit}

The number of $z$$\ge$8 objects selected behind MACS0717 is lower than what has been found behind the two first FFs clusters. We computed the expected number of $z$$\sim$8 galaxies detected at 5$\sigma$ in the MACS0717 FF data assuming the UV LF evolution published in \citet{2015ApJ...803...34B} and the mass model provided by the CATS team. Taking into account uncertainties on the LF parameters, the number of $z$$>$7.5 objects should be 2.98$^{+5.55}_{-1.14}$, showing that at least 1 object should be detected on the FFs images. However, the area effectively covered at very high-$z$ redshift by \textit{HST} images is small enough to be strongly affected by Cosmic Variance (CV). We used the method described in \citet{2008ApJ...676..767T} to account for CV in the expected number of objects. Based on the intervalle of $z$$>$8 objects detected at more than 5$\sigma$ in our data, the CV enlarges the range of expected objects to between 0 to 10.6 such that an absence of any $z$$>8$ candidates behind MACS0717 cluster is possible.

 \subsection{The Star Formation Rate Density}
\label{SFRd}

One can constrain the role played by the first galaxies during the epoch of reionization by estimating the densities of UV photons they produced and how these densities evolve with redshift (e.g., \citealt{2015arXiv150308228B}). This quantity is related to the SFRd occurring as a function of redshift, and is deduced from the following equation :
\begin{equation}
\rho_{SFR}=1.25 \times 10^{-28} \int^{\infty}_{0.03L^{\star}_{z=3}} \Phi(L_{1500})dL_{1500}
\end{equation}
where $ \Phi(L_{1500})$ is the UV LF estimated in the previous section(e.g., \citealt{2005ApJ...619L..47S} ). Thanks to the magnification applied by lensing clusters, we can integrate the UV LF down to 0.03$L^{\star}_{z=3}$ (i.e. $M_{1500}$$\sim$-17).  

We corrected these densities for dust attenuation following the method described in \citet{2005ApJ...619L..47S} with the $\beta$ slopes published in \citet{2012ApJ...754...83B}. In order to have a homogeneous determination of the star formation rate densities, we used previous UV LF parameterizations in several redshift intervals published in \citet{2005ApJ...619L..15W}, \citet{2010A&A...523A..74V}, \citet{2009MNRAS.395.2196M}, \citet{2010ApJ...725L.150O},\citet{2009ApJ...692..778R}, \citet{2012ApJ...759..135O}, \citet{2015ApJ...803...34B}, \citet{2014arXiv1412.1472M}. We deduced 1$\sigma$ errors bars on each density based on uncertainties on the Schechter parameters, however in cases where the parameters are fixed to a given value, we assumed uncertainties of 0.20, 0.20 or 20\% of the values respectively for $\alpha$, $M^{\star}$ and $\Phi^{\star}$. 

The densities computed using half of the full FF data are in good agreement with previous results at $z<$8, and confirm the change of the slope in the evolution of the SFRd as function of redshift beyond $z\sim$8 (\citealt{2011Natur.469..504B}, \citealt{2014ApJ...786..108O}, \citealt{2015ApJ...799...12I}). The evolution of the SFRd as a function of redshift could be well fitted by the equation given in \citet{2001MNRAS.326..255C} up to $z\sim$8 and given by: 
\begin{equation}
\rho_{SFR}(z) = \frac{a+bz}{1+\Big(\frac{z}{c}\Big)^d}h
\end{equation}
where we estimated (a,b,c,d)=(0.0,0.05,2.55,3.30) using $\chi^2$ minimization.

However the previous parameterization does not take into account the slope change at $z>8$, that is well fitted by equation 39 of \citet{2015ApJ...799...12I} given by :
\begin{equation}
\rho_{SFR}(z) = \frac{2\rho_{UV,z=8}}{10^{a(z-8)}+10^{b(z-8)}}
\end{equation}
where (a,b)=(0.21,0.58) were estimated by $\chi^2$ minimization. 

\begin{figure*}
   \centering   
           \includegraphics[width=15.5cm]{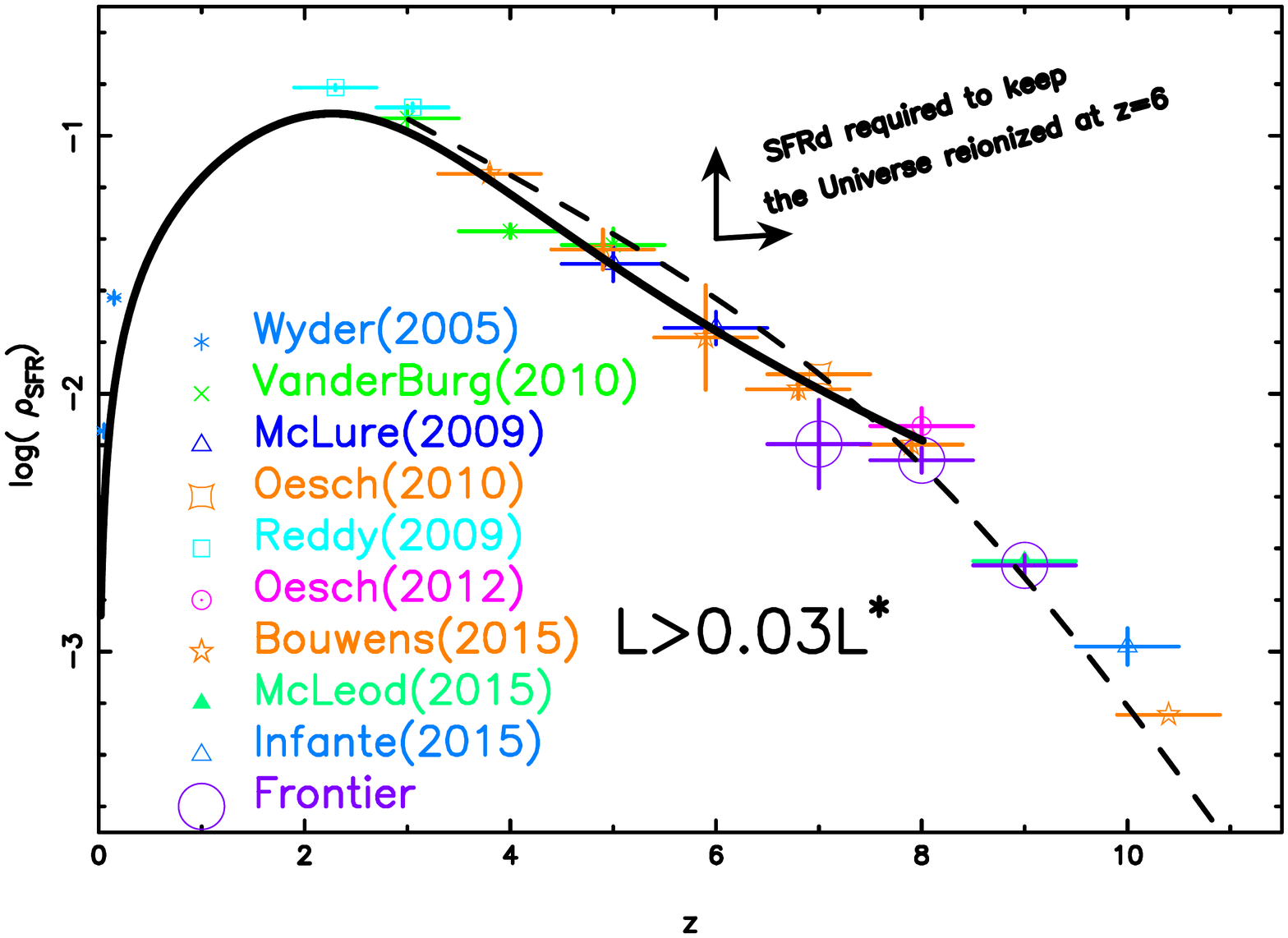}
      \caption{\label{sfrd_fit} Evolution of the SFRd including densities deduced from the half Frontier Fields dataset. We compared these results with previous measurement published in \citet{2005ApJ...619L..15W}, \citet{2010A&A...523A..74V}, \citet{2009MNRAS.395.2196M}, \citet{2010ApJ...725L.150O}, \citet{2009ApJ...692..778R}, \citet{2012ApJ...759..135O}, \citet{2015ApJ...803...34B}, \citet{2014arXiv1412.1472M}. Two parameterizations are over-plotted: the solid-line shows the shape published in \citet{2001MNRAS.326..255C} and the dashed-line displays the evolution as seen by \citet{2015ApJ...799...12I}. }
   \end{figure*}

 Figure \ref{sfrd_fit} shows this evolution compared with the SFRd required to keep the Universe reionized as deduced from \citet{1999ApJ...514..648M}. We computed this limit using a clumping factor of $C=6$ according to \citet{2009MNRAS.394.1812P} and consistent with recent simulations published by \citet{2015ApJ...810..154K}. The escape fraction was estimated following \citet{2013MNRAS.431.2826F}  $f_{esc}\sim$0.08, which is in good agreement with the recent upper limit published by \citet{2015arXiv151108504B}. We corrected this value for dust extinction, which is neglected in the \citet{1999ApJ...514..648M} equation following the method described above.
We noticed that the SFRd observed for galaxies at $z$$\sim$6  with $L_{1500}$$>$0.03$L^{\star}_{z=3}$ is still lower than what is expected to keep the Universe reionized. However, if we used extreme values of the two parameters, $f_{esc}$$\sim$0.13 and $C$$\sim$2 we start to reconcile the observed SFRd produced by L$>$0.03$L^{\star}_{z=3}$ galaxies with the SFRd required to keep the Universe reionized. 


\section{Conclusions}

After 1.5 years of observations, the FF program has already provided extremely deep data around 4 galaxies clusters, Abell 2744, MACSJ0416-2403, MACSJ0717+3745 and MACS1149.5+2223, helping to increase the number of $z$$>$6 objects currently known. In this study, we selected 39 $z$$>$6 objects using the Lyman Break technique in the two datasets provided by this legacy program (cluster and parallel fields). We confirmed the non-detection at optical wavelength of our candidates by using an optical $\chi^2$ method that takes into account the position of our objects in the cluster fields. A comparison between our samples and those published using shallower optical data (e.g. CLASH) demonstrates the crucial role played by extremely deep optical data to remove extreme mid-$z$ interlopers. In this way, we have been able to identify 4 mid-$z$ interlopers. The size of our sample at $z$$\sim$6-7 is comparable to previous findings, however the number of $z$$>8$ objects is much lower than what has been found in the two first frontier fields clusters and could be explained by Cosmic Variance. 

We combined the $z$$>$6 objects selected on MACS0717 datasets with all objects previously selected on the two first FF clusters, increasing the number of candidates to 100. We computed photometric redshifts for our candidates from two independent approaches, $\chi^2$ minimization and a Bayesian method, and demonstrated that the results are in good agreement. Based on SED-fitting, we deduced physical properties of our candidates, such as the SFR, the reddening, the stellar mass and Age, and studied the relationship between several properties. Thus we confirmed the trend observed previously in the evolution of  SFR as function of  galaxy mass as well as in the evolution of the size of galaxies as a function of the UV luminosity at very high-redshift. 

Thanks to the amplification of the light by the cluster, the majority of sources are faint and give us an opportunity to add robust constraints on the faint-end of the UV LF at very high-$z$. We confirmed the shape of the UV LF at $z$$\sim$7 and 8 up to $M_{1500}$=-16.5. However, due to the absence of $z$$>$8.5 objects behind MACS0717 and the small number of candidates selected on the two previous FF dataset, we confirmed that the evolution of the UV LF from $z$$\sim$8 to 9 could be stronger than what is observed between $z\sim$7 and 8. We used the LF parameterization to estimate the SFR densities produced by the galaxies up to $z$$\sim$10, and confirmed the change in the evolution of SFRd between $z$$\sim$8 and 10.

All objects discussed in these papers have been selected from photometric datasets carried out with the HST. We discussed in section \ref{contaminants} the contamination rate of our sample, and demonstrated that to date it appears difficult to identify which objects could be mid-$z$ interlopers without spectroscopic observations. However few targets identified behind MACSJ0717.5+3745 are bright enough to be observed with current NIR facilities (e.g. MOSFIRE/Keck, EMIR/GTC). Spectroscopic confirmation is absolutely essential to assess the photometrically based conclusion obtained to date, particularly in light of the small number of objects currently confirmed by spectroscopy (\citealt{2015ApJ...804L..30O},  \citealt{2013Natur.502..524F}).

\acknowledgements
Authors thank the anonymous referee for his/her useful comments that strongly improve the quality of the paper. We acknowledge support from CONICYT-Chile grants Basal-CATA PFB-06/2007 (NL, LI, FEB, SK ), Gemini-CONICYT \#32120003 (NL), "EMBIGGEN" Anillo ACT1101 (FEB), FONDECYT 1141218 (FEB), FONDECYT Postdoctorado 3160122 (NL),  3140542 (PT)  and Project IC120009 ``Millennium Institute of Astrophysics (MAS)'' of the Iniciativa Cient\'{\i}fica Milenio del Ministerio de Econom\'{\i}a, Fomento y Turismo (FEB) and the French Agence Nationale de la Recherche bearing the reference ANR-09-BLAN-0234 (RP, DB). A.M. acknowledges the financial support of the Brazilian funding agency FAPESP (Post-doc fellowship - process number 2014/11806-9). This work been  supported  by  award  AR-13279  from  the  Space  Telescope Science Institute (STScI), which is operated by the Association of Universities for Research in Astronomy, Inc. under  NASA  contract  NAS  5-26555. IC acknowledges the support from Smithsonian Astrophysical Observatory Telescope Data Center and from the grants MD-7355.2015.2 by the research council of the president of the Russian Federation,15-32-21062 and 15-52-15050 by the Russian Foundation for Basic Research.
This work is based on observations made with the NASA/ESA Hubble Space Telescope, obtained at the Space Telescope Science Institute (STScI), which is operated by the Association of Universities for Research in Astronomy, Inc., under NASA contract NAS 5-26555. The HST image mosaics were produced by the Frontier Fields Science Data Products Team at STScI. This work is based in part on observations made with the Spitzer Space Telescope, which is operated by the Jet Propulsion Laboratory, California Institute of Technology under a contract with NASA.
This work utilizes gravitational lensing models produced by PIs Bradac, Ebeling, Merten \& Zitrin, Sharon, and Williams funded as part of the HST Frontier Fields program conducted by STScI. STScI is operated by the Association of Universities for Research in Astronomy, Inc. under NASA contract NAS 5-26555. The lens models were obtained from the Mikulski Archive for Space Telescopes (MAST).

\begin{table*}
\footnotesize
\centering                          
\caption{\label{cluster_sample} $6 \leq z \leq 8$ objects selected on the cluster field. }
\begin{tabular}{l | c c | c  c c  c c | c c | c }        
\hline\hline                 
ID 	& RA 	& DEC 	& F814W & F105W 	& F125W  &  F140W & F160W  & 3.6$\mu$m & 4.5$\mu$m & $\chi^2_{opt}$\\    
         &[J2000]  & [J2000]  & [AB] 	& [AB]  	& [AB]     	& [AB]	& [AB] & [AB] & [AB]  	  &    \\          
 \hline                        

2927	&109.3892755	&37.7248568	&29.19		&26.87		&26.71		&26.73		&26.74		&25.49			&25.43 			&  -0.38\\
	&		&		&$\pm$0.21	&$\pm$0.05	&$\pm$0.04	&$\pm$0.05	&$\pm$0.05			&$\pm$0.14		&$\pm$0.10		&	\\
9313	&109.381054	&37.7316083	&28.83		&26.91		&26.87		&27.04		&27.16		&\textit{blended}	&\textit{blended} 	& -0.62	\\
	&		&		&$\pm$0.15	&$\pm$0.05	&$\pm$0.05	&$\pm$0.06	&$\pm$0.07			&-				&-				&	\\
12325	&109.4136628	&37.7346385	&29.61		&26.72		&26.68		&26.81		&26.96	&$>$26.60		&$>$26.60 		& 0.03	\\
	&		&		&$\pm$0.31	&$\pm$0.05	&$\pm$0.04	&$\pm$0.05	&$\pm$0.06			&-				&-				&	\\
13963	&109.3770165	&37.7364332	&29.29		&26.77		&26.64		&26.86		&26.80	&$>$26.60		&$>$26.60 		& 0.13	\\
	&		&		&$\pm$0.23	&$\pm$0.05	&$\pm$0.04	&$\pm$0.05	&$\pm$0.05			&-				&-				&	\\
25550	&109.4159	&37.7467276	&$> 30.2$		&27.73		&27.56		&27.83		&27.77	&$>$26.60		&$>$26.60		& 0.08	\\
	&		&		&-		&$\pm$0.11	&$\pm$0.10	&$\pm$0.13	&$\pm$0.13				&-				&-				&	\\
29413	&109.3814076	&37.7503301	&29.09		&26.67		&26.84		&26.64		&26.56	&\textit{blended}	&\textit{blended} 	& 0.09 	\\
	&		&		&$\pm$0.19	&$\pm$0.04	&$\pm$0.05	&$\pm$0.04	&$\pm$0.04			&-				&-				&	\\
30458	&109.3862351	&37.7519202	&28.98		&26.45		&26.30		&26.39		&26.44	&\textit{blended}	&\textit{blended}	& 0.00	\\
	&		&		&$\pm$0.18	&$\pm$0.04	&$\pm$0.03	&$\pm$0.04	&$\pm$0.04			&-				&-				&	\\
33447	&109.4090663	&37.7546801	&28.39		&26.21		&26.27		&26.29		&26.41	&25.21			&25.61 			& -0.14	\\
	&		&		&$\pm$0.10	&$\pm$0.03	&$\pm$0.03	&$\pm$0.03	&$\pm$0.04			&$\pm$0.14		&$\pm$0.13		&	\\
46206	&109.3990963	&37.7649606	&29.16		&25.98		&25.79		&25.92		&25.89	&\textit{blended}	&\textit{blended}	& -0.87	\\
	&		&		&$\pm$0.21	&$\pm$0.02	&$\pm$0.02	&$\pm$0.02	&$\pm$0.02			&-				&-				&	\\
3119	&109.3854632		&37.7251234	&29.91		&27.87		&28.28		&27.98		&27.69	&$>$26.60		&$>$26.60 		& -1.41	 \\
	&		&		&$\pm$0.41	&$\pm$0.13	&$\pm$0.19	&$\pm$0.15	&$\pm$0.12			&-				&-				&	\\
25990	&109.3694898	&37.7470086	&30.43		&27.83		&28.08		&28.03		&28.00	&$>$26.60		&$>$26.60 		& 0.17	\\
	&		&		&$\pm$0.67	&$\pm$0.13	&$\pm$0.16	&$\pm$0.16	&$\pm$0.16			&-				&-				&	\\
15440 	&109.39233   	&37.738083	&27.22		&26.51 		&26.60 		&26.67 		&26.90	&\textit{blended}	&\textit{blended}	&  0.19 \\
	&			&			&$\pm$0.02	&$\pm$0.01	&$\pm$0.01	&$\pm$0.01	&$\pm$0.01    		&-				&-				& \\
21962	&109.40773	&37.742736	&27.99		&26.69		&26.74		&26.77		&26.90	&\textit{blended}	&\textit{blended}	& 0.20  \\
		&			&			&$\pm$0.04	&$\pm$0.01	&$\pm$0.01	&$\pm$0.01	&$\pm$0.01 &-				&-				& \\   
46005	&109.3988091	&37.7650708	&29.22		&26.86		&26.94		&27.05		&27.08	&$>$26.60		&$>$26.60 		& 0.13	\\
	&		&		&$\pm$0.22	&$\pm$0.05	&$\pm$0.06	&$\pm$0.07	&$\pm$0.07			&-				&-				&	\\ \hline
802	&109.3864548		&37.7346659	&$> 30.2$		&27.09		&27.14		&26.69		&26.73	&$>$26.60		&$>$26.60		& 0.26	\\
	&		&		&-			&$\pm$0.06	&$\pm$0.07	&$\pm$0.05	&$\pm$0.05	&-		&-	&	\\
16621	&109.4186495	&37.7387916	&29.74		&27.86		&28.05		&27.86		&28.24	&$>$26.60		&$>$26.60		& 0.24	\\
	&		&		&$\pm$0.35	&$\pm$0.13	&$\pm$0.15	&$\pm$0.14	&$\pm$0.19	&-		&-	&	\\
47376	&109.4008562	&37.7662314	&29.57		&27.95		&27.86		&27.46		&27.81	&$>$26.60		&$>$26.60 		& 0.21	\\
	&		&		&$\pm$0.30	&$\pm$0.14	&$\pm$0.13	&$\pm$0.10	&$\pm$0.13	&-		&-	&	\\
15756	&109.3790232	&37.7383872	&30.09		&28.07		&28.00		&28.32		&28.20	&$>$26.60		&$>$26.60 		& 0.84	\\
	&		&		&$\pm$0.49	&$\pm$0.16	&$\pm$0.15	&$\pm$0.21	&$\pm$0.19	&-		&-	&	\\ 
17265	&109.3912133	&37.7391643	&29.63		&26.90		&26.74		&26.70		&26.73	&\textit{blended}	&\textit{blended} 	& 1.46	\\
	&		&		&$\pm$0.32	&$\pm$0.05	&$\pm$0.05	&$\pm$0.05	&$\pm$0.05	&-		&-	&	\\
20756	&109.3776056	&37.7417947	&28.73		&26.61		&26.68		&26.37		&26.46	&$>$26.60		&$>$26.60		& 0.66 	\\
	&		&		&$\pm$0.14	&$\pm$0.04	&$\pm$0.04	&$\pm$0.04	&$\pm$0.04	&-		&-	&	\\
28748	&109.3854382	&37.7499249	&29.16		&27.40		&27.44		&27.23		&27.23	&$>$26.60		&$>$26.60		& 0.88	\\
	&		&		&$\pm$0.21	&$\pm$0.08	&$\pm$0.09	&$\pm$0.08	&$\pm$0.08	&-		&-	&	\\
45614	&109.3950838	&37.7644073	&30.20		&27.53		&27.90		&27.32		&27.46	&$>$26.60		&$>$26.60 		& 0.96	\\
	&		&		&$\pm$0.54	&$\pm$0.10	&$\pm$0.13	&$\pm$0.08	&$\pm$0.09	&-		&-	&	\\
12402	&109.4128542	&37.7338042	&29.70		&26.51		&26.46		&26.65		&26.65	&25.78		&25.59 		& 0.62	\\
	&		&		&$\pm$0.34	&$\pm$0.04	&$\pm$0.04	&$\pm$0.05	&$\pm$0.04	&$\pm$0.42		&$\pm$0.35	&	\\
13806	&109.3803311	&37.7366722	&29.80		&28.19		&28.37		&28.78		&28.41	&$>$26.60		&$>$26.60 		& 0.73	\\
	&		&		&$\pm$0.37	&$\pm$0.17	&$\pm$0.21	&$\pm$0.32	&$\pm$0.23	&-		&-	&	\\
14977	&109.4132994	&37.7374793	&30.44		&28.07		&28.14		&28.03		&28.29	&$>$26.60		&$>$26.60 		& 1.45	\\
	&		&		&$\pm$0.68	&$\pm$0.16	&$\pm$0.17	&$\pm$0.16	&$\pm$0.20	&-		&-	&	\\
26338	&109.3657244	&37.7474107	&$> 30.2$		&27.95		&27.87		&28.14		&27.89	&$>$26.60		&$>$26.60 		& 4.56	\\
	&		&		&-		&$\pm$0.14	&$\pm$0.13	&$\pm$0.18	&$\pm$0.14		&-		&-	&	\\
28488	&109.3698122	&37.7486357	&30.16		&27.69		&27.33		&27.75		&27.06	&$>$26.60		&$>$26.60 		& 1.21	\\
	&		&		&$\pm$0.52	&$\pm$0.11	&$\pm$0.08	&$\pm$0.12	&$\pm$0.07	&-		&-	&	\\
45217	&109.3968464	&37.7630624	&28.40		&27.26		&27.37		&27.07		&27.04	&$>$26.60		&$>$26.60 		& 0.33	\\
	&		&		&$\pm$0.10	&$\pm$0.07	&$\pm$0.08	&$\pm$0.07	&$\pm$0.06	&-		&-	&	\\
\hline
\hline                                   
\end{tabular}
\small

All error bars are computed from noise measured in 0.4'' diameter apertures distributed over each object. The last column displays the $\chi^2_{opt}$, all objects above the solid line have a $\chi^2_{opt}$ consistent with a non-detection in optical.
\end{table*}

\begin{table*}
\caption{\label{parallel_z8_sample}  $z \gtrsim 8$ objects selected in the parallel field.}
\footnotesize
\centering                          
\begin{tabular}{l | c c | c  c c  c  | c c | c }        
\hline\hline                 
ID 	& RA 	& DEC 	&  F105W 	& F125W  &  F140W & F160W  &  3.6$\mu$m & 4.5$\mu$m & $\chi^2_{opt}$\\    
         &[J2000]  & [J2000]  & [AB]  	& [AB]     	& [AB]	& [AB]  & [AB]  & [AB]  	  &    \\          
 \hline                        
44317	&109.3234246	&37.8278453	&29.14		&28.59		&28.07		&28.85 &$>$26.60		&$>$26.60 		& -2.53 \\
	&		&		&$\pm$0.12	&$\pm$0.10	&$\pm$0.05	&$\pm$0.18 &-		&-	& \\
30169	&109.3245233	&37.8237391	&28.67		&27.51		&27.67		&27.70 &$>$26.60		&$>$26.60 		& -0.21 \\
	&		&		&$\pm$0.08	&$\pm$0.04	&$\pm$0.04	&$\pm$0.06 &-		&-	& \\
39832	&109.3397497	&37.8269393	&29.75		&28.86		&28.66		&29.25 &$>$26.60		&$>$26.60 		& -0.11 \\
	&		&		&$\pm$0.21	&$\pm$0.13	&$\pm$0.09	&$\pm$0.27 &-		&-	& \\  \hline
30759	&109.3320764	&37.8234454	&28.09		&27.32		&27.27		&27.37 &$>$26.60		&$>$26.60 		& 5.36 \\
	&		&		&$\pm$0.05	&$\pm$0.03	&$\pm$0.02	&$\pm$0.05 &-		&-	& \\
87051	&109.3312151	&37.8458071	&27.76		&27.25		&26.99		&27.28 &$>$26.60		&$>$26.60 		& 0.48 \\
	&		&		&$\pm$0.03	&$\pm$0.03	&$\pm$0.02	&$\pm$0.04 &-		&-	& \\
7588	&109.3288254	&37.8149487	&28.37		&27.47		&27.64		&27.92 &$>$26.60		&$>$26.60 		& 1.96 \\
	&		&		&$\pm$0.06	&$\pm$0.04	&$\pm$0.03	&$\pm$0.08 &-		&-	& \\
49505	&109.3437961	&37.829237	&29.45		&27.83		&27.66		&27.92 &$>$26.60		&$>$26.60 		& 0.96 \\
	&		&		&$\pm$0.16	&$\pm$0.05	&$\pm$0.04	&$\pm$0.08 &-		&-	& \\
81697	&109.32343	&37.8432187	&28.30		&27.78		&27.66		&27.57 &$>$26.60		&$>$26.60 		& 1.27 \\
	&		&		&$\pm$0.06	&$\pm$0.05	&$\pm$0.04	&$\pm$0.06 &-		&-	& \\
\hline
\hline                                   
\end{tabular}
\small

All error bars are computed from noise measured in 0.4'' diameter apertures distributed over each object. The last column displays the $\chi^2_{opt}$, all objects above the solid line have a $\chi^2_{opt}$ consistent with a non-detection in optical.
\end{table*}

\begin{table*}
\caption{\label{parallel_z6_sample} $6 \leq z \leq 8$ objects selected in the parallel field.}
\tiny
\centering                          
\begin{tabular}{l | c c | c  c c  c c | c c | c }        
\hline\hline                 
ID 	& RA 	& DEC 	& F814W & F105W 	& F125W  &  F140W & F160W  & 3.6$\mu$m & 4.5$\mu$m &  $\chi^2_{opt}$\\    
         &[J2000]  & [J2000]  & [AB] 	& [AB]  	& [AB]     	& [AB]	& [AB]  	 & [AB]  	 & [AB]  	  &    \\          
 \hline                        
2035	&109.3133037	&37.8101138	&$> 30.5$		&27.33		&27.07		&27.15		&27.27	&$>$26.60		&$>$26.60 		& -0.09 \\ 
	&			&			&-			&$\pm$0.02	&$\pm$0.03	&$\pm$0.02	&$\pm$0.04	&- &- & \\ 
6576	&109.3167913	&37.8144653	&$> 30.5$			&28.36		&28.03		&27.92		&27.80	&$>$26.60		&$>$26.60 		& -0.51 \\
	&		&		&-		&$\pm$0.06	&$\pm$0.06	&$\pm$0.04	&$\pm$0.07	&- &- & \\ 
10738	&109.3218622	&37.8167184	&29.51		&27.78		&27.92		&28.19		&27.89	&$>$26.60		&$>$26.60 		& 0.05 \\
	&		&		&$\pm$0.21	&$\pm$0.03	&$\pm$0.06	&$\pm$0.06	&$\pm$0.08	&- &- & \\ 
24830	&109.3045523	&37.8218034	&$> 30.5$			&27.36		&27.31		&27.34		&27.20	&$>$26.60		&$>$26.60 		& -0.16  \\
	&		&		&-		&$\pm$0.02	&$\pm$0.03	&$\pm$0.03	&$\pm$0.04	&- &- & \\ 
26762	&109.3026294	&37.8226524	&$> 30.5$			&27.44		&27.71		&27.52		&27.49	&$>$26.60		&$>$26.60 		& -0.15  \\
	&		&		&-		&$\pm$0.03	&$\pm$0.05	&$\pm$0.03	&$\pm$0.05	&- &- & \\ 
32445	&109.3265426	&37.8243045	&29.17		&26.95		&27.02		&27.12		&27.09	&$>$26.60		&$>$26.60 		& -0.43 \\
	&		&		&$\pm$0.16	&$\pm$0.02	&$\pm$0.02	&$\pm$0.02	&$\pm$0.04	&- &- & \\ 
33421	&109.3168345	&37.8243211	&$> 30.5$			&27.28		&27.09		&27.23		&27.33	&$>$26.60		&$>$26.60 		& -0.07 \\
	&		&		&-		&$\pm$0.02	&$\pm$0.03	&$\pm$0.02	&$\pm$0.05	&- &- & \\ 
37890	&109.3431884	&37.8259591	&29.58		&27.87		&27.94		&28.11		&28.03	&$>$26.60		&$>$26.60 		& -0.39  \\
	&		&		&$\pm$0.23	&$\pm$0.04	&$\pm$0.06	&$\pm$0.05	&$\pm$0.09	&- &- & \\ 
39809	&109.348495	&37.8265778	&$> 30.5$			&27.55		&27.01		&27.07		&27.11	&$>$26.60		&$>$26.60 		& -0.06 \\
	&		&		&-		&$\pm$0.03	&$\pm$0.02	&$\pm$0.02	&$\pm$0.04	&- &- &\\
42718	&109.3511417	&37.827195	&$> 30.5$			&26.93		&26.86		&27.03		&27.11	& 25.86 	& 27.23	& -0.07 \\	
	&		&		&-		&$\pm$0.02	&$\pm$0.02	&$\pm$0.02	&$\pm$0.04	& 0.15	&0.74 	&\\
43555	&109.3323968	&37.8275686	&$> 30.5$			&27.27		&27.18		&26.98		&26.98	&\textit{blended}	&\textit{blended} 	& -0.11 \\
	&		&		&-		&$\pm$0.02	&$\pm$0.03	&$\pm$0.02	&$\pm$0.03	&- &- &\\
46175	&109.3288319	&37.828303	&$> 30.5$			&27.74		&27.30		&27.99		&27.66	&$>$26.60		&$>$26.60 		& -0.06 \\
	&		&		&-		&$\pm$0.03	&$\pm$0.03	&$\pm$0.05	&$\pm$0.06	&- &- &\\
46719	&109.3072346	&37.8284		&$> 30.5$			&27.65		&27.84		&28.09		&28.05	&$>$26.60		&$>$26.60 		& 0.02 \\
	&		&		&-		&$\pm$0.03	&$\pm$0.05	&$\pm$0.05	&$\pm$0.09	&- &- &\\
50815	&109.3202576	&37.8295874	&29.27		&28.31		&28.26		&27.84		&27.87	&$>$26.60		&$>$26.60 		& -0.27 \\
	&		&		&$\pm$0.17	&$\pm$0.06	&$\pm$0.08	&$\pm$0.04	&$\pm$0.07	&- &- &\\
58730	&109.3475986	&37.8316658	&29.78		&26.24		&25.99		&26.06		&26.03	&\textit{blended}	&\textit{blended} 	& -1.30  \\
	&		&		&$\pm$0.27	&$\pm$0.01	&$\pm$0.01	&$\pm$0.01	&$\pm$0.01	&- &- & \\
66722	&109.3337826	&37.836336	&29.76		&28.47		&28.66		&28.97		&28.83	&$>$26.60		&$>$26.60 		& -1.01 \\
	&		&		&$\pm$0.27	&$\pm$0.06	&$\pm$0.11	&$\pm$0.12	&$\pm$0.18	&- &- & \\
91692	&109.3254561	&37.848214	&$> 30.5$			&28.49		&28.12		&28.26		&28.12	&$>$26.60		&$>$26.60 		& 0.08 \\
	&		&		&-		&$\pm$0.07	&$\pm$0.07	&$\pm$0.06	&$\pm$0.09	&- &- & \\
7406	&109.3273542	&37.8146525	&28.84		&26.30		&26.42		&26.24		&26.20	&\textit{blended}	&\textit{blended} 	& -0.06 \\
	&		&		&$\pm$0.12	&$\pm$0.01	&$\pm$0.01	&$\pm$0.01	&$\pm$0.02	&- &- & \\
17548	&109.3241209	&37.8190172	&28.59		&27.04		&27.10		&27.16		&27.02	&\textit{blended}	&\textit{blended} 	& -0.07 \\
	&		&		&$\pm$0.09	&$\pm$0.02	&$\pm$0.03	&$\pm$0.02	&$\pm$0.03	&- &- & \\
28313	&109.3082952	&37.8231455	&29.18		&27.35		&27.24		&27.51		&27.45	&$>$26.60		&$>$26.60 		& -0.11 \\
	&		&		&$\pm$0.16	&$\pm$0.02	&$\pm$0.03	&$\pm$0.03	&$\pm$0.05	&- &- & \\
49274	&109.3115613	&37.8291645	&$> 30.5$			&28.15		&27.92		&28.20		&28.27	&$>$26.60		&$>$26.60 		& -0.29 \\
	&		&		&-		&$\pm$0.05	&$\pm$0.06	&$\pm$0.06	&$\pm$0.11	&- &- & \\
70084	&109.3288391	&37.8376963	&$> 30.5$			&27.88		&27.22		&27.32		&26.78	&\textit{blended}	&\textit{blended} 	& -0.40 \\ 
	&		&		&-		&$\pm$0.04	&$\pm$0.03	&$\pm$0.03	&$\pm$0.03	&- &- & \\\hline
3014		&109.3241258	&37.8114557	&29.66		&27.85		&27.61		&27.63		&27.31	&$>$26.60		&$>$26.60 		& 0.65 \\
	&		&		&$\pm$0.25	&$\pm$0.04	&$\pm$0.04	&$\pm$0.03	&$\pm$0.04	&- &- & \\
58664	&109.3438555	&37.8320906	&29.64		&26.97		&26.99		&27.00		&27.16	&$>$26.60		&$>$26.60 		& 0.44  \\
	&		&		&$\pm$0.24	&$\pm$0.02	&$\pm$0.02	&$\pm$0.02	&$\pm$0.04	&- &- & \\
51380	&109.3420084	&37.8295913	&29.52		&27.37		&27.34		&27.19		&27.07	&$>$26.60		&$>$26.60 		& 0.39  \\
	&		&		&$\pm$0.21	&$\pm$0.02	&$\pm$0.03	&$\pm$0.02	&$\pm$0.04	&- &- &\\
32892	&109.3267033	&37.8244667	&29.42		&27.44		&27.37		&27.32		&27.34	&$>$26.60		&$>$26.60 		& 0.53 \\
	&		&		&$\pm$0.20	&$\pm$0.03	&$\pm$0.03	&$\pm$0.03	&$\pm$0.05	&- &- & \\ 
32407	&109.3210662	&37.8240635	&29.63		&26.75		&26.35		&26.32		&26.47	& 25.55	& 25.85	& 3.42 \\
	&		&		&$\pm$0.24	&$\pm$0.01	&$\pm$0.01	&$\pm$0.01	&$\pm$0.02	&$\pm$0.16  &$\pm$0.21 & \\
47840	&109.3335154	&37.8282783	&29.97		&28.59		&28.83		&29.14		&28.66	&$>$26.60		&$>$26.60 		& 1.28  \\
	&		&		&$\pm$0.33	&$\pm$0.07	&$\pm$0.13	&$\pm$0.14	&$\pm$0.15	&- &- & \\
56519	&109.3428398	&37.8316497	&29.36		&28.08		&27.93		&28.14		&28.28	&$>$26.60		&$>$26.60 		& 22.04 \\
	&		&		&$\pm$0.19	&$\pm$0.04	&$\pm$0.06	&$\pm$0.05	&$\pm$0.11	&- &- & \\
61782	&109.3227636	&37.8336724	&$> 30.5$			&29.26		&28.92		&29.52		&29.16	&$>$26.60		&$>$26.60 		& 1.45  \\
	&		&		&-		&$\pm$0.13	&$\pm$0.14	&$\pm$0.20	&$\pm$0.24	&- &- & \\
75653	&109.3184585	&37.8402601	&28.69		&27.38		&27.08		&27.21		&26.99	&$>$26.60		&$>$26.60 		& 1.50  \\
	&		&		&$\pm$0.10	&$\pm$0.02	&$\pm$0.03	&$\pm$0.02	&$\pm$0.03	&- &- & \\
83324	&109.30879	&37.8438485	&$> 30.5$			&27.16		&26.81		&26.64		&26.55	& 26.30 	& 26.84	& 2.07  \\
	&		&		&-		&$\pm$0.02	&$\pm$0.02	&$\pm$0.01	&$\pm$0.02	&$\pm$0.22  &$\pm$0.61 & \\
53875	&109.3309648	&37.8302942	&29.49		&27.47		&27.32		&27.15		&27.17	&$>$26.60		&$>$26.60 		& 5.64  \\
	&		&		&$\pm$0.21	&$\pm$0.03	&$\pm$0.03	&$\pm$0.02	&$\pm$0.04	&- &- & \\
79925	&109.3120545	&37.8423928	&$> 30.5$			&28.43		&28.37		&28.32		&27.92	&$>$26.60		&$>$26.60 		& 1.87 \\
	&		&		&-		&$\pm$0.06	&$\pm$0.08	&$\pm$0.07	&$\pm$0.08	&- &- & \\\hline
\hline                                   
\end{tabular}
\small

All error bars are computed from noise measured in 0.4'' diameter apertures distributed over each object. The last column displays the $\chi^2_{opt}$, all objects above the solid line have a $\chi^2_{opt}$ consistent with a non-detection in optical.
\end{table*}

   \begin{figure*}
   \centering
           \includegraphics[width=15cm]{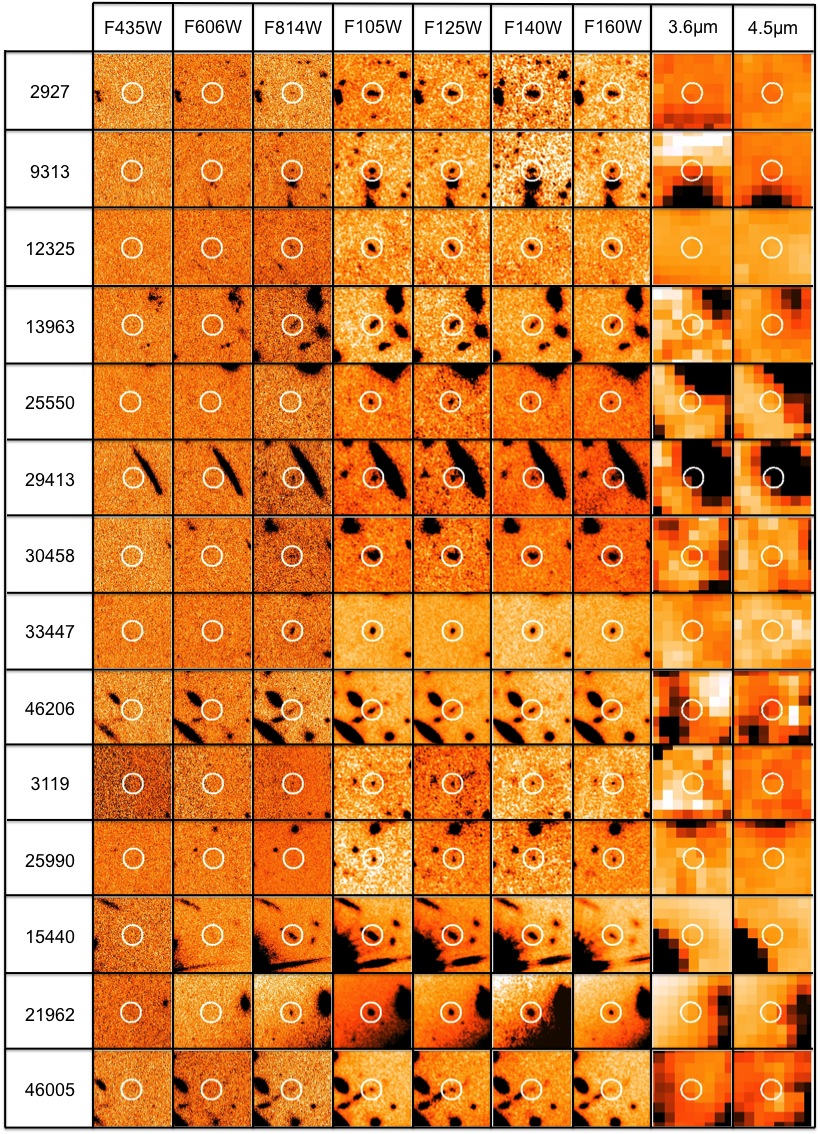}
      \caption{\label{stamps_cluster} Thumbnail images of $z\sim$7 candidates selected behind the cluster field. Each stamp is 5''$\times$5'' and the position of each candidate is displayed by a white 0.6'' radius circle.}
   \end{figure*}
 
   \begin{figure*}
   \centering
           \includegraphics[width=15cm]{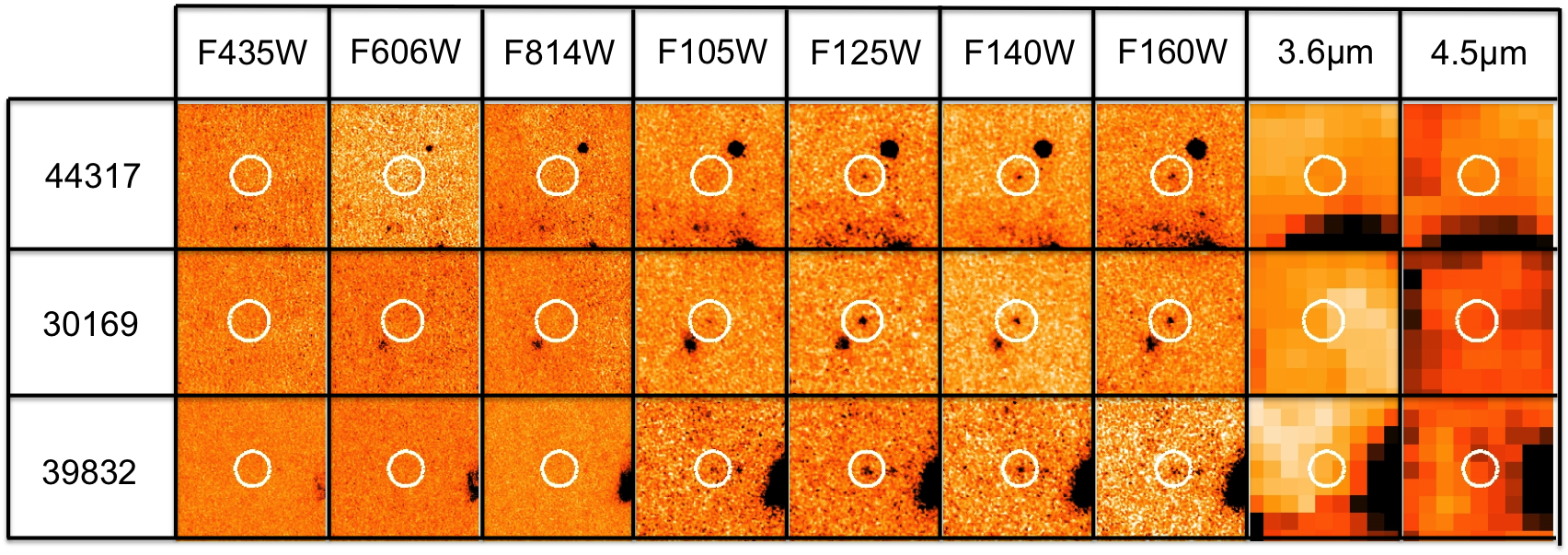}
      \caption{\label{stamps_cluster} Thumbnail images of $z\sim$8 candidates selected on the parallel field. Each stamp is 5''$\times$5'' and the position of each candidate is displayed by a white 0.6'' radius circle.}
   \end{figure*}

\begin{table*}
\caption{Photometric redshift computed in two ways: $\chi^2$ minimization and Bayesian approach.}
\label{photo_z}      
\begin{center}                          
\begin{tabular}{l | c c | c c | c  }        
\hline\hline                 
ID 	& $z_{\textit{Hyperz}}$ & 1$\sigma$ &  $z_{BPZ}$ & 1$\sigma$  & $\mu$ \\   
        	 	&	  & 	  & 	 	&  & \\    \hline
2927		& 6.38 & [5.88: 6.62] & 6.40 & [6.31: 6.45] & 2.01 $\pm$ 0.25	\\
9313		& 6.31 & [6.10: 6.57] & 6.42 & [6.34: 6.48] & 2.22 $\pm$ 0.60	\\
12325	& 6.74 & [6.50: 6.96] & 6.69 & [6.60: 6.74] & 6.68 $\pm$ 1.75	\\
13963$^{*}$		& 6.53 & [6.32: 6.85] & 6.65 & [6.59: 6.73] & 2.51 $\pm$ 0.84	\\
25550	& 6.95 & [5.22: 7.47] & 6.89 & [6.77: 7.09] & 5.17 $\pm$ 1.28	\\
29413	& 6.38 & [6.14: 6.55] & 6.44 & [6.35: 6.53] & 3.51 $\pm$ 0.89	\\
30458	& 6.53 & [6.35: 6.77] & 6.68 & [6.59: 6.72] & 3.22 $\pm$ 1.00	\\
33447$^{\star}$	& 6.44 & [6.26: 6.57] & 6.51 & [6.45: 6.53] & 6.97 $\pm$ 2.16	\\
46206	& 6.74 & [6.59: 6.95] &  6.80 & [6.75: 6.87]  & 3.11 $\pm$ 0.85	\\
3119		& 6.26 & [5.41: 6.77] &  6.34 & [6.03: 6.53]  & 1.85 $\pm$ 0.28	 \\
25990	& 6.47 & [5.57: 7.13] & 6.47 & [6.27: 6.71]& 4.17 $\pm$ 1.09	\\
15440	& 5.69 & [5.64: 5.75] & 5.73 & [5.67: 5.75] & 18.7 $\pm$ 7.5 \\
21962$^{\star}$	& 5.98 & [5.89: 6.07] & 6.10 & [6.07: 6.14] & 27.9 $\pm$ 8.8 \\
46005	& 6.52 & [6.24: 6.73] & 6.53 & [6.44: 6.59] & 3.06 $\pm$ 0.84	\\ \hline
44317	& 2.11 & [1.69: 2.27] & 7.45 & [7.10: 7.68] & \textit{1.1}\\
30169	& 8.02 & [7.82: 8.15] & 7.94 & [7.87: 8.00] & \textit{1.1} \\
39832	& 8.09 & [6.89: 8.47] & 7.82 & [7.38: 8.05] & \textit{1.1}\\
2035		& 7.11 & [6.84: 7.29] & 7.25 & [7.10: 7.33]  & \textit{1.1}	\\ 
6576		& 6.88 & [3.02: 7.46] & 6.66 & [6.53: 6.97] & \textit{1.1}	 \\
10738	& 6.26 & [6.00: 6.54] & 7.05 & [6.97: 7.09] & \textit{1.1}	 \\
24830	& 6.48 & [5.42: 7.00] & 6.93 & [6.78: 7.10] & \textit{1.1}	  \\
26762	& 6.29 & [6.00: 6.56] & 6.28 & [6.18: 6.38] & \textit{1.1}	 \\
32445	& 6.41 & [6.25: 6.61] & 6.85 & [6.76: 6.91] & \textit{1.1}	 \\
33421	& 7.05 & [6.75: 7.15] & 6.70 & [6.60: 6.77] & \textit{1.1}	 \\
37890	& 6.21 & [5.91: 6.59] & 6.49 & [6.43: 6.54] & \textit{1.1}	  \\
39809	& 7.51 & [7.24: 7.64] & 6.89 & [6.85: 6.97] & \textit{1.1}	 \\
42718	& 6.67 & [6.47: 6.93] & 6.28 & [6.18: 6.41] & \textit{1.1}	 \\	
43555	& 6.23 & [5.74: 6.68] & 7.43 & [7.34: 7.48] & \textit{1.1}	 \\
46175	& 7.09 & [6.94: 7.23] & 6.77 & [6.75: 6.85] & \textit{1.1}	 \\
46719	& 6.36 & [5.53: 6.57] & 6.84 & [6.78: 6.94] & \textit{1.1}	\\
50815	& 0.89 & [0.00: 1.79] & 6.45 & [6.39: 6.52] & \textit{1.1}	 \\
58730	& 7.14 & [6.86: 7.18] & 0.79 & [0.67: 0.93] & \textit{1.1}	 \\
66722	& 6.04 & [5.66: 6.49] & 7.02 & [6.80: 7.05] & \textit{1.1}	 \\
91692	& 7.27 & [4.15: 7.57] & 6.03 & [5.86: 6.24] & \textit{1.1}	 \\
7406		& 6.35 & [6.31: 6.43] & 7.13 & [6.70: 7.24] & \textit{1.1}	 \\
17548	& 6.11 & [5.95: 6.25] & 6.52 & [6.44: 6.56] & \textit{1.1}	 \\
28313	& 6.22 & [6.05: 6.57] & 6.20 & [6.13: 6.25] & \textit{1.1}	 \\
49274	& 6.98 & [6.47: 7.23] & 6.45 & [6.32: 6.54] & \textit{1.1}	 \\
70084	& 4.67 & [4.58: 4.72] & 6.83 & [6.69: 6.93] & \textit{1.1}	 \\ 
 \hline                        
\hline                                   
\end{tabular}
\end{center}
Columns: (1) Object ID, (2) (3) photometric redshift from \textit{Hyperz} with the corresponding 1$\sigma$ error, (4) (5) photometric redshift from \textit{BPZ} with the corresponding 1$\sigma$ error, (6) amplification (for objects selected in parallel field, we set the amplification at $\mu$=1.1 \\
$\star$: objects confirmed by spectroscopy at $z$=6.39 in \citet{2014ApJ...783L..12V} \\
$*$: object confirmed by HST-spectroscopy at $z$=6.51 in \citet{Schmidt15} \\
\end{table*}


\begin{table*}
\centering
\scriptsize
\caption{Physical properties of candidates selected on MACS0717 data.}
\label{properties_macs0417}
\begin{tabular}{lcccccc}
\hline\hline
Galaxy ID & Redshift & $\rm log M_*$   	& log SFR  	& AGE		&Av		& Cluster \\
          &          & $\rm [M_{\odot}]$  & $\rm [M_{\odot} yr^{-1}]$  &$\rm [Gyr]$    &   &  \\
\hline
\hline
                2035 &   7.05 &   8.5$\rm ^{+  0.2 }_{-  0.3 } $ &   0.5$\rm ^{+  0.1 }_{-  0.1 } $ &   0.2$\rm ^{+  0.3 }_{-  0.1 } $ &   0.1$\rm ^{+  0.1 }_{-  0.1 } $ &  M0717 \\
                6576 &   6.93 &   8.8$\rm ^{+  0.3 }_{-  0.3 } $ &   0.6$\rm ^{+  0.2 }_{-  0.2 } $ &   0.4$\rm ^{+  0.3 }_{-  0.3 } $ &   0.6$\rm ^{+  0.5 }_{-  0.3 } $ &  M0717 \\
               10738 &   6.28 &   8.2$\rm ^{+  0.3 }_{-  0.4 } $ &   0.1$\rm ^{+  0.1 }_{-  0.1 } $ &   0.2$\rm ^{+  0.3 }_{-  0.2 } $ &   0.1$\rm ^{+  0.1 }_{-  0.1 } $ &  M0717 \\
               24830 &   6.85 &   8.6$\rm ^{+  0.1 }_{-  0.1 } $ &   1.1$\rm ^{+  0.3 }_{-  0.3 } $ &   0.1$\rm ^{+  0.2 }_{-  0.1 } $ &   1.2$\rm ^{+  0.4 }_{-  0.4 } $ &  M0717 \\
               26762 &   6.70 &   9.2$\rm ^{+  0.1 }_{-  0.2 } $ &   0.4$\rm ^{+  0.3 }_{-  2.1 } $ &   0.4$\rm ^{+  0.3 }_{-  0.3 } $ &   0.8$\rm ^{+  0.6 }_{-  0.6 } $ &  M0717 \\
               32445 &   6.49 &   8.4$\rm ^{+  0.3 }_{-  0.3 } $ &   0.5$\rm ^{+  0.1 }_{-  0.1 } $ &   0.2$\rm ^{+  0.3 }_{-  0.1 } $ &   0.1$\rm ^{+  0.1 }_{-  0.1 } $ &  M0717 \\
               33421 &   6.89 &   8.7$\rm ^{+  0.1 }_{-  0.1 } $ &   0.5$\rm ^{+  0.1 }_{-  0.1 } $ &   0.3$\rm ^{+  0.3 }_{-  0.2 } $ &   0.1$\rm ^{+  0.1 }_{-  0.1 } $ &  M0717 \\
               37890 &   6.28 &   8.3$\rm ^{+  0.3 }_{-  0.4 } $ &   0.2$\rm ^{+  0.2 }_{-  0.1 } $ &   0.3$\rm ^{+  0.3 }_{-  0.2 } $ &   0.1$\rm ^{+  0.2 }_{-  0.1 } $ &  M0717 \\
               39809 &   7.43 &   8.7$\rm ^{+  0.2 }_{-  0.2 } $ &   0.6$\rm ^{+  0.1 }_{-  0.1 } $ &   0.3$\rm ^{+  0.2 }_{-  0.2 } $ &   0.1$\rm ^{+  0.1 }_{-  0.1 } $ &  M0717 \\
               42718 &   6.77 &   9.2$\rm ^{+  0.1 }_{-  0.1 } $ &  -2.8$\rm ^{+  1.4 }_{-  2.9 } $ &   0.1$\rm ^{+  0.1 }_{-  0.1 } $ &   0.2$\rm ^{+  0.3 }_{-  0.1 } $ &  M0717 \\
               43555 &   6.60 &   9.5$\rm ^{+  0.1 }_{-  0.1 } $ &  -1.7$\rm ^{+  0.1 }_{-  0.1 } $ &   0.2$\rm ^{+  0.1 }_{-  0.1 } $ &   0.2$\rm ^{+  0.1 }_{-  0.1 } $ &  M0717 \\
               46175 &   6.84 &   8.6$\rm ^{+  0.2 }_{-  0.1 } $ &   0.4$\rm ^{+  0.4 }_{-  0.2 } $ &   0.3$\rm ^{+  0.3 }_{-  0.2 } $ &   0.2$\rm ^{+  1.0 }_{-  0.1 } $ &  M0717 \\
               46719 &   6.45 &   9.3$\rm ^{+  0.1 }_{-  0.1 } $ &  -1.9$\rm ^{+  0.1 }_{-  0.2 } $ &   0.3$\rm ^{+  0.1 }_{-  0.1 } $ &   0.1$\rm ^{+  0.3 }_{-  0.1 } $ &  M0717 \\
               50815 &   7.79 &   8.3$\rm ^{+  0.4 }_{-  0.5 } $ &   0.5$\rm ^{+  0.4 }_{-  0.1 } $ &   0.1$\rm ^{+  0.3 }_{-  0.1 } $ &   0.2$\rm ^{+  0.2 }_{-  0.1 } $ &  M0717 \\
               58730 &   7.02 &   8.9$\rm ^{+  0.1 }_{-  0.1 } $ &   1.0$\rm ^{+  0.1 }_{-  0.1 } $ &   0.2$\rm ^{+  0.1 }_{-  0.1 } $ &   0.1$\rm ^{+  0.1 }_{-  0.1 } $ &  M0717 \\
               66722 &   6.03 &   8.1$\rm ^{+  0.3 }_{-  0.4 } $ &  -0.1$\rm ^{+  0.2 }_{-  0.1 } $ &   0.4$\rm ^{+  0.3 }_{-  0.3 } $ &   0.2$\rm ^{+  0.3 }_{-  0.1 } $ &  M0717 \\
               91692 &   7.13 &   8.5$\rm ^{+  0.3 }_{-  0.3 } $ &   0.3$\rm ^{+  0.2 }_{-  0.2 } $ &   0.3$\rm ^{+  0.3 }_{-  0.2 } $ &   0.3$\rm ^{+  0.5 }_{-  0.2 } $ &  M0717 \\
                7406 &   6.52 &   8.4$\rm ^{+  0.4 }_{-  0.2 } $ &   1.1$\rm ^{+  0.3 }_{-  0.2 } $ &   0.1$\rm ^{+  0.2 }_{-  0.1 } $ &   0.2$\rm ^{+  0.1 }_{-  0.1 } $ &  M0717 \\
               17548 &   6.20 &   8.6$\rm ^{+  0.2 }_{-  0.3 } $ &   0.5$\rm ^{+  0.1 }_{-  0.1 } $ &   0.3$\rm ^{+  0.3 }_{-  0.2 } $ &   0.1$\rm ^{+  0.1 }_{-  0.1 } $ &  M0717 \\
               28313 &   6.45 &   8.4$\rm ^{+  0.2 }_{-  0.4 } $ &   0.4$\rm ^{+  0.1 }_{-  0.1 } $ &   0.2$\rm ^{+  0.3 }_{-  0.2 } $ &   0.1$\rm ^{+  0.1 }_{-  0.1 } $ &  M0717 \\
               49274 &   6.83 &   8.3$\rm ^{+  0.3 }_{-  0.3 } $ &   0.2$\rm ^{+  0.1 }_{-  0.1 } $ &   0.3$\rm ^{+  0.3 }_{-  0.2 } $ &   0.1$\rm ^{+  0.2 }_{-  0.1 } $ &  M0717 \\
               70084 &   7.41 &   9.0$\rm ^{+  0.3 }_{-  0.4 } $ &   1.2$\rm ^{+  0.3 }_{-  0.2 } $ &   0.2$\rm ^{+  0.3 }_{-  0.1 } $ &   0.7$\rm ^{+  0.3 }_{-  0.2 } $ &  M0717 \\
               44317 &   7.45 &   8.7$\rm ^{+  0.3 }_{-  0.3 } $ &   0.4$\rm ^{+  0.3 }_{-  0.2 } $ &   0.4$\rm ^{+  0.2 }_{-  0.2 } $ &   0.6$\rm ^{+  0.5 }_{-  0.3 } $ &  M0717 \\
               30169 &   7.94 &   8.6$\rm ^{+  0.2 }_{-  0.3 } $ &   0.5$\rm ^{+  0.2 }_{-  0.1 } $ &   0.3$\rm ^{+  0.2 }_{-  0.2 } $ &   0.2$\rm ^{+  0.4 }_{-  0.1 } $ &  M0717 \\
               39832 &   7.82 &   8.4$\rm ^{+  0.3 }_{-  0.3 } $ &   0.2$\rm ^{+  0.3 }_{-  0.2 } $ &   0.3$\rm ^{+  0.2 }_{-  0.2 } $ &   0.5$\rm ^{+  0.5 }_{-  0.3 } $ &  M0717 \\
                2927 &   6.40 &   8.8$\rm ^{+  0.1 }_{-  0.1 } $ &   0.5$\rm ^{+  0.1 }_{-  0.1 } $ &   0.3$\rm ^{+  0.3 }_{-  0.2 } $ &   0.4$\rm ^{+  0.3 }_{-  0.2 } $ &  M0717 \\
                9313 &   6.42 &   8.4$\rm ^{+  0.1 }_{-  0.1 } $ &   0.3$\rm ^{+  0.2 }_{-  0.2 } $ &   0.3$\rm ^{+  0.3 }_{-  0.2 } $ &   0.1$\rm ^{+  0.1 }_{-  0.1 } $ &  M0717 \\
               12325 &   6.69 &   7.9$\rm ^{+  0.5 }_{-  0.4 } $ &  -0.1$\rm ^{+  0.7 }_{-  0.7 } $ &   0.2$\rm ^{+  0.3 }_{-  0.2 } $ &   0.1$\rm ^{+  0.1 }_{-  0.1 } $ &  M0717 \\
               13963 &   6.65 &   8.4$\rm ^{+  0.1 }_{-  0.1 } $ &   0.3$\rm ^{+  0.2 }_{-  0.3 } $ &   0.3$\rm ^{+  0.3 }_{-  0.2 } $ &   0.1$\rm ^{+  0.1 }_{-  0.1 } $ &  M0717 \\
               25550 &   6.89 &   8.0$\rm ^{+  0.4 }_{-  0.3 } $ &  -0.2$\rm ^{+  0.4 }_{-  0.5 } $ &   0.3$\rm ^{+  0.3 }_{-  0.2 } $ &   0.3$\rm ^{+  0.5 }_{-  0.2 } $ &  M0717 \\
               29413 &   6.44 &   8.5$\rm ^{+  0.3 }_{-  0.3 } $ &   0.3$\rm ^{+  0.3 }_{-  0.4 } $ &   0.3$\rm ^{+  0.3 }_{-  0.2 } $ &   0.3$\rm ^{+  0.3 }_{-  0.1 } $ &  M0717 \\
               30458 &   6.68 &   8.4$\rm ^{+  0.3 }_{-  0.2 } $ &   0.4$\rm ^{+  0.4 }_{-  0.4 } $ &   0.2$\rm ^{+  0.3 }_{-  0.2 } $ &   0.1$\rm ^{+  0.1 }_{-  0.1 } $ &  M0717 \\
               33447 &   6.51 &   8.0$\rm ^{+  0.6 }_{-  0.5 } $ &   0.1$\rm ^{+  0.7 }_{-  0.7 } $ &   0.2$\rm ^{+  0.3 }_{-  0.1 } $ &   0.1$\rm ^{+  0.1 }_{-  0.1 } $ &  M0717 \\
               46206 &   6.80 &   8.3$\rm ^{+  0.3 }_{-  0.3 } $ &   0.6$\rm ^{+  0.4 }_{-  0.4 } $ &   0.1$\rm ^{+  0.1 }_{-  0.1 } $ &   0.1$\rm ^{+  0.1 }_{-  0.1 } $ &  M0717 \\
                3119 &   6.34 &   8.5$\rm ^{+  0.1 }_{-  0.2 } $ &   0.2$\rm ^{+  0.1 }_{-  0.1 } $ &   0.4$\rm ^{+  0.3 }_{-  0.3 } $ &   0.5$\rm ^{+  0.5 }_{-  0.3 } $ &  M0717 \\
               25990 &   6.47 &   8.1$\rm ^{+  0.3 }_{-  0.2 } $ &  -0.2$\rm ^{+  0.3 }_{-  0.4 } $ &   0.4$\rm ^{+  0.3 }_{-  0.2 } $ &   0.4$\rm ^{+  0.5 }_{-  0.3 } $ &  M0717 \\
               15440 &   5.73 &   7.6$\rm ^{+  1.3 }_{-  1.3 } $ &  -1.7$\rm ^{+  1.3 }_{-  1.3 } $ &   0.1$\rm ^{+  0.1 }_{-  0.1 } $ &   0.1$\rm ^{+  0.1 }_{-  0.1 } $ &  M0717 \\
               21962 &   6.10 &   7.0$\rm ^{+  1.2 }_{-  1.3 } $ &  -0.8$\rm ^{+  1.4 }_{-  1.4 } $ &   0.1$\rm ^{+  0.1 }_{-  0.1 } $ &   0.1$\rm ^{+  0.1 }_{-  0.1 } $ &  M0717 \\
               46005 &   6.53 &   8.2$\rm ^{+  0.2 }_{-  0.1 } $ &   0.1$\rm ^{+  0.3 }_{-  0.4 } $ &   0.3$\rm ^{+  0.3 }_{-  0.2 } $ &   0.1$\rm ^{+  0.1 }_{-  0.1 } $ &  M0717 \\
\hline\hline                 
\end{tabular}

 The following quantities are reported: ID (1), photo-$z$ (2), $\log$($M_{^{\star}}$) (3),  $\log$(SFR) (4), Age (5), A$_v$ (6) and FF (7). \\
 All values are corrected by their magnification factor and errors are shown at 1sigma confidence level.
\end{table*}

\begin{table*}
\centering
\scriptsize
\caption{Physical properties of all candidates selected in the first three Frontier Fields}
{\centering
\label{table_pp}
\begin{tabular}{lccccccl}
\hline\hline
Galaxy ID & Redshift & $\rm log M_*$   	& log SFR  	& AGE		&Av		& Cluster  & Reference\\
          &          & $\rm [M_{\odot}]$  & $\rm [M_{\odot} yr^{-1}]$  &$\rm [Gyr]$    &   &   &   \\
\hline
\hline
                 YD1 &   8.64 &   8.8$\rm ^{+  0.3 }_{-  0.2 } $ &   0.7$\rm ^{+  0.2 }_{-  0.1 } $ &   0.3$\rm ^{+  0.2 }_{-  0.2 } $ &   0.6$\rm ^{+  0.5 }_{-  0.3 } $ &  A2744 & \citet{2014ApJ...795...93Z}  \\
                 YD2 &   8.26 &   8.6$\rm ^{+  0.2 }_{-  0.2 } $ &   0.4$\rm ^{+  0.1 }_{-  0.1 } $ &   0.3$\rm ^{+  0.2 }_{-  0.2 } $ &   0.5$\rm ^{+  0.5 }_{-  0.3 } $ &  A2744 & \citet{2014ApJ...795...93Z}  \\
                 YD3 &   8.56 &   8.3$\rm ^{+  0.1 }_{-  0.1 } $ &   0.1$\rm ^{+  0.1 }_{-  0.2 } $ &   0.3$\rm ^{+  0.2 }_{-  0.2 } $ &   0.6$\rm ^{+  0.5 }_{-  0.3 } $ &  A2744 & \citet{2014ApJ...795...93Z} \\
                 YD4 &   8.51 &   9.4$\rm ^{+  0.1 }_{-  0.2 } $ &   1.2$\rm ^{+  0.1 }_{-  0.1 } $ &   0.3$\rm ^{+  0.2 }_{-  0.2 } $ &   0.6$\rm ^{+  0.4 }_{-  0.3 } $ &  A2744 & \citet{2014ApJ...795...93Z} \\
                 YD5 &   8.45 &   8.4$\rm ^{+  0.1 }_{-  0.1 } $ &   0.3$\rm ^{+  0.1 }_{-  0.2 } $ &   0.3$\rm ^{+  0.2 }_{-  0.2 } $ &   0.6$\rm ^{+  0.5 }_{-  0.3 } $ &  A2744 & \citet{2014ApJ...795...93Z} \\
                 YD6 &   8.25 &   9.5$\rm ^{+  0.2 }_{-  0.2 } $ &   1.2$\rm ^{+  0.3 }_{-  0.2 } $ &   0.4$\rm ^{+  0.2 }_{-  0.2 } $ &   0.9$\rm ^{+  0.5 }_{-  0.4 } $ &  A2744 & \citet{2014ApJ...795...93Z} \\
                 YD7 &   8.24 &   9.1$\rm ^{+  0.1 }_{-  0.1 } $ &   1.1$\rm ^{+  0.1 }_{-  0.1 } $ &   0.2$\rm ^{+  0.2 }_{-  0.2 } $ &   0.3$\rm ^{+  0.2 }_{-  0.1 } $ &  A2744 & \citet{2014ApJ...795...93Z} \\
                 YD8 &   8.15 &   8.8$\rm ^{+  0.1 }_{-  0.1 } $ &   0.8$\rm ^{+  0.1 }_{-  0.1 } $ &   0.2$\rm ^{+  0.2 }_{-  0.2 } $ &   0.3$\rm ^{+  0.3 }_{-  0.1 } $ &  A2744 & \citet{2014ApJ...795...93Z} \\
                 YD9 &   8.26 &   8.4$\rm ^{+  0.2 }_{-  0.2 } $ &   0.3$\rm ^{+  0.1 }_{-  0.1 } $ &   0.3$\rm ^{+  0.2 }_{-  0.2 } $ &   0.5$\rm ^{+  0.5 }_{-  0.3 } $ &  A2744 & \citet{2014ApJ...795...93Z} \\
                YD10 &   8.25 &   8.7$\rm ^{+  0.2 }_{-  0.2 } $ &   0.6$\rm ^{+  0.1 }_{-  0.1 } $ &   0.3$\rm ^{+  0.2 }_{-  0.2 } $ &   0.5$\rm ^{+  0.5 }_{-  0.3 } $ &  A2744 & \citet{2014ApJ...795...93Z} \\
                YD11 &   8.27 &   8.0$\rm ^{+  0.2 }_{-  0.2 } $ &  -0.1$\rm ^{+  0.2 }_{-  0.3 } $ &   0.3$\rm ^{+  0.2 }_{-  0.2 } $ &   0.7$\rm ^{+  0.5 }_{-  0.3 } $ &  A2744 & \citet{2014ApJ...795...93Z} \\
                 ZD1 &   8.67 &   8.6$\rm ^{+  0.3 }_{-  0.2 } $ &   0.5$\rm ^{+  0.2 }_{-  0.1 } $ &   0.3$\rm ^{+  0.2 }_{-  0.2 } $ &   0.6$\rm ^{+  0.6 }_{-  0.3 } $ &  A2744 & \citet{2014ApJ...795...93Z} \\
                 ZD2 &   7.85 &   9.4$\rm ^{+  0.1 }_{-  0.2 } $ &   1.4$\rm ^{+  0.1 }_{-  0.1 } $ &   0.2$\rm ^{+  0.2 }_{-  0.1 } $ &   0.4$\rm ^{+  0.2 }_{-  0.1 } $ &  A2744 & \citet{2014ApJ...795...93Z} \\
                 ZD3 &   7.70 &   9.2$\rm ^{+  0.2 }_{-  0.2 } $ &   1.1$\rm ^{+  0.1 }_{-  0.1 } $ &   0.3$\rm ^{+  0.2 }_{-  0.2 } $ &   0.5$\rm ^{+  0.4 }_{-  0.2 } $ &  A2744 & \citet{2014ApJ...795...93Z} \\
                 ZD4 &   7.83 &   8.7$\rm ^{+  0.3 }_{-  0.3 } $ &   0.5$\rm ^{+  0.2 }_{-  0.1 } $ &   0.3$\rm ^{+  0.2 }_{-  0.2 } $ &   0.6$\rm ^{+  0.5 }_{-  0.3 } $ &  A2744 & \citet{2014ApJ...795...93Z} \\
                 ZD5 &   7.63 &   8.6$\rm ^{+  0.1 }_{-  0.2 } $ &   0.4$\rm ^{+  0.1 }_{-  0.1 } $ &   0.3$\rm ^{+  0.3 }_{-  0.2 } $ &   0.3$\rm ^{+  0.4 }_{-  0.2 } $ &  A2744 & \citet{2014ApJ...795...93Z} \\
                 ZD6 &   7.48 &   9.3$\rm ^{+  0.2 }_{-  0.3 } $ &   1.2$\rm ^{+  0.2 }_{-  0.1 } $ &   0.3$\rm ^{+  0.3 }_{-  0.2 } $ &   0.7$\rm ^{+  0.4 }_{-  0.2 } $ &  A2744 & \citet{2014ApJ...795...93Z} \\
                 ZD7 &   7.32 &   8.2$\rm ^{+  0.5 }_{-  0.4 } $ &   0.1$\rm ^{+  0.5 }_{-  0.6 } $ &   0.3$\rm ^{+  0.3 }_{-  0.2 } $ &   0.3$\rm ^{+  0.4 }_{-  0.1 } $ &  A2744 & \citet{2014ApJ...795...93Z} \\
                 ZD8 &   7.52 &   7.5$\rm ^{+  0.8 }_{-  0.8 } $ &  -0.7$\rm ^{+  0.8 }_{-  0.9 } $ &   0.3$\rm ^{+  0.2 }_{-  0.2 } $ &   0.5$\rm ^{+  0.5 }_{-  0.3 } $ &  A2744 & \citet{2014ApJ...795...93Z} \\
                 ZD9 &   6.93 &   8.4$\rm ^{+  0.3 }_{-  0.2 } $ &   0.4$\rm ^{+  0.3 }_{-  0.4 } $ &   0.2$\rm ^{+  0.3 }_{-  0.2 } $ &   0.3$\rm ^{+  0.2 }_{-  0.1 } $ &  A2744 & \citet{2014ApJ...795...93Z} \\
                ZD10 &   6.86 &   7.5$\rm ^{+  0.6 }_{-  0.6 } $ &  -0.8$\rm ^{+  0.7 }_{-  0.7 } $ &   0.4$\rm ^{+  0.3 }_{-  0.3 } $ &   0.5$\rm ^{+  0.5 }_{-  0.3 } $ &  A2744 & \citet{2014ApJ...795...93Z} \\
                ZD11 &   6.96 &   7.8$\rm ^{+  0.4 }_{-  0.3 } $ &  -0.2$\rm ^{+  0.5 }_{-  0.6 } $ &   0.2$\rm ^{+  0.3 }_{-  0.2 } $ &   0.1$\rm ^{+  0.1 }_{-  0.1 } $ &  A2744 & \citet{2014ApJ...795...93Z} \\
            HFF1P-i1 &   7.29 &   8.8$\rm ^{+  0.3 }_{-  0.3 } $ &   0.8$\rm ^{+  0.1 }_{-  0.1 } $ &   0.2$\rm ^{+  0.3 }_{-  0.2 } $ &   0.1$\rm ^{+  0.1 }_{-  0.1 } $ &  A2744 & \citet{2015ApJ...804..103K}\\
            HFF1P-i2 &   6.36 &   8.6$\rm ^{+  0.3 }_{-  0.4 } $ &   0.5$\rm ^{+  0.1 }_{-  0.1 } $ &   0.2$\rm ^{+  0.3 }_{-  0.2 } $ &   0.1$\rm ^{+  0.1 }_{-  0.1 } $ &  A2744 & \citet{2015ApJ...804..103K}\\
            HFF1P-i3 &   6.16 &   9.0$\rm ^{+  0.2 }_{-  0.3 } $ &   0.8$\rm ^{+  0.2 }_{-  0.2 } $ &   0.4$\rm ^{+  0.3 }_{-  0.2 } $ &   0.5$\rm ^{+  0.4 }_{-  0.2 } $ &  A2744 & \citet{2015ApJ...804..103K}\\
            HFF1P-i4 &   7.10 &   8.9$\rm ^{+  0.3 }_{-  0.3 } $ &   0.7$\rm ^{+  0.2 }_{-  0.2 } $ &   0.3$\rm ^{+  0.3 }_{-  0.2 } $ &   0.3$\rm ^{+  0.4 }_{-  0.2 } $ &  A2744 & \citet{2015ApJ...804..103K}\\
            HFF1P-i5 &   7.85 &   8.3$\rm ^{+  0.5 }_{-  0.3 } $ &   0.9$\rm ^{+  0.5 }_{-  0.2 } $ &   0.1$\rm ^{+  0.2 }_{-  0.1 } $ &   0.1$\rm ^{+  0.1 }_{-  0.1 } $ &  A2744 & \citet{2015ApJ...804..103K}\\
            HFF1P-i6 &   6.70 &   8.7$\rm ^{+  0.3 }_{-  0.4 } $ &   0.5$\rm ^{+  0.2 }_{-  0.1 } $ &   0.3$\rm ^{+  0.3 }_{-  0.2 } $ &   0.1$\rm ^{+  0.2 }_{-  0.1 } $ &  A2744 & \citet{2015ApJ...804..103K}\\
            HFF1P-i7 &   6.83 &   8.7$\rm ^{+  0.3 }_{-  0.4 } $ &   0.5$\rm ^{+  0.2 }_{-  0.1 } $ &   0.3$\rm ^{+  0.3 }_{-  0.2 } $ &   0.2$\rm ^{+  0.3 }_{-  0.1 } $ &  A2744 & \citet{2015ApJ...804..103K}\\
            HFF1P-i8 &   6.06 &   8.6$\rm ^{+  0.3 }_{-  0.4 } $ &   0.4$\rm ^{+  0.1 }_{-  0.1 } $ &   0.3$\rm ^{+  0.4 }_{-  0.2 } $ &   0.1$\rm ^{+  0.2 }_{-  0.1 } $ &  A2744 & \citet{2015ApJ...804..103K}\\
            HFF1P-i9 &   6.62 &   8.6$\rm ^{+  0.3 }_{-  0.4 } $ &   0.5$\rm ^{+  0.2 }_{-  0.1 } $ &   0.3$\rm ^{+  0.3 }_{-  0.2 } $ &   0.2$\rm ^{+  0.2 }_{-  0.1 } $ &  A2744 & \citet{2015ApJ...804..103K}\\
           HFF1P-i10 &   5.98 &   8.8$\rm ^{+  0.3 }_{-  0.4 } $ &   0.5$\rm ^{+  0.2 }_{-  0.2 } $ &   0.4$\rm ^{+  0.3 }_{-  0.3 } $ &   0.3$\rm ^{+  0.4 }_{-  0.2 } $ &  A2744 & \citet{2015ApJ...804..103K}\\
           HFF1P-i11 &   5.86 &   8.7$\rm ^{+  0.3 }_{-  0.4 } $ &   0.4$\rm ^{+  0.2 }_{-  0.2 } $ &   0.4$\rm ^{+  0.3 }_{-  0.3 } $ &   0.4$\rm ^{+  0.5 }_{-  0.2 } $ &  A2744 & \citet{2015ApJ...804..103K}\\
           HFF1P-i12 &   7.81 &   8.8$\rm ^{+  0.3 }_{-  0.5 } $ &   0.8$\rm ^{+  0.3 }_{-  0.2 } $ &   0.2$\rm ^{+  0.3 }_{-  0.2 } $ &   0.3$\rm ^{+  0.4 }_{-  0.2 } $ &  A2744 & \citet{2015ApJ...804..103K}\\
           HFF1P-i13 &   6.45 &   8.8$\rm ^{+  0.3 }_{-  0.4 } $ &   0.5$\rm ^{+  0.3 }_{-  0.2 } $ &   0.4$\rm ^{+  0.3 }_{-  0.3 } $ &   0.4$\rm ^{+  0.5 }_{-  0.2 } $ &  A2744 & \citet{2015ApJ...804..103K}\\
           HFF1P-i14 &   5.86 &   8.7$\rm ^{+  0.3 }_{-  0.4 } $ &   0.4$\rm ^{+  0.2 }_{-  0.2 } $ &   0.4$\rm ^{+  0.3 }_{-  0.3 } $ &   0.4$\rm ^{+  0.4 }_{-  0.2 } $ &  A2744 & \citet{2015ApJ...804..103K}\\
           HFF1P-i16 &   6.39 &   8.6$\rm ^{+  0.3 }_{-  0.4 } $ &   0.3$\rm ^{+  0.3 }_{-  0.2 } $ &   0.4$\rm ^{+  0.3 }_{-  0.3 } $ &   0.3$\rm ^{+  0.5 }_{-  0.2 } $ &  A2744 & \citet{2015ApJ...804..103K}\\
            HFF1P-Y1 &   7.53 &   9.0$\rm ^{+  0.2 }_{-  0.3 } $ &   0.9$\rm ^{+  0.2 }_{-  0.2 } $ &   0.3$\rm ^{+  0.2 }_{-  0.2 } $ &   0.3$\rm ^{+  0.4 }_{-  0.2 } $ &  A2744 & \citet{2015ApJ...804..103K}\\
            HFF1P-Y2 &   7.63 &   9.0$\rm ^{+  0.3 }_{-  0.4 } $ &   0.8$\rm ^{+  0.3 }_{-  0.2 } $ &   0.3$\rm ^{+  0.2 }_{-  0.2 } $ &   0.5$\rm ^{+  0.5 }_{-  0.3 } $ &  A2744 & \citet{2015ApJ...804..103K}\\
                8958 &   9.88 &   8.4$\rm ^{+  0.9 }_{-  1.0 } $ &  -0.9$\rm ^{+  0.2 }_{-  0.3 } $ &   0.4$\rm ^{+  0.1 }_{-  0.1 } $ &   1.0$\rm ^{+  0.8 }_{-  0.5 } $ &  M0416 & \citet{Infante2015}  \\
                1859 &   9.46 &   8.3$\rm ^{+  0.2 }_{-  0.2 } $ &   0.3$\rm ^{+  0.3 }_{-  0.4 } $ &   0.2$\rm ^{+  0.2 }_{-  0.2 } $ &   0.5$\rm ^{+  0.5 }_{-  0.3 } $ &  M0416 & \citet{Infante2015}  \\
                8364 &   9.17 &   9.1$\rm ^{+  0.1 }_{-  0.1 } $ &   1.0$\rm ^{+  0.1 }_{-  0.1 } $ &   0.2$\rm ^{+  0.2 }_{-  0.2 } $ &   0.4$\rm ^{+  0.4 }_{-  0.2 } $ &  M0416 & \citet{Infante2015}  \\
                 491 &   8.42 &   8.9$\rm ^{+  0.1 }_{-  0.1 } $ &   1.0$\rm ^{+  0.1 }_{-  0.1 } $ &   0.2$\rm ^{+  0.2 }_{-  0.1 } $ &   0.2$\rm ^{+  0.1 }_{-  0.1 } $ &  M0416 & \citet{Infante2015}  \\
                8428 &   8.29 &  11.0$\rm ^{+  0.2 }_{-  0.2 } $ &   1.7$\rm ^{+  0.2 }_{-  0.2 } $ &   0.6$\rm ^{+  0.1 }_{-  0.1 } $ &   2.5$\rm ^{+  0.1 }_{-  0.1 } $ &  M0416 & \citet{Infante2015}  \\
                1213 &   8.27 &   8.3$\rm ^{+  0.1 }_{-  0.1 } $ &   0.2$\rm ^{+  0.1 }_{-  0.1 } $ &   0.3$\rm ^{+  0.2 }_{-  0.2 } $ &   0.4$\rm ^{+  0.5 }_{-  0.2 } $ &  M0416 & \citet{Infante2015}  \\
                4008 &   7.71 &   8.3$\rm ^{+  0.1 }_{-  0.1 } $ &   0.2$\rm ^{+  0.1 }_{-  0.2 } $ &   0.3$\rm ^{+  0.2 }_{-  0.2 } $ &   0.2$\rm ^{+  0.3 }_{-  0.1 } $ &  M0416 & \citet{Infante2015}  \\
                3687 &   9.36 &   8.5$\rm ^{+  0.4 }_{-  0.4 } $ &   0.5$\rm ^{+  0.4 }_{-  0.3 } $ &   0.2$\rm ^{+  0.2 }_{-  0.2 } $ &   0.6$\rm ^{+  0.6 }_{-  0.3 } $ &  M0416 & \citet{Infante2015}  \\
                4177 &   9.34 &   8.7$\rm ^{+  0.2 }_{-  0.3 } $ &   0.7$\rm ^{+  0.2 }_{-  0.1 } $ &   0.2$\rm ^{+  0.2 }_{-  0.1 } $ &   0.2$\rm ^{+  0.2 }_{-  0.1 } $ &  M0416 & \citet{Infante2015}  \\
                3076 &   9.13 &   7.6$\rm ^{+  0.1 }_{-  0.1 } $ &   0.5$\rm ^{+  0.7 }_{-  0.1 } $ &   0.1$\rm ^{+  0.1 }_{-  0.1 } $ &   0.1$\rm ^{+  0.1 }_{-  0.1 } $ &  M0416 & \citet{Infante2015}  \\
                5296 &   8.28 &  11.0$\rm ^{+  0.1 }_{-  0.1 } $ &   2.6$\rm ^{+  0.1 }_{-  0.4 } $ &   0.4$\rm ^{+  0.2 }_{-  0.2 } $ &   1.2$\rm ^{+  0.5 }_{-  0.2 } $ &  M0416 & \citet{Infante2015}  \\
                1301 &   8.35 &   8.1$\rm ^{+  0.2 }_{-  0.3 } $ &   0.2$\rm ^{+  0.1 }_{-  0.1 } $ &   0.2$\rm ^{+  0.2 }_{-  0.1 } $ &   0.1$\rm ^{+  0.1 }_{-  0.1 } $ &  M0416 & \citet{Infante2015}  \\
                3814 &   7.88 &   8.8$\rm ^{+  0.4 }_{-  0.4 } $ &   0.6$\rm ^{+  0.3 }_{-  0.2 } $ &   0.3$\rm ^{+  0.2 }_{-  0.2 } $ &   0.7$\rm ^{+  0.5 }_{-  0.3 } $ &  M0416 & \citet{Infante2015}  \\
                1241 &   7.90 &   8.3$\rm ^{+  0.2 }_{-  0.3 } $ &   0.2$\rm ^{+  0.1 }_{-  0.1 } $ &   0.3$\rm ^{+  0.2 }_{-  0.2 } $ &   0.1$\rm ^{+  0.1 }_{-  0.1 } $ &  M0416 & \citet{Infante2015}  \\
                3790 &   7.79 &   8.6$\rm ^{+  0.2 }_{-  0.2 } $ &   0.8$\rm ^{+  0.1 }_{-  0.1 } $ &   0.1$\rm ^{+  0.2 }_{-  0.1 } $ &   0.1$\rm ^{+  0.1 }_{-  0.1 } $ &  M0416 & \citet{Infante2015}  \\
                4125 &   7.40 &   7.3$\rm ^{+  0.1 }_{-  0.1 } $ &   0.8$\rm ^{+  0.1 }_{-  0.1 } $ &   0.1$\rm ^{+  0.1 }_{-  0.1 } $ &   0.1$\rm ^{+  0.1 }_{-  0.1 } $ &  M0416 & \citet{Infante2015}  \\
                6999 &   7.54 &   8.6$\rm ^{+  0.2 }_{-  0.3 } $ &   0.5$\rm ^{+  0.1 }_{-  0.1 } $ &   0.3$\rm ^{+  0.2 }_{-  0.2 } $ &   0.1$\rm ^{+  0.1 }_{-  0.1 } $ &  M0416 & \citet{Infante2015}  \\
                7361 &   7.52 &   7.4$\rm ^{+  0.1 }_{-  0.1 } $ &   0.3$\rm ^{+  0.4 }_{-  0.1 } $ &   0.1$\rm ^{+  0.1 }_{-  0.1 } $ &   0.1$\rm ^{+  0.1 }_{-  0.1 } $ &  M0416 & \citet{Infante2015}  \\
                1331 &   7.25 &   7.8$\rm ^{+  0.2 }_{-  0.2 } $ &   0.6$\rm ^{+  0.4 }_{-  0.1 } $ &   0.1$\rm ^{+  0.1 }_{-  0.1 } $ &   0.1$\rm ^{+  0.1 }_{-  0.1 } $ &  M0416 & \citet{Infante2015}  \\
                1386 &   7.24 &   8.2$\rm ^{+  0.2 }_{-  0.3 } $ &   0.3$\rm ^{+  0.1 }_{-  0.1 } $ &   0.2$\rm ^{+  0.3 }_{-  0.1 } $ &   0.1$\rm ^{+  0.1 }_{-  0.1 } $ &  M0416 & \citet{Infante2015}  \\
                 146 &   7.25 &   8.3$\rm ^{+  0.2 }_{-  0.3 } $ &   0.3$\rm ^{+  0.1 }_{-  0.1 } $ &   0.2$\rm ^{+  0.2 }_{-  0.1 } $ &   0.1$\rm ^{+  0.1 }_{-  0.1 } $ &  M0416 & \citet{Infante2015}  \\
                1513 &   6.66 &  10.2$\rm ^{+  0.1 }_{-  0.1 } $ &  -2.1$\rm ^{+  2.1 }_{-  2.6 } $ &   0.8$\rm ^{+  0.1 }_{-  0.2 } $ &   1.2$\rm ^{+  0.2 }_{-  0.2 } $ &  M0416 & \citet{Infante2015}  \\
\hline\hline                 
\end{tabular}}

Same as in table \ref{properties_macs0417}
\end{table*}

\begin{table*}
\caption{ \label{LF_densities} Number densities of $z$$>$6.5 objects computed using half of the Frontier Fields survey}
\centering                          
\begin{tabular}{|c | c c ||c | c c || c | c c |    }        
\hline\hline                 
$<$z$>$ 	& M$_{1500}$ 	& $\Phi(M_{1500})$ & $<$z$>$ 	& M$_{1500}$ 	& $\Phi(M_{1500})$ & $<$z$>$ 	& M$_{1500}$ 	& $\Phi(M_{1500})$\\   
		&			& $\times$10$^{-4}$ Mpc$^{-3}$ & 	&			& $\times$10$^{-4}$ Mpc$^{-3}$ & 	&			& $\times$10$^{-4}$ Mpc$^{-3}$\\ \hline
\multirow{8}*{7} & -21.75$\pm$0.50 & $<$0.27 & \multirow{8}*{8} 	& 			 	&				& \multirow{8}*{9}	& 			 &	\\
			& -20.75$\pm$0.50 & 0.64$\pm$0.53 &			& -21.25$\pm$0.50	 & $<$ 0.30		&				& 	& \\
			& -20.00$\pm$0.25 & 2.85$\pm$1.40 &			& -20.50$\pm$0.25 & 1.09$\pm$0.70	&				& -20.75$\pm$0.50 	& $<$0.33	\\	
			& -19.50$\pm$0.25 & 4.21$\pm$1.85 &			& -19.75$\pm$0.50 & 2.58$\pm$1.37	&				&  -19.75$\pm$0.50	& 0.57$\pm$0.46 \\
			& -19.00$\pm$0.25 & 7.06$\pm$2.38&			& -18.75$\pm$0.50 & 7.25$\pm$4.04	&				& -18.75$\pm$0.50	& 2.27$\pm$1.82\\
			& -18.50$\pm$0.25 & 13.7$\pm$7.56&			& -17.75$\pm$0.50 & 14.7$\pm$13.9	&				& -17.75$\pm$0.50	& 16.6$\pm$13.3\\
			& -18.00$\pm$0.25 & 22.7$\pm$12.4&			& -16.75$\pm$0.50 & 67.7$\pm$65.9	&				&  	& \\
			&-17.25$\pm$0.50 & 47.4$\pm$41.8&	&	& & & &\\
 \hline                        
\hline                                   
\end{tabular}
\end{table*}


\begin{table*}
\caption{\label{LF_parameters}  Parameterization of the UV Luminosity Function}
\centering                          
\begin{tabular}{|c | c c c c |    }        
\hline\hline                 
$<$z$>$ 		&Reference 					& M$^{\star}$				& $\Phi^{\star}$ 						& $\alpha$ \\  \hline 
\multirow{4}*{7} & This work					& -20.33$_{-0.47}^{+0.37}$	& 0.37$^{+0.12}_{-0.11}$$\times$10$^{-3}$	& -1.91$^{+0.26}_{-0.27}$ \\
			 & \citet{2015ApJ...803...34B}		& -20.87$\pm$0.26			& 0.29$^{+0.21}_{-0.12}$$\times$10$^{-3}$	&-2.06$\pm$0.13 \\
			 & \citet{2013MNRAS.432.2696M}	& -19.90$_{-0.28}^{+0.23}$	& 1.09$^{+0.56}_{-0.45}$$\times$10$^{-3}$	& -1.90$^{+0.22}_{-0.23}$ \\
			 & \citet{2014MNRAS.440.2810B}	& -20.3					& 0.36$\times$10$^{-3}$					& -2.1 \\ \hline
\multirow{4}*{8} & This work					& -20.32$_{-0.26}^{+0.49}$	& 0.30$^{+0.85}_{-0.19}$$\times$10$^{-3}$	& -1.95$^{+0.43}_{-0.40}$ \\
			 & \citet{2015ApJ...803...34B}		& -20.63$\pm$0.36			& 0.21$^{+0.23}_{-0.11}$$\times$10$^{-3}$	&-2.02$\pm$0.23 \\
			 & \citet{2013MNRAS.432.2696M}	& -20.12$_{-0.48}^{+0.37}$	& 0.45$^{+0.35}_{-0.29}$$\times$10$^{-3}$	& -2.02$^{+0.22}_{-0.47}$ \\
			 & \citet{2012ApJ...760..108B}		& -20.26$_{-0.34}^{+0.29}$	& 0.43$^{+0.35}_{-0.23}$$\times$10$^{-3}$	& -1.98$^{+0.23}_{-0.22}$ \\ \hline
\multirow{3}*{9} & This work					& -20.45 (\textit{fixed})		& 0.70$^{+0.30}_{-0.30}$$\times$10$^{-4}$	& -2.17$^{+0.41}_{-0.43}$ \\
			 & \citet{2014arXiv1412.1472M}	& -20.1					& 2.51$^{+1.46}_{-1.39}$$\times$10$^{-4}$	& -2.02( \textit{fixed})\\
			 & \citet{2015arXiv150601035B}	& -20.45					& 1.0$\times$10$^{-4}$					& -2.3 \\
 \hline                        
\hline                                   
\end{tabular}
\end{table*}

\bibliographystyle{aa}
   \bibliography{M0717_arxiv.bib} 

\end{document}